\documentclass[a4paper,11pt]{article} 
\pdfoutput=1 
\usepackage{jinstpub} 

\usepackage{caption}
\usepackage{subcaption}
\usepackage{mathrsfs}
\newcommand{\plotpath}{plots} 
\graphicspath{{figures/}{/plots}} 

\usepackage{xspace}

\newcommand {\ie}{\mbox{i.e.}\xspace}     
\newcommand {\eg}{\mbox{e.g.}\xspace}     

\title{Beam tests of an integrated prototype of the ATLAS Forward Proton detector}

\author[a,1]{J.~Lange\note{Corresponding author.}} 
\author[b]{L.~Adamczyk}
\author[c]{G.~Avoni}
\author[d]{E.~Banas}
\author[e]{A.~Brandt}
\author[c]{M.~Bruschi}
\author[d,f]{P.~Buglewicz}
\author[a]{E.~Cavallaro}
\author[g]{D.~Caforio}
\author[h]{G.~Chiodini}
\author[i]{L.~Chytka}
\author[d,f]{K.~Cie\'{s}la}
\author[j]{P.M.~Davis}
\author[b]{M.~Dyndal}
\author[a,k]{S.~Grinstein}
\author[b]{K.~Janas}
\author[i]{K.~Jirakova}
\author[l]{M.~Kocian}
\author[d]{K.~Korcyl}
\author[a]{I.~Lopez Paz}
\author[m]{D.~Northacker}
\author[i]{L.~Nozka}
\author[m,n]{M.~Rijssenbeek}
\author[o]{L.~Seabra}
\author[d]{R.~Staszewski}
\author[d,f]{P.~\'{S}wierska  }
\author[p]{T.~Sykora}

\affiliation[a]{Institut de F\'{i}sica d'Altes Energies (IFAE), The Barcelona Institute of Science and Technology (BIST), 08193 Bellaterra (Barcelona), Spain}

\affiliation[b]{AGH University of Science and Technology, Faculty of Physics and Applied Computer Science, al. Mickiewicza 30, PL-30059 Cracow, Poland}

\affiliation[c]{INFN Bologna and Universit\`{a}  di Bologna, Dipartimento di Fisica, viale C. Berti Pichat, 6/2, IT - 40127 Bologna, Italy}

\affiliation[d]{H. Niewodniczanski Institute of Nuclear Physics PAN, Cracow, Poland}

\affiliation[e]{The University of Texas at Arlington, Department of Physics, Box 19059, Arlington, TX 76019, United States of America}
\affiliation[f]{Cracow University of Technology, Cracow, Poland}

\affiliation[g]{Czech Technical University in Prague, Zikova 4, CZ - 166 35 Praha 6, Czech Republic}
\affiliation[h]{INFN Lecce and Universit\`{a}  del Salento, Dipartimento di Fisica, Via Arnesano IT - 73100 Lecce, Italy}
\affiliation[i]{Palack\'{y} University, SLO\&RCPTM, Olomouc, Czech Republic}
\affiliation[j]{Centre for Particle Physics, Department of Physics, University of Alberta, Edmonton, AB T6G 2G7, Canada}

\affiliation[k]{ICREA, Pg. Llu\'{i}s Companys 23, 08010 Barcelona, Spain}

\affiliation[l]{Stanford Linear Accelerator Center, Stanford, California 94309, United States of America}
\affiliation[m]{Stony Brook University, Dept. of Physics and Astronomy, Nicolls Road, Stony Brook, NY 11794-3800, United States of America}
\affiliation[n]{European Laboratory for Particle Physics (CERN), Geneva, Switzerland}
\affiliation[o]{Laborat\'{o}rio de Instrumenta\c{c}\~{a}o e F\'{i}sica Experimental de Part\'{i}culas, LIP, Av. Elias Garcia 14, Lisbon, Portugal}

\affiliation[p]{Charles University in Prague, Faculty of Mathematics and Physics, Institute of Particle and Nuclear Physics, V Holesovickach 2, CZ - 18000 Praha 8, Czech Republic}

\emailAdd{joern.lange@cern.ch} 

\abstract{ 

The ATLAS Forward Proton (AFP) detector is intended to measure protons scattered at small angles from the ATLAS interaction point. To this end, a combination of 3D Silicon pixel tracking modules and Quartz-Cherenkov time-of-flight (ToF) detectors is installed 210\,m away from the interaction point at both sides of ATLAS. Beam tests with an AFP prototype detector combining tracking and timing sub-detectors and a common readout have been performed at the CERN-SPS test-beam facility in November 2014 and September 2015 to complete the system integration and to study the detector performance. The successful tracking-timing integration was demonstrated. Good tracker hit efficiencies above 99.9\% at a sensor tilt of 14$^{\circ}$, as foreseen for AFP, were observed. Spatial resolutions in the short pixel direction with 50\,$\mu$m pitch of $5.5 \pm 0.5\,\mu$m per pixel plane and of $2.8 \pm 0.5\,\mu$m for the full four-plane tracker at 14$^{\circ}$ were found, largely surpassing the AFP requirement of 10\,$\mu$m. The timing detector showed also good hit efficiencies above 99\%, and a full-system time resolution of $35\pm6$\,ps was found for the ToF prototype detector with two Quartz bars in-line (half the final AFP size) without dedicated optimisation, fulfilling the requirements for initial low-luminosity AFP runs.

}


\begin{document}

\maketitle 
\flushbottom 

\section{Introduction}
\label{sec:intro}

The ATLAS collaboration~\cite{bib:ATLAS} at the Large Hadron Collider (LHC) at the European Laboratory for Particle Physics (CERN) is installing the ATLAS Forward Proton (AFP) detector to measure very forward protons (p) scattered at small angles from the ATLAS interaction point (IP)~\cite{bib:AFPreference1}. To this end, a combination of high-resolution pixel tracking modules for fractional-energy loss and momentum measurements and fast time-of-flight (ToF) detectors for event pile-up removal is placed at about 210\,m from the IP at both sides of ATLAS and only 2--3\,mm away from the outgoing Large Hadron Collider (LHC) beam. Roman pots are used as the beam interface. The approved AFP scenario foresees an initial low-luminosity operation with a low pile-up (number of interactions per bunch crossing $\mu\lesssim 1$) during short dedicated LHC runs. At a later stage, the system might be also operated at standard LHC luminosities during a large part of the regular LHC runs if a safe operation under these conditions has been demonstrated. The installation is performed in two stages: an AFP tracking system at one side (``one-arm'') of the IP was already installed during the end-of-year 2015-2016 shutdown. The full two-arm system with both tracking and timing detectors at both sides of the IP is planned to be completed during the extended end-of-year 2016-2017 shutdown.

Parts of the individual AFP detector components and sub-systems have been tested in the past. 
However, it is critical to demonstrate that the separate detector components can be operated together as an integrated system. 
To this end, a first unified AFP prototype has been developed, which combines tracking and ToF prototype detectors and a common trigger and readout (excluding the Roman-pot housing at this stage). To verify its operability and measure its performance, beam tests have been carried out at the CERN-Super Proton Synchrotron (SPS) with 120\,GeV pions in November 2014 and September 2015. 

In section~\ref{sec:AFPdetector} a short overview on the design of the AFP detector, its components and readout is given, as well as a description of the AFP prototype and the beam-test setup. Section~\ref{sec:operation} describes the operation during beam tests including calibration, triggering and data taking. The beam-test results including the measured performance of the tracking and ToF detectors are presented in section~\ref{sec:performance}. Summary and conclusions are given in section~\ref{sec:conclusions}.

\section{The AFP detector and the beam-test prototype}
\label{sec:AFPdetector}

\begin{figure}[hbtp]
	\centering
	\includegraphics[width=10cm]{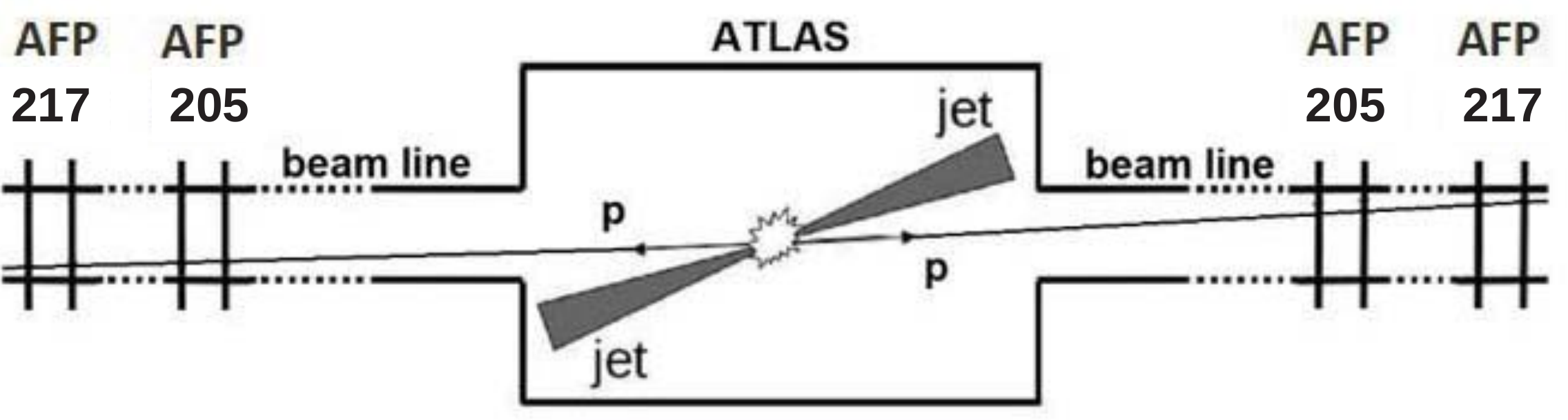}
	 \includegraphics[width=10cm]{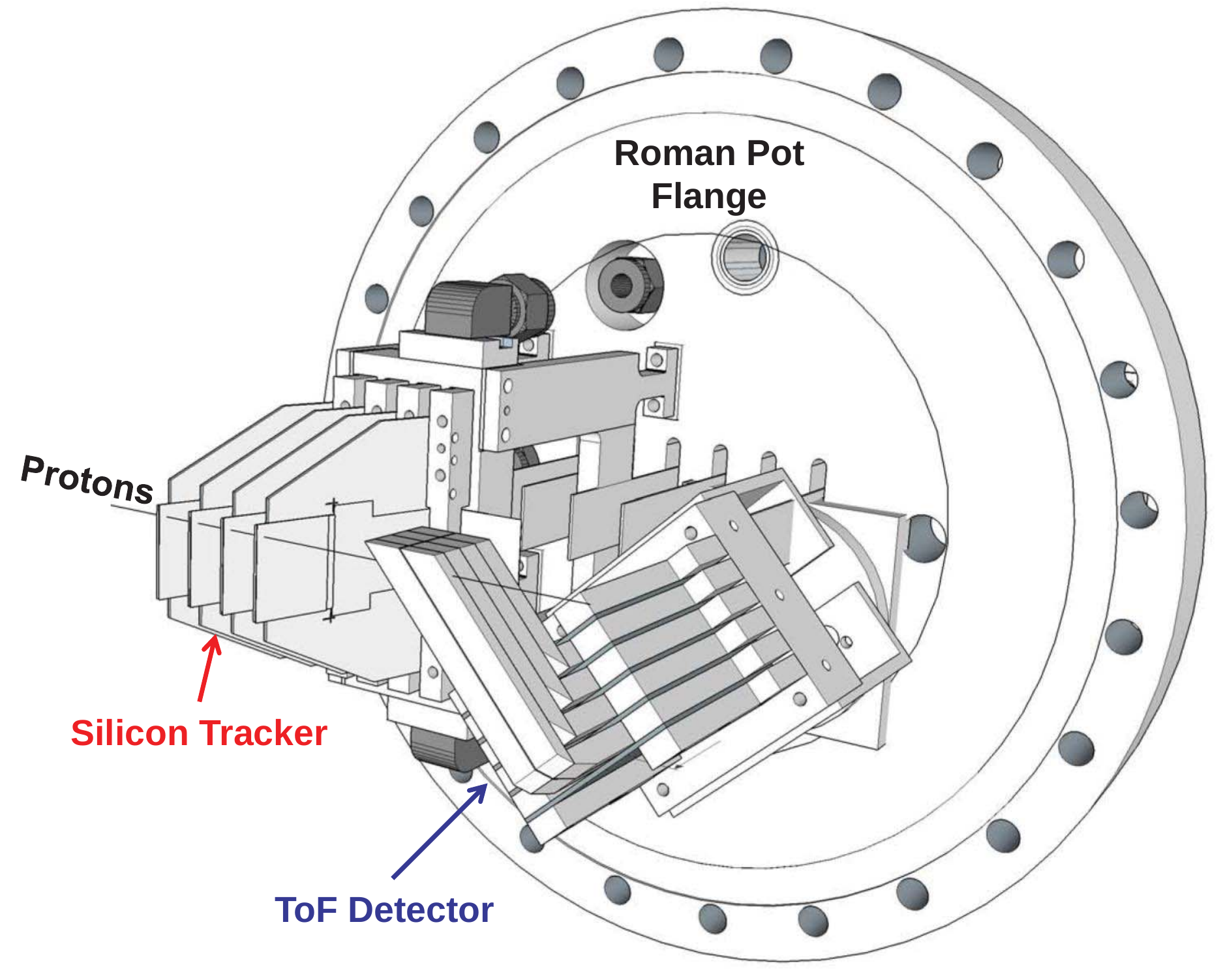}
	\caption{Top: Layout of the AFP stations at both sides of the ATLAS IP at 205 and 217\,m. Bottom: Design of the 217\,m AFP detector including tracking and time-of-flight systems (only two LQbars per train are shown as used in the beam test; the final version will comprise four LQbars per train).}
	\label{fig:AFPsketch}
\end{figure}


\begin{figure}[hbtp]
	\centering
	 \includegraphics[width=15cm]{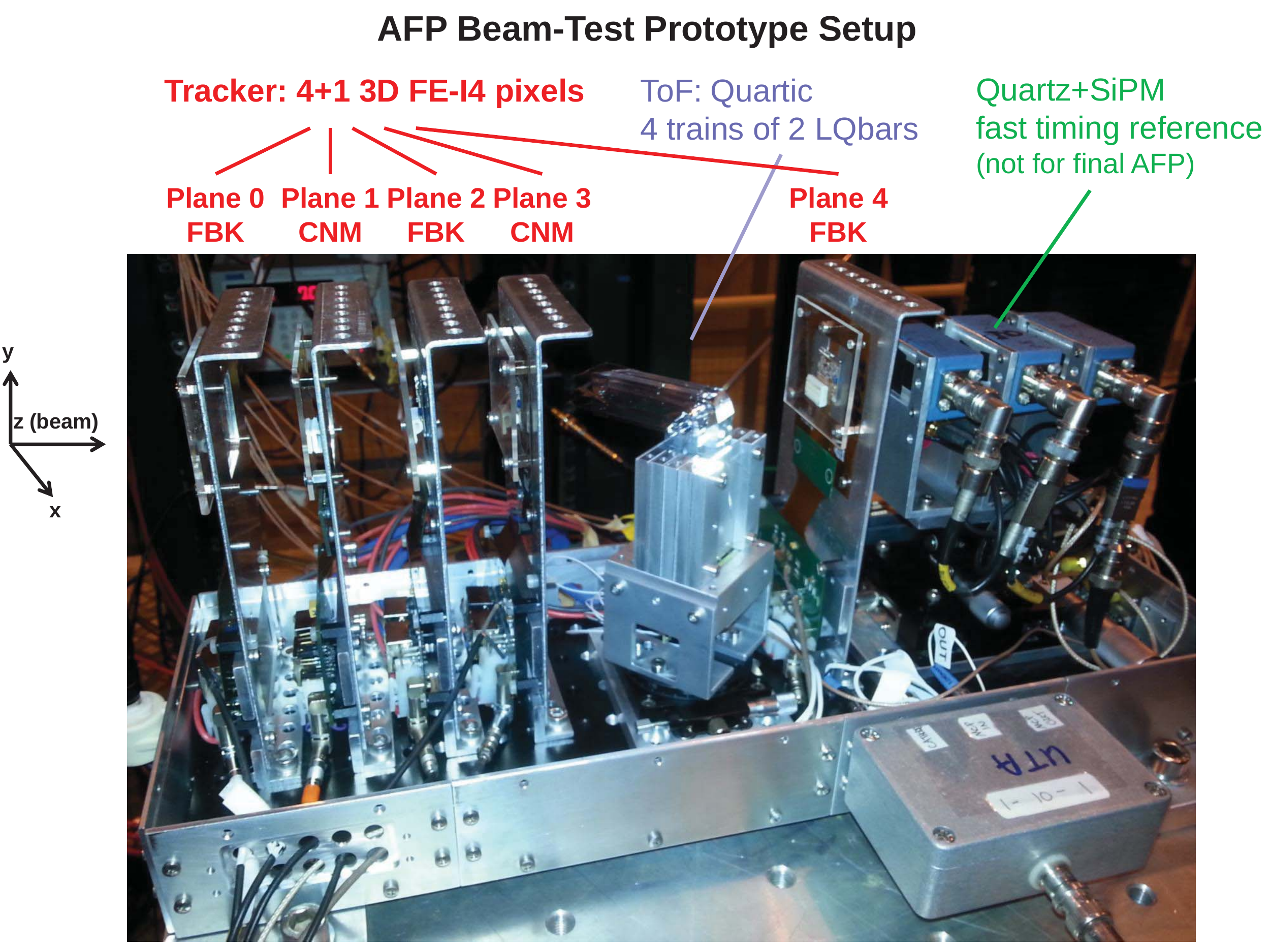}
	\caption{The integrated AFP beam-test prototype with tracking planes at perpendicular incidence.}
	\label{fig:AFPbeamTestSetup}
\end{figure}

In this section, the design of the final AFP detector is briefly described, as well as the AFP beam-test prototype and setup. More details of the final design can be found in the AFP Technical Design Report~\cite{bib:AFPreference1}. 

The final AFP detector will consist of two Roman-pot stations at each side of the ATLAS IP, at 205 and 217\,m away from it, as sketched in figure~\ref{fig:AFPsketch} (top). The station closer to the IP will include a tracker of four pixel planes. The station further away (shown in figure~\ref{fig:AFPsketch}, bottom) will comprise an identical tracker and in addition a time-of-flight (ToF) detector of four trains of four L-shaped Cherenkov-radiating Quartz bars (LQbars) each (only two LQbars per train are shown in the sketch). 

The AFP prototype for the beam tests is shown in  figure~\ref{fig:AFPbeamTestSetup}. It was designed to be similar to the final AFP layout, but exhibited some differences explained in the following. It was built of five pixel planes and a ToF system of four trains of two LQbars each (\ie half of the final AFP ToF system). 

Note that the coordinate system during these beam tests differs from the standard AFP convention used in Ref.~\cite{bib:AFPreference1}. In the beam tests, the short pixel direction with 50\,$\mu$m pitch was oriented along the $y$ direction and the long pixel direction with 250\,$\mu$m pitch in the $x$ direction (see figures~\ref{fig:AFPbeamTestSetup} and~\ref{fig:PixelModule}). The beam axis was in $z$ direction. The coordinate origin was placed in the centre of tracker plane 0. This coordinate system will be used in this paper unless noted otherwise. 

\subsection{Tracking system}
\subsubsection{AFP design and requirements}
\label{sec:tracker}

The purpose of the AFP tracker is the measurement of the position and angle of the scattered protons, which in combination with the LHC magnet system between the IP and AFP will allow the determination of their fractional energy loss and momentum. The tracker is required to exhibit a position resolution of 10\,$\mu$m per four-plane station in the direction horizontal to the LHC tunnel floor and 30\,$\mu$m in the vertical one. Furthermore, its proximity to the beam implies the need for slim edges of about 100--200\,$\mu$m to minimise dead material, and it has to be able to withstand a highly non-uniform irradiation (with expected maximum fluence levels of about $3\times10^{15}\,n_{eq}$/cm$^2$ for 100\,fb$^{-1}$ integrated luminosity if AFP is operated in its measurement position close to the beam). Each AFP tracking station comprises four pixel modules, each made of a 3D silicon pixel sensor interconnected to an FE-I4 front-end chip~\cite{bib:FEI4}. The modules are placed with a pitch of 9\,mm as shown in figure~\ref{fig:AFPsketch} (bottom). The pixels have a size of 50 and $250\,\mu$m in the horizontal and vertical directions, respectively. In the horizontal (short pixel) direction, the tracker is oriented with a small tilt of 14$^{\circ}$ between the sensor normal and the beam to enhance the hit efficiency and horizontal resolution, whereas in the vertical (long pixel) direction, a staggering of about $60\,\mu$m between successive planes is planned to improve the vertical resolution.

Silicon pixel sensors based on the 3D technology~\cite{bib:3D}, in which the electrodes penetrate the sensor bulk as columns perpendicular to the surface, are chosen for AFP due to an excellent radiation hardness together with a low depletion voltage and their maturity proven by successful production runs for the ATLAS Insertable B-Layer (IBL)~\cite{bib:IBL3Dprod}. The vendors include FBK (Fondazione Bruno Kessler, Trento, Italy)~\cite{bib:FBKIBLProduction} and CNM (Centro Nacional de Microelectronica, Barcelona, Spain)~\cite{bib:CNMIBLProduction}. The sensors of both vendors are produced on a 230\,$\mu$m thick p-type substrate, but FBK sensors have 3D columns fully passing through, whereas CNM 3D columns stop about 20\,$\mu$m before reaching the other side. For edge termination, FBK uses a 3D guard fence of ohmic columns, whereas CNM in addition implements a 3D guard ring. For the first stage of AFP, CNM produced already a dedicated run of 3D sensors with 180\,$\mu$m slim edge~\cite{bib:LangeVertex}.

The read-out is performed by the FE-I4 front-end chip (version FE-I4B) with $336\times 80$ pixels with a pixel size of $50\times 250\,\mu$m$^2$, comprising a total active area of $1.68\times2.00$\,cm$^2$. It operates with a clock at 40\,MHz consistent with the nominal LHC bunch crossing rate. The chip contains pre-amplifiers and a discriminator for each pixel with adjustable signal threshold (typically in the range of 1.5--3\,ke$^{-}$) and time-over-threshold (ToT). The ToT is recorded with a resolution of 4 bits in units of clock cycles (25\,ns) and is related to the measured charge. The tuning of threshold and ToT as well as the calibration of the ToT-to-charge relation is performed with a charge-injection circuit using an injection capacitor and an adjustable voltage step pulse. The FE-I4 chip provides a so-called HitOr output signal if at least one pixel fires, which is the logical OR of the discriminator signals of all pixels and can be used for triggering.

During the IBL development and qualification, 3D FE-I4 pixel modules have demonstrated a hit efficiency above 97\% after uniform proton irradiation to $5\times10^{15}~n_{eq}$/cm$^2$~\cite{bib:IBLprototypes}. The efficiency was observed to be about 1\% higher in case of a sensor tilt of about 15$^{\circ}$ as low-field and dead regions from the 3D columns lose their impact when a particle does not traverse perpendicularly. The spatial resolution depends on a number of different parameters such as the beam incidence angle in combination with the sensor thickness (determining the degree of charge sharing between neighbouring pixels), the operational parameters (voltage, tuning points), the charge or ToT resolution and the cluster-centre algorithm. For perpendicular incidence, the spatial resolution was found to be 12\,$\mu$m in the short pixel direction for the 98\% of the events with pixel-cluster size 1 and 2 (degrading to 15\,$\mu$m including all cluster sizes due to delta rays)~\cite{bib:Animma2015}. In the long direction, the overall resolution was measured to be 73\,$\mu$m. As demonstrated in the AFP beam tests of this study, for sensors tilted by 14$^{\circ}$ with respect to the short pixel direction, as planned for AFP, the resolution in that direction improves to about 6\,$\mu$m for cluster size 1 and 2 due to enhanced charge sharing and interpolation (see section~\ref{sec:resolution}). Furthermore, for a full AFP station of four pixel modules, an improvement of position resolution over the single module is expected as discussed further in section~\ref{sec:resolution}.

In view of the application of 3D pixel modules in AFP, it was verified in dedicated studies that slim edges with a remaining insensitive width of only 15--200\,$\mu$m can be produced without affecting the current, noise and edge efficiency and that the modules can withstand a highly non-uniform irradiation up to fluences expected after running for 100\,fb$^{-1}$ at standard LHC luminosity~\cite{bib:AFP3D1,bib:AFP3D2}. 

\subsubsection{Prototype tracking system}
\label{sec:prototypeTracking}

\begin{figure}[hbt]
	\centering
	 \includegraphics[width=15cm]{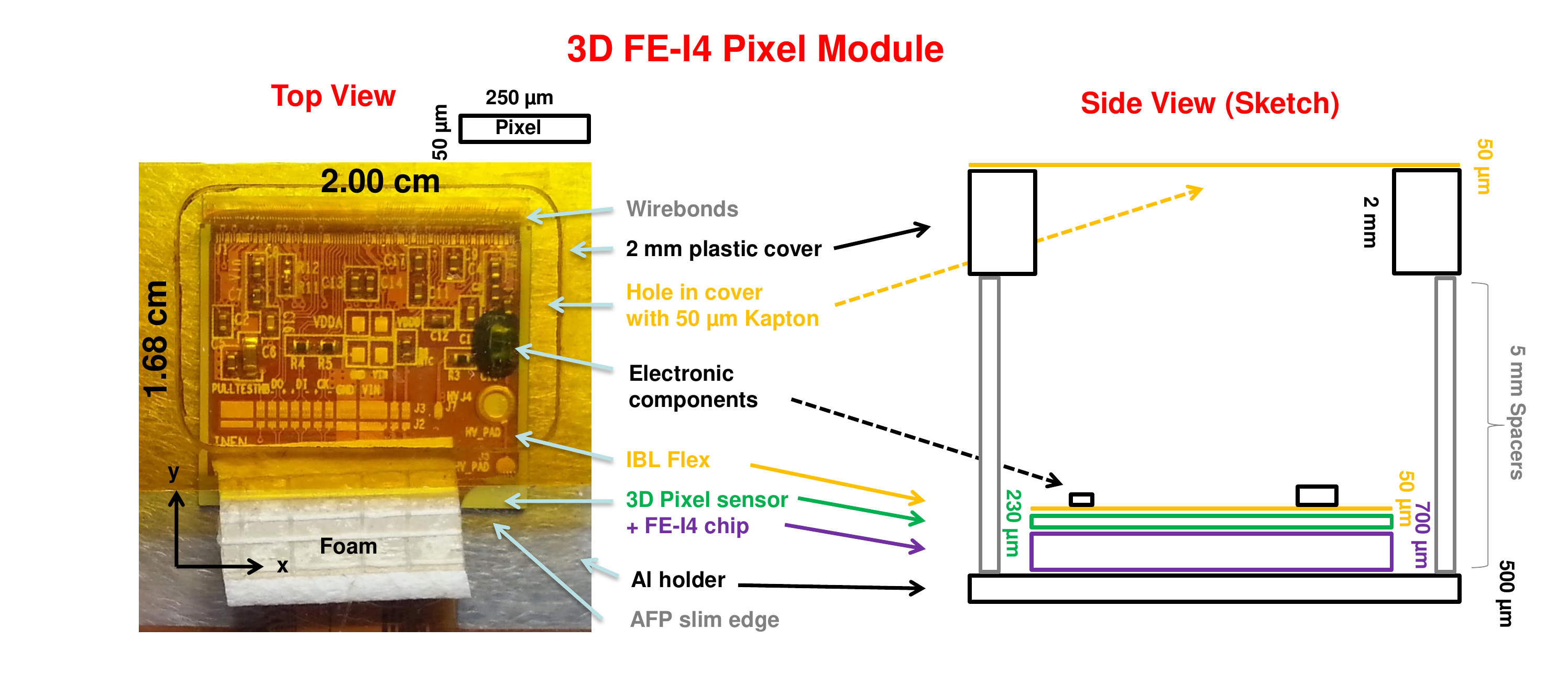}
	\caption{The 3D FE-I4 pixel-module prototype in top view (left) and a side-view sketch (right).}
	\label{fig:PixelModule}
\end{figure}

For the AFP beam tests, the pixel modules used consisted of spare 3D sensors from the IBL production (three by FBK, two by CNM) bump-bonded to the FE-I4B readout chip and assembled on an IBL-type flexible circuit board (flex) as shown in figure~\ref{fig:PixelModule}. All sensors except plane 3 had already AFP-compatible slim edges. To protect the modules, especially the wirebonds, a 2\,mm thick plastic cover was used with a distance of 5\,mm to the sensor. In the 2014 beam test, the plastic cover extended over the full sensor area, whereas in 2015 holes covered with Kapton tape were made over the upper part of the sensor to minimise material in front of the sensor (see section~\ref{sec:material}). The modules were mounted on Aluminium frames with a thickness of 0.5\,mm under the sensor. The frames were placed on a base plate in two different configurations:

\begin{enumerate}

	\item Facing the beam under normal incidence (0$^{\circ}$ between the sensor normal and the beam axis). Because of the easy mounting of the pixel modules, this was the standard beam-test configuration for the AFP integration tests and measuring the performance of the ToF detector (see figure~\ref{fig:AFPbeamTestSetup}). Four modules (number 0 to 3) were placed with a pitch of 3.75\,cm in front of the timing system (similarly to the final AFP configuration). An additional module (number~4) was placed behind the timing system with a distance of 13.75\,cm to module 3 to improve the reconstructed-track precision at the position of the ToF detector and to allow a monitoring of particle interactions in the Quartz material. 
	
	\item With a tilt of 14$^{\circ}$ between the sensor normal and the beam axis with respect to the short pixel direction ($y$ coordinate in the beam test). This configuration was used to study the tracker performance under more realistic AFP conditions. In this case, all five planes were placed with an equidistant pitch of 5\,cm. The ToF system was not included.

\end{enumerate}

\subsection{Time-of-flight system}
\subsubsection{AFP design and requirements}


In the final two-arm AFP detector, the ToF system is designed to reject combinatoric background from pile-up by precisely measuring the arrival times of the two protons to determine if they both come from the primary vertex as identified by the central tracker.
For the initial low-luminosity runs with a low pile-up of $\mu\lesssim1$, a time resolution of about 30\,ps is required, whereas for runs at standard LHC luminosity with more than 50 pile-up events 10\,ps are envisaged in order to give a primary-vertex constraint of 2\,mm for sufficient background rejection.

As timing detectors in the 217\,m stations, a system based on a set of L-shaped Cherenkov-radiating Quartz or fused silica bars (LQbar) is foreseen~\cite{bib:AFPreference1}, a Roman-pot-compatible modification of the original Quartic detector~\cite{Albrow:2008pn,bib:Albrow2012,bib:ToFOverviewPinfold}. The baseline consists of 16 LQbars organised into four rows (called trains) of four LQbars each (see also figure~\ref{fig:ToFprototype} for the half-size prototype version). The Cherenkov radiator bars are oriented with the Cherenkov angle of 48$^\circ$ with respect to the beam axis. Thus, the created Cherenkov light passes along the long direction of the radiators to the 90$^\circ$ bend, where the light is reflected so that it continues along the perpendicular light-guide bars to a multi-pixel Micro-Channel-Plate Photomultiplier (MCP-PMT). The output signals of all MCP-PMT channels are amplified with a low-noise preamplifier (PreAmp) and discriminated using Constant Fraction Discriminators (CFD). The CFDs apply two types of thresholds: a fixed threshold that determines the minimum level for the signal amplitude to be accepted; and a fractional threshold, set to a certain fraction of the signal amplitude, which determines the starting and end time of the CFD output signal, thereby compensating time-walk effects. Finally, the time of the rising edge of the CFD output is digitised with a 12-channel High-Precision Time-to-Digital Converter (HPTDC) board,  
which includes three HPTDC chips~\cite{bib:ToFOverviewPinfold,bib:Rijssenbeek2014,HPTDCspecs}. The finest time bin of the HPTDC chip is 24.4\,ps. Laboratory tests indicate an intrinsic HPTDC time resolution of about 13\,ps.

Previous beam tests were primarily focused on the performance of straight bar (Qbar) detectors with various bar configurations and dimensions. Single-Qbar + MCP-PMT combinations were measured with a 6\,GHz LeCroy oscilloscope (without the HPTDC) and determined to have a resolution of about 20\,ps for bar heights ranging from 2 to 5\,mm.
Multiple-Qbar configurations showed that each additional bar in the train improves the measurement despite non-trivial correlations between the bars. For example at a previous beam test, a six-Qbar train was measured to have a resolution of about 14\,ps including the HPTDC resolution, closing in on the 10\,ps target. The morphing of the straight bar into the LQbar was studied and shows that the LQbar-based detector could match or even exceed the Qbar detector with careful optimisation~\cite{bib:qsim}.

\subsubsection{Prototype time-of-flight system}
\label{sec:prototypeToF}

\begin{figure}[hbt]
	\centering
	 \includegraphics[width=8cm]{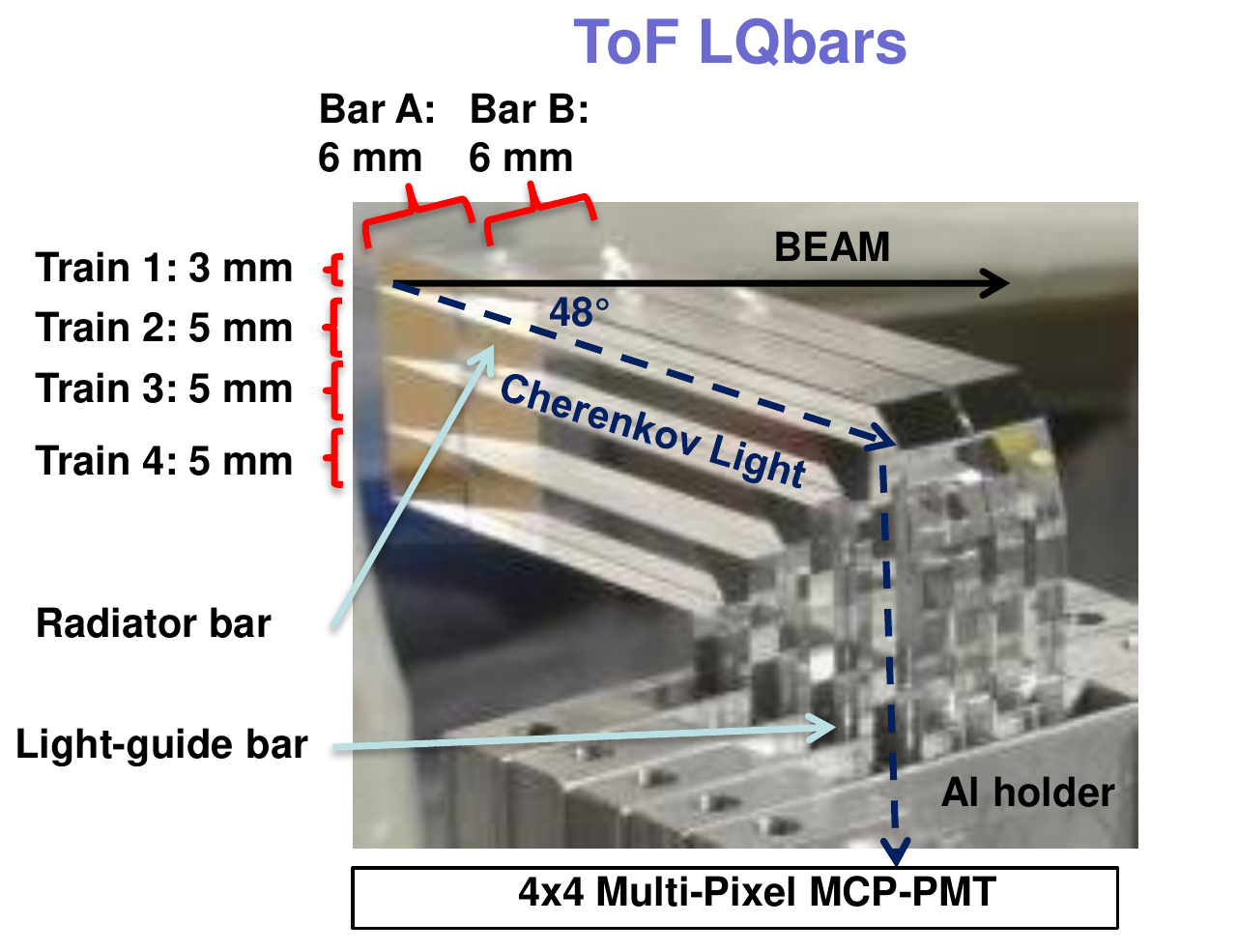}
	\caption{The LQbar ToF prototype detector with four trains of two LQbars each (half of the final number of LQbars per train).}
	\label{fig:ToFprototype}
\end{figure}

The AFP beam-test prototype ToF system consisted of four rows (trains 1--4) of two LQbars (A and B) as shown in figures~\ref{fig:AFPbeamTestSetup} and~\ref{fig:ToFprototype}, oriented with the Cherenkov angle to the beam (half the number of LQbars per train with respect to the baseline design). The radiators of the upper train were 3\,mm high in $y$, those of the lower three trains 5\,mm. The length of the radiator bars in the short horizontal direction was 6\,mm, the length in the long horizontal direction ranged from 35 to 57\,mm. At the side opposite to the kink, the radiators were cut such that this edge was parallel to the beam, giving an effective edge length of about 8\,mm. The LQbars were mounted in an Aluminium holder with 1\,mm thick isolation plates between the light-guide bars of different trains and 125\,$\mu$m thin spacer wires between those of the same train. At the radiator level, the different trains were optically isolated using mylar foils. The ends of the LQbars were brought into contact (without the use of optical grease) with a 10\,$\mu$m-pore mini-Planacon MCP-PMT by Photonis with 4x4 anode pixels of 6x6\,mm$^{2}$ size with a space of 0.25\,mm between adjacent pixels. 

In addition to the LQbar timing system, three fast timing reference detectors consisting of straight Quartz bars ($3\times3$\,mm$^2$ cross section, 3\,cm long in beam direction) coupled to Silicon Photomultipliers (SiPMs) by STMicroelectronics (Catania, Italy) were used~\cite{bib:Albrow2012}. They were placed under perpendicular beam incidence (\ie not oriented with the Cherenkov angle) behind the AFP prototype for testing purposes. 

The signals of all timing detectors were amplified, discriminated with CFDs and digitised with the HPTDC board as described above. 


In the 2015 beam test, also other LQbar types and configurations were tested, such as single bars, matt bars, spatial gaps between bars of the same train, the addition of optical grease, as well as different types of MCP-PMTs. However, the analysis of these different ToF configurations is still on-going. In this paper, only the above-described standard configuration will be covered. A publication with a more detailed description of the production and properties of the AFP ToF detector is in preparation. 

\subsection{Readout and trigger}
\subsubsection{AFP design and requirements}

\begin{figure}[hbt]
	\centering
	 \includegraphics[width=15cm]{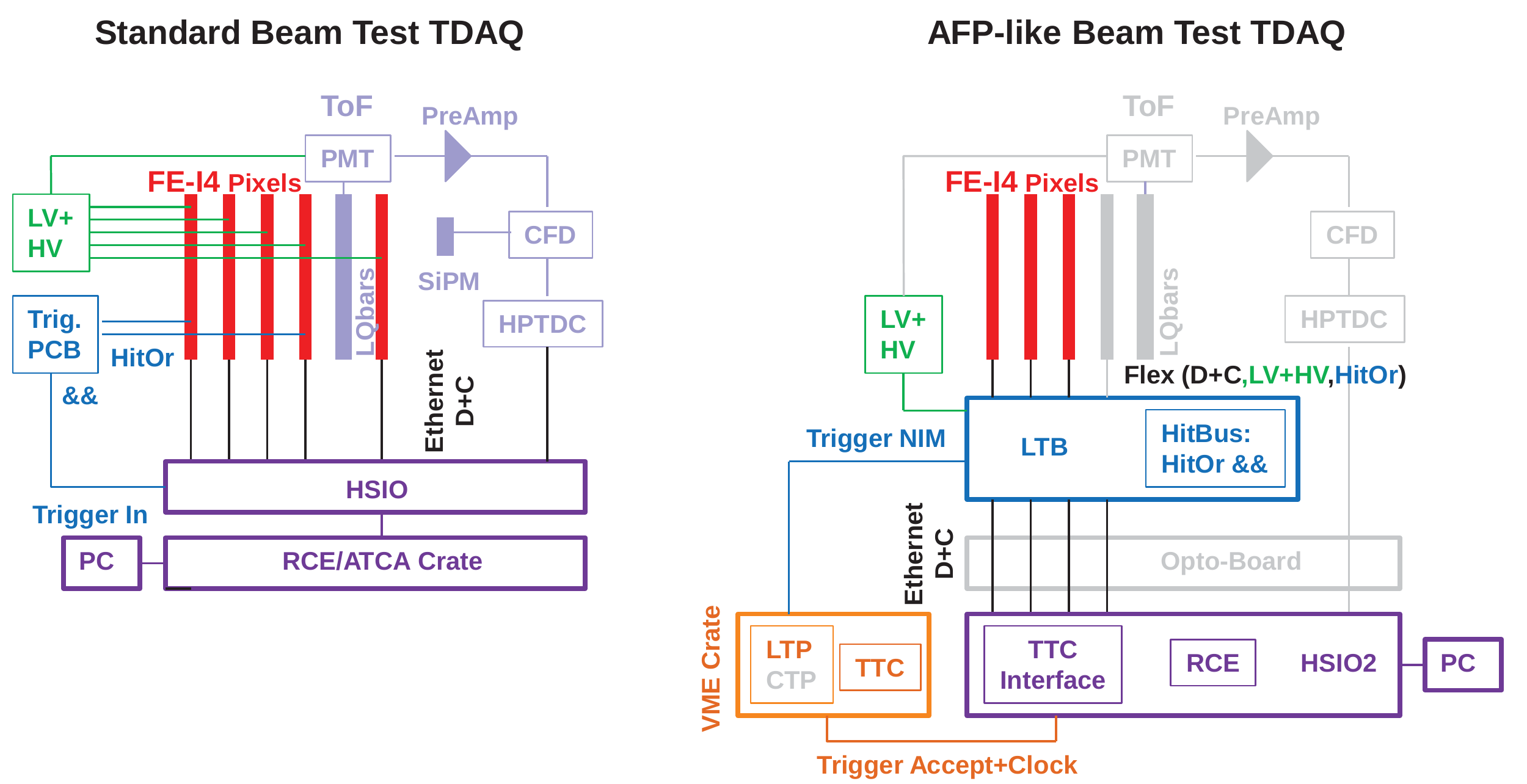}
	\caption{The readout and TDAQ schematics for most of the runs (left) and for special runs to test a TDAQ system close to the final AFP one (right); grey components were not included in these special runs.}
	\label{fig:AFPTDAQ}
\end{figure}

For the readout of both tracking and ToF modules, the Reconfigurable Cluster Elements (RCE) system~\cite{bib:RCE1} 
is used, which is based on an Advanced Telecommunications Computing Architecture (ATCA)~\cite{bib:ATCA-Specs} standard. It consists of the RCE boards that generate commands and receive data (C+D), the Cluster Interconnect Modules (CIM) that work as control units and communication interfaces, as well as of the High-Speed Input-Output (HSIO) board that interfaces to the pixel front-ends and HPTDC chips and is responsible for data decoding, buffering and routing. The data and commands between the RCE system on the one hand and the FE-I4 and HPTDC chips on the other hand are sent via optical fiber using an opto-board close to the detectors as a converter between electrical and optical signals.

In the initial one-arm AFP phase, the trigger is taken from the coincidence of several tracking planes using the FE-I4 HitOr signal (see section~\ref{sec:tracker}). The logical processing of the HitOr trigger signals from different pixel planes is performed on the HitBus chip~\cite{bib:hitbus}. For the full AFP detector including the ToF system, the trigger is planned to be extracted from a coincidence of LQbar signals in one train to reduce the trigger dead time (the HitOr dead time is discussed in section~\ref{sec:TDAQ}) and to include a coarse position information by distinguishing which train is hit. 

The trigger signals are then converted to the Nuclear Instrumentation Module (NIM) logic standard and sent to the ATLAS Central Trigger Processor (CTP)~\cite{bib:CTPandTTC} for combining it with trigger signals from other ATLAS sub detectors. The signal from the ATLAS central trigger informing that an event is accepted, as well as the ATLAS central clock, are distributed to the AFP RCE system via the Timing, Trigger and Control (TTC) unit~\cite{bib:CTPandTTC}. The readout is sketched in figure~\ref{fig:AFPTDAQ} (right).

\subsubsection{Prototype readout and trigger}
\label{sec:prototypeTrigger}

Different versions of the RCE system were used in the beam tests: In 2014, the first version with the HSIO1 board connected via optical fiber to an RCE ATCA crate outside the beam area; and in 2015 the second version with the HSIO2 board which includes already an RCE component on-board, making an external crate unnecessary. 
Whereas the operation of the RCE system with the FE-I4 chip had been extensively proven already before, \eg in the IBL stave integration~\cite{bib:RCE2}, its operation with the HPTDC system still had to be implemented and optimised before and during the November 2014 beam test. 

The trigger of the combined system was given by signals from the tracking system as planned for the initial one-arm AFP phase.
Two different configurations were used (see figure~\ref{fig:AFPTDAQ} left and right):

\begin{enumerate}

	\item During large parts of both beam tests (for tracking-timing integration and detector performance studies), a custom-made electronic circuit board was used to combine several FE-I4 HitOr signals from different planes to form a coincidence trigger signal in the Transistor-Transistor-Logic (TTL) format. This trigger signal was then directly fed into the RCE HSIO board to trigger the readout of the FE-I4 pixel devices and the HPTDC. The data and commands between the FE-I4 pixel devices or the HPTDC and the RCE HSIO were sent electrically via shielded Ethernet cables. The clock was provided internally by the RCE system. Low and high voltage (LV/HV) was provided separately and directly for each pixel module.
	
	\item In dedicated tests, a configuration more similar to the final AFP Trigger and Data Acquisition (TDAQ) system was studied. Three FE-I4 pixel modules were connected via a flexible cable to a Local Trigger Board (LTB) that provided low and high voltage for each module, as well as the data and command interface to the RCE HSIO2 board via Ethernet cables. This board included also the HitBus chip for different trigger logic processing of the HitOr signals from up to three FE-I4 pixel modules. The HitBus trigger output was an LVDS signal that was converted into NIM standard and sent to a Local Trigger Processor (LTP) in an external Versa Module Eurocard bus (VMEbus) crate, which was used to locally test the compatibility with the ATLAS CTP system. The LTP created a trigger-accept signal, which was then sent via a TTC system in the same VMEbus crate and optical fiber cable to the TTC interface board on the RCE HSIO2 board. Also the clock was externally provided by the TTC module to the RCE HSIO2 board. 
	
	\end{enumerate}

In addition to the full HPTDC-RCE readout system, for testing purposes and specific time-resolution studies without the HPTDC contributions, the ToF signals were also recorded with a LeCroy SDA760ZI oscilloscope. 

\section{Beam test operation}
\label{sec:operation}

The AFP beam tests took place at the H6B and H6A beam lines of the CERN-SPS with 120\,GeV pions for one week in November 2014 and two weeks in September 2015, respectively. 

\subsection{Operational parameters and calibration}
\label{sec:calib}
Before the data taking, the operational parameters were set and the pixel devices and the HPTDC calibrated. 

By default, bias voltages of 10\,V were applied to the 3D pixel sensors. Each FE-I4 pixel was tuned to a threshold of 2\,ke$^{-}$ (3\,ke$^{-}$ in 2014) and a ToT of 10 (in units of 25\,ns clock cycles) at an injected charge $Q$ of 20\,ke$^{-}$ (referred to as 10@20\,ke$^{-}$ in the following). For this, the internal charge-injection mechanism was used with the same calibration parameters (such as injection capacitance) assumed for all five pixel planes (in reality, the charge-injection calibration parameters have a chip-to-chip spread of 15\%~\cite{bib:Backhaus}, but the exact values were not known for the IBL spare devices used in these beam tests). 

The MCP-PMT voltage, which determines the gain, was set to 1900\,V by default (corresponding to a gain of $2\times10^{5}$), the SiPM voltage to 30.7\,V. The CFD fixed threshold was set to 100\,mV, the fractional threshold to 24\% of the signal amplitude.

In dedicated runs also the following variations of the values of some of these parameters were studied (default values marked in \textbf{bold}):
\begin{itemize}
	\item Pixel sensor bias voltage [V]: 1, 2, 4, 7, \textbf{10}, 20
	\item Pixel threshold [ke$^{-}$]:    1.5, \textbf{2.0}, 2.5, 3.0
	\item Pixel ToT:                     10@16\,ke$^{-}$, \textbf{10@20\,ke$^{-}$}, 5@20\,ke$^{-}$
	\item MCP-PMT voltage [V] (gain):    1750 ($0.7\times10^{5}$), 1800 ($1.0\times10^{5}$), 1850 ($1.7\times10^{5}$), \textbf{1900 ($2.0\times10^{5}$)}
\end{itemize}

The ToT-to-charge relation of the pixel devices can in principle be obtained for each device and each pixel separately by scanning the injected charge per pixel and measuring the ToT response~\cite{bib:Backhaus}. This was, however, not possible with the setup during the beam test. Instead, as a rough approximation, a global relation was obtained previously in stand-alone studies from the average over all pixels of a similar pixel device. The relation is approximately parabolic with a fit giving 

\begin{equation}
Q [\mbox{e}^-] = 1909 + 363\times(ToT-1) + 141\times(ToT-1)^2
\label{eq:ToTtoQ}
\end{equation}
for the standard tuning.\footnote{Here and in all following plots, $ToT$ is the decoded ToT information from the front-end chip in units of 25\,ns clock cycles, ranging from 1 for events just above the threshold to 14 as overflow bin.} In addition, a global charge calibration factor of 1.4 was applied as obtained from the measurement of the gamma lines of Am-241 and Cd-109 sources, consistent with earlier observations~\cite{bib:Backhaus}.

The calibration of the HPTDC is primarily to characterise the non-linearities in the timing measurement due to variations of the time-bin size. Uncorrected these non-linearities can introduce a timing error up to 150\,ps. Calibration requires an input uncorrelated to the reference clock, usually a free-running oscillator. A normalised histogram of these events over the number of TDC bins gives a measurement of the size of each bin allowing a look-up table to be generated to correct for the accumulation of the error in bin size. This calibration can then either be applied online or offline. Once calibrated, measurements remain fairly stable although degradation with aging, radiation damage, and temperature variation is expected requiring periodic re-calibration. Variation in absolute timing vs. temperature is measured at 2.5\,ps/K. Chip temperatures vary little during operation and each is instrumented with a thermistor to monitor this.

\subsection{Trigger and data taking}
\label{sec:TDAQ}
As mentioned in section~\ref{sec:prototypeTrigger}, for most of the runs, a custom-made PCB was used to provide the trigger from the hit coincidence of several tracking planes. In 2014, planes 0, 3 and 4 were used for triggering, in 2015 only planes 0 and 3. In dedicated runs, the more realistic TDAQ system with the HitBus chip, LTP and TTC was successfully tested. In those runs, the first three planes were included in the HitBus-chip trigger logic, which was configured with the RCE system to give a trigger signal if all three planes fire or, alternatively, if two out of the three planes fire. 

\begin{figure}[bt]
	\centering
	\includegraphics[width=7.5cm]{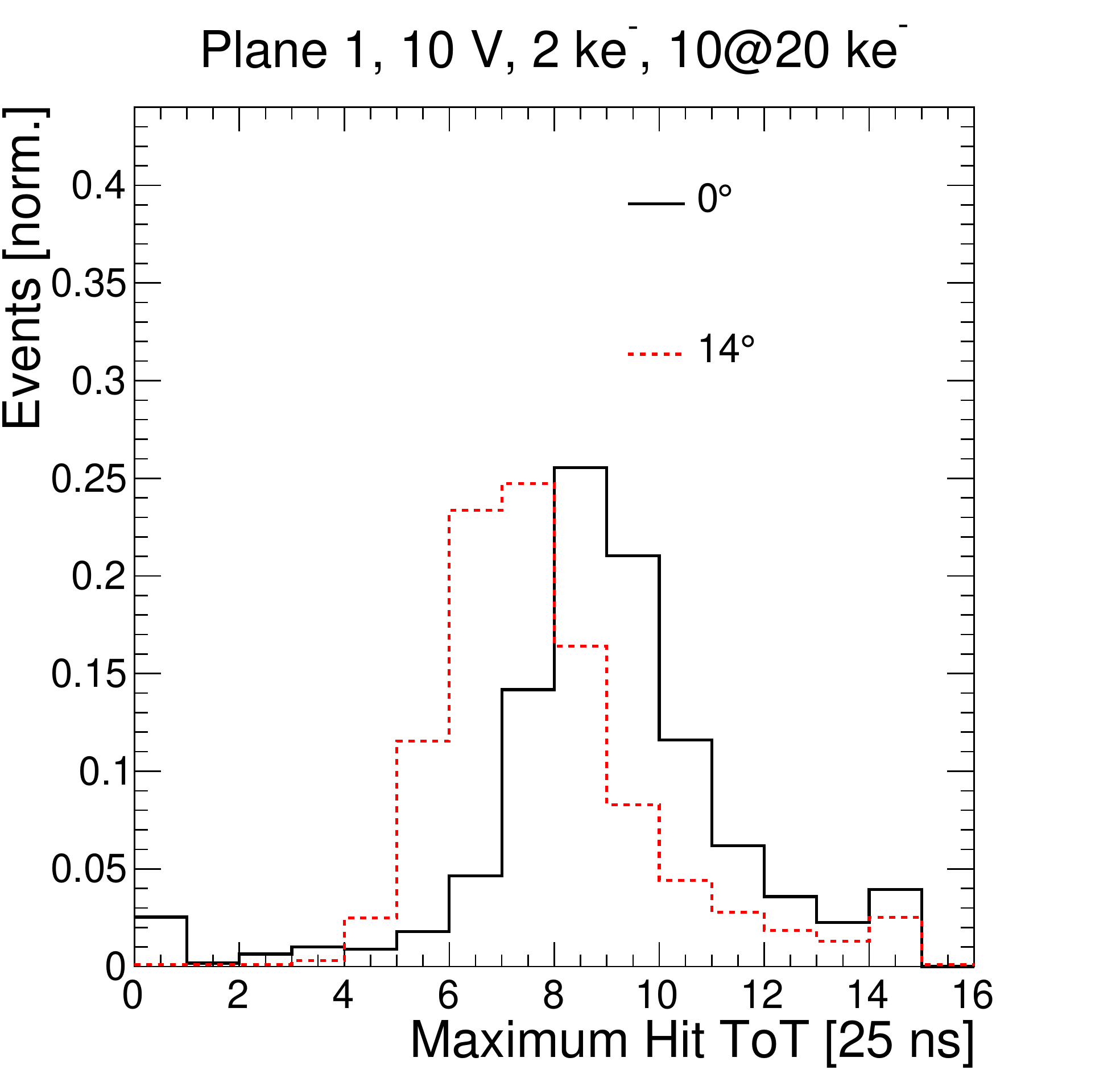}
	\includegraphics[width=7.5cm]{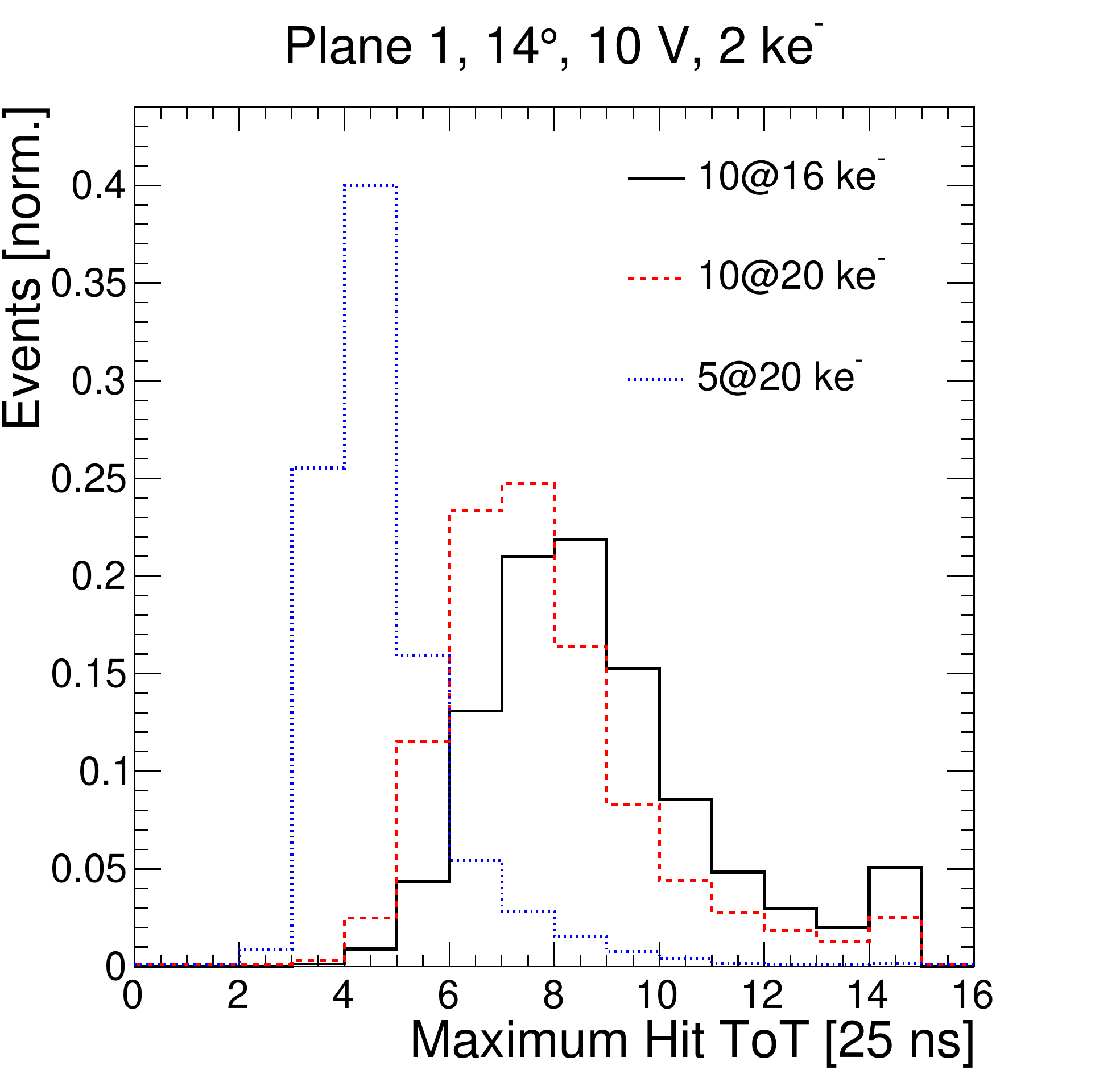}
	\caption{Distribution of the maximum hit ToT in an event in plane 1 (indicative for the HitOr duration) for different angles (left) and ToT tunings (right) at otherwise standard operational parameters. The bin at 14 includes entries for ToT$\geq$14 (overflow).}
	\label{fig:hitOr}
\end{figure}

Of special interest for the integration of the AFP trigger into the ATLAS TDAQ system are the latency (\ie the delay of the arrival of the trigger signal after the particle crossing) and the duration of the trigger signal, which were measured with the oscilloscope (using the SiPM signals as fast timing reference). The latency including the FE-I4 HitOr processing, the HitBus-chip processing, as well as the NIM conversion was found to be about 100\,ns, \ie 4 nominal LHC bunch-crossing spacings of 25\,ns, which can be accommodated into the AFP trigger-latency requirements. The duration of the NIM trigger signal was found to be about 200\,ns with a significant spread of about 50\,ns for the standard operational parameters and at perpendicular beam incidence. This can be explained with the HitOr signal being the logical OR of the discriminator signals of all pixels in a chip, \ie its duration is typically the ToT of the pixel with the highest signal in an event (referred to as \emph{maximum hit ToT} in the following). This distribution is shown in figure~\ref{fig:hitOr}. For a most probable deposited charge of about 17\,ke$^-$ in 230\,$\mu$m silicon at perpendicular incidence and a standard ToT tuning of 10@20\,ke$^-$, this leads to a peak of 8 clock cycles of 25\,ns with a significant spread due to Landau fluctuations and charge sharing. The AND between the HitOr signal of different planes in the HitBus chip is then dominated by the shortest HitOr signal among all planes. 

The large duration of the HitOr trigger signal has two consequences. Firstly, the ATLAS CTP would assign triggers for each bunch crossing for which the trigger signal is high, although a particle might have crossed the detector only during the first bunch crossing. Hence for the final AFP TDAQ system, the trigger-signal duration will be reduced to 25\,ns before being fed into the CTP. Secondly, it implies a trigger dead time since no new trigger can be given while the HitOr signal is still high from a previous hit. It will depend on the run conditions whether this has significant implications on the trigger efficiency. It is possible that the first dedicated low-luminosity AFP runs will be operated not with the nominal LHC bunch-crossing spacing of 25\,ns, but at \eg 100\,ns. In addition, the AFP detector occupancy per minimum-bias interaction is expected to be only about 2--4\%~\cite{bib:AFPreference1}. Hence, for low-pile up conditions with \eg only one interaction per bunch crossing, a 200\,ns trigger dead time in combination with 100 (25)\,ns bunch-crossing spacing will lead to trigger efficiencies of 96 (85)\% in case of 2\% occupancy per interaction or 92 (72)\% in case of 4\% occupancy. It should be noted, however, that this is only a rough calculation neglecting beam-related backgrounds (which are not yet known precisely) and the exact distribution of pile-up multiplicity and of the trigger dead time. 

In any case, in order to reduce the trigger dead time and improve the efficiency, the dependence of the maximum-hit-ToT distribution (as indication of the HitOr duration) was studied for different angles and tunings. Figure~\ref{fig:hitOr} shows that the maximum hit ToT is reduced for the AFP tilt of 14$^{\circ}$ (due to enhanced charge sharing) and for a ToT tuning of 5@20\,ke$^-$. Tuning to even lower ToT values at 20\,ke$^-$ was not successful during the beam test but might be achieved with more care. However, reducing the ToT for a given charge compromises the position resolution for charge-interpolating algorithms, as discussed in section~\ref{sec:resolution}. A compromise could be to tune two of the four AFP planes to \eg 5@20\,ke$^-$ for efficient triggering, and the remaining two to \eg 10@20\,ke$^-$ for improved position resolution. No strong dependence of the maximum hit ToT on bias voltage and threshold tuning was observed.

\begin{figure}[bt]
	\centering
	\includegraphics[width=15cm]{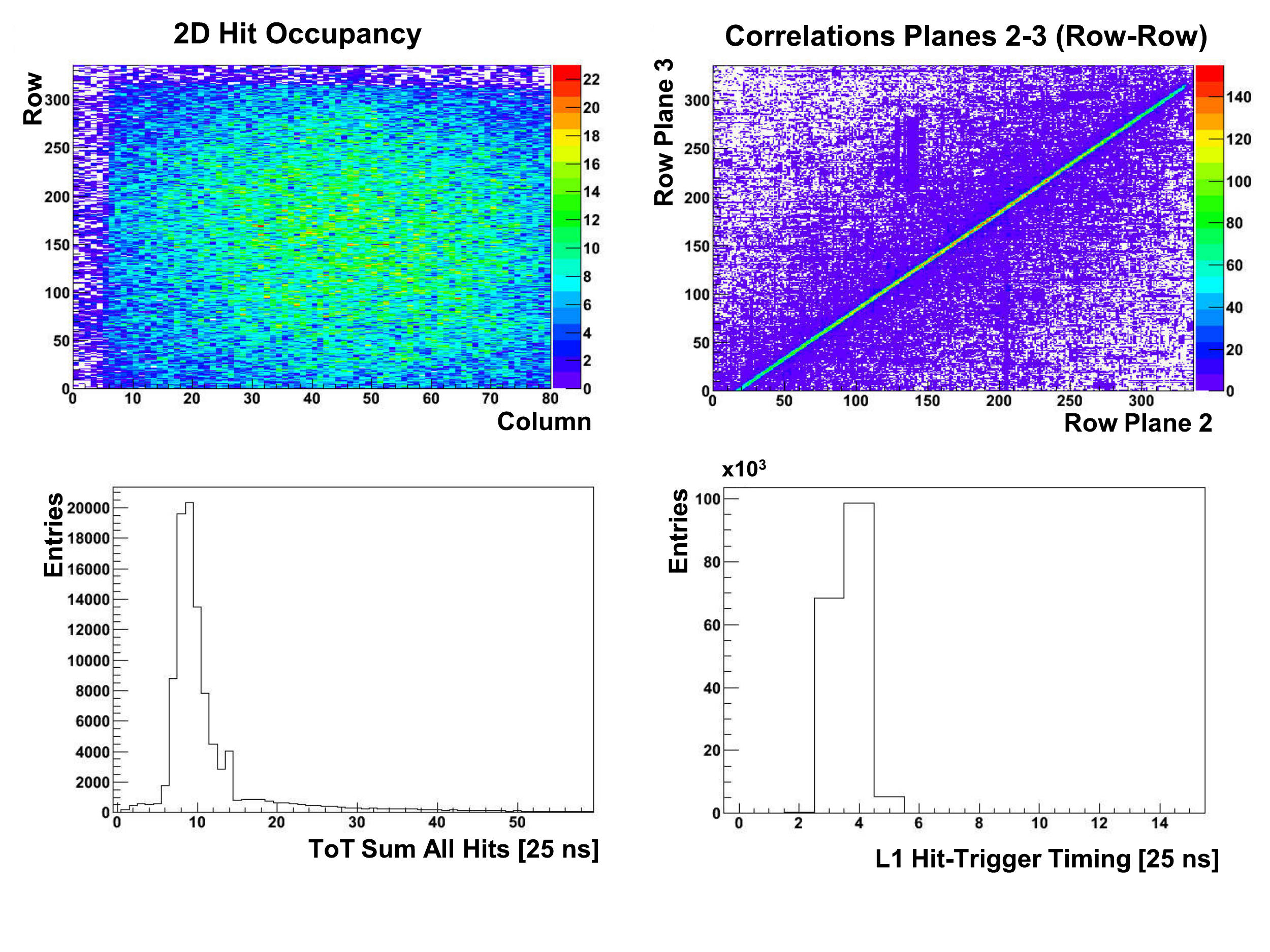}
	\caption{Online-monitoring distributions. Top left: the 2D hit occupancy map of one pixel sensor. Top right: the correlation between the rows of two consecutive pixel planes. Bottom left: The ToT sum of all hits in the event (simple cluster-charge distribution). Bottom right: The timing of the recorded hits with respect to the trigger signal (Level 1 or L1 distribution).}
	\label{fig:onlineMon}
\end{figure}

After triggering, the signals of the pixel detectors, \ie the addresses and ToTs of the pixels above threshold, were recorded for typically 16 (5) consecutive clock cycles in 2014 (2015). The timing distribution of the recorded hits with respect to the trigger signal (so-called Level 1 distribution, in units of 25\,ns clock cycles) is shown in figure~\ref{fig:onlineMon}, bottom right. The clear peak indicates low noise levels and a good synchronisation between the recorded hits and the trigger. 

The system operation during data taking was stable, \eg half-day runs without user interaction were possible. 
Altogether, 38\,M events were collected in 2014 and 210\,M events in 2015, at typical average rates of a few hundred Hz.


Many parameters of the tracking system could be monitored at the online level (see figure~\ref{fig:onlineMon} for examples). In addition to the Level-1 distribution already discussed, the 2D hit occupancy maps for each tracking plane were monitored and used for performing the alignment of the beam with respect to the detectors. 
Good correlations between the hit column (and row) numbers of two consecutive pixel planes, respectively, indicated that real tracks were recorded and that the inter-plane alignment precision was at the mm level. Moreover, the ToT sum of all hits (roughly indicating the cluster charge) was monitored online and behaved as expected.


\begin{figure}[hp]
\flushleft
	\includegraphics[width=7.5cm]{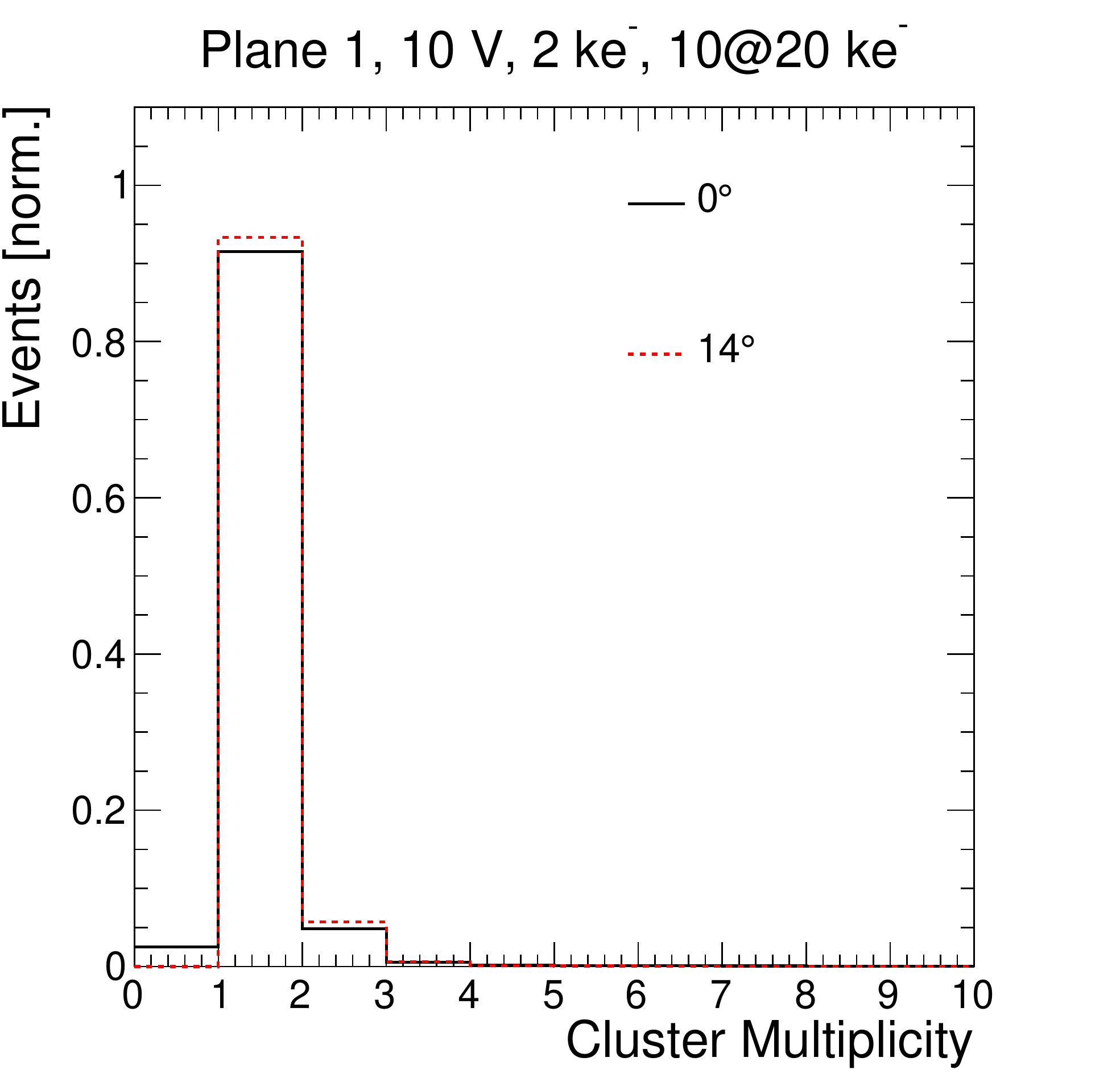}
	\newline
	\includegraphics[width=7.5cm]{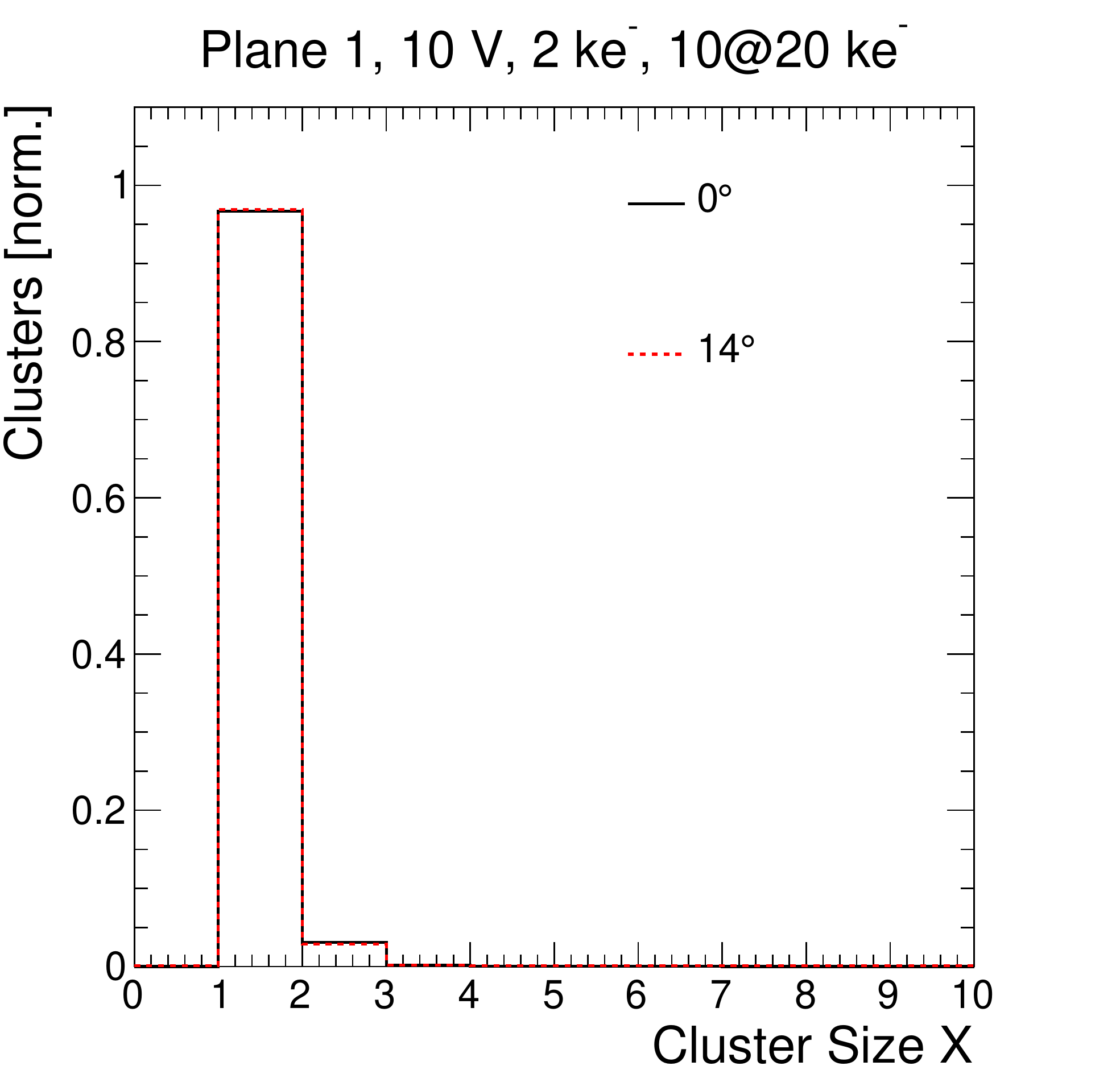}
	\includegraphics[width=7.5cm]{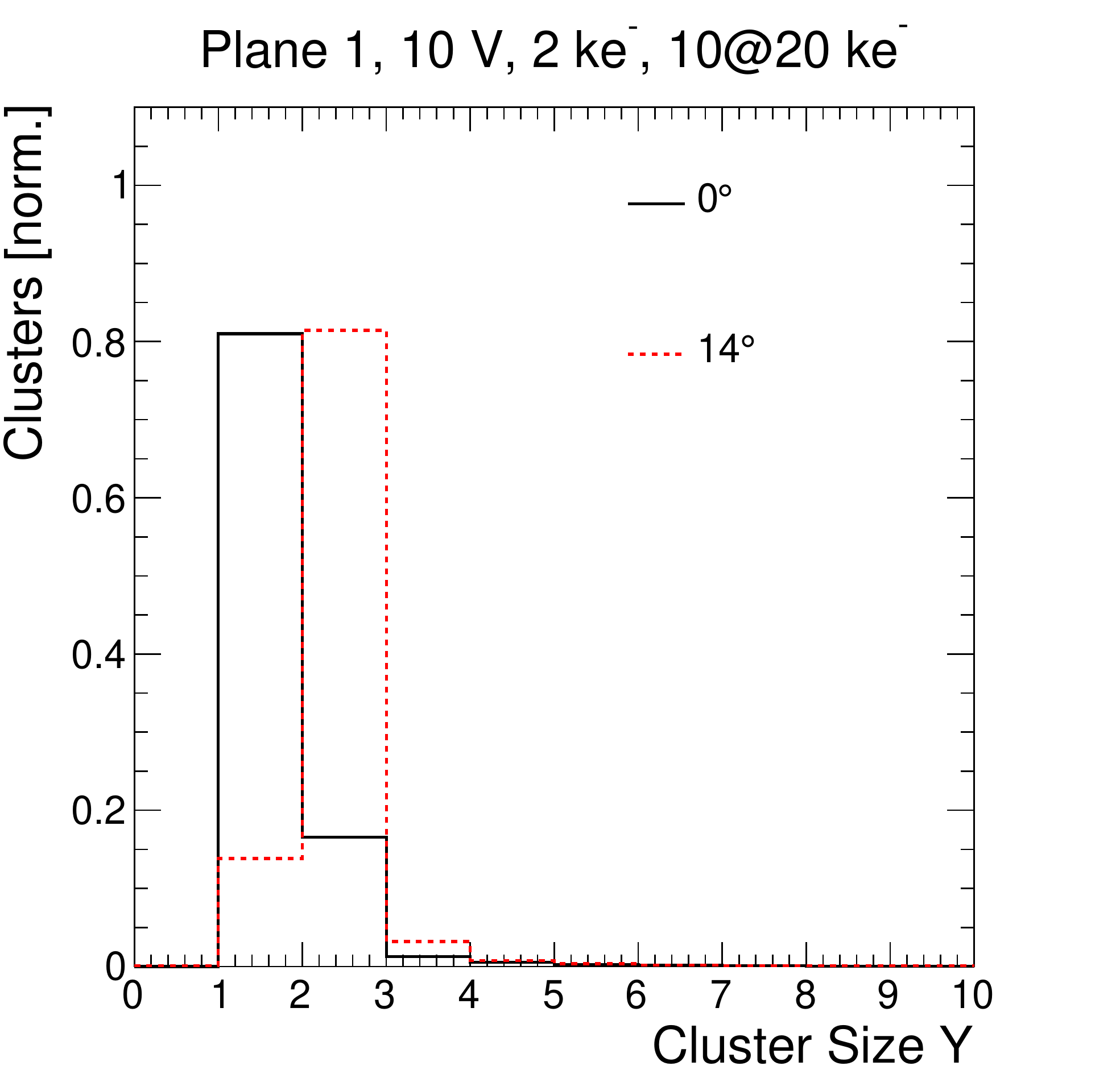}
	\includegraphics[width=7.5cm]{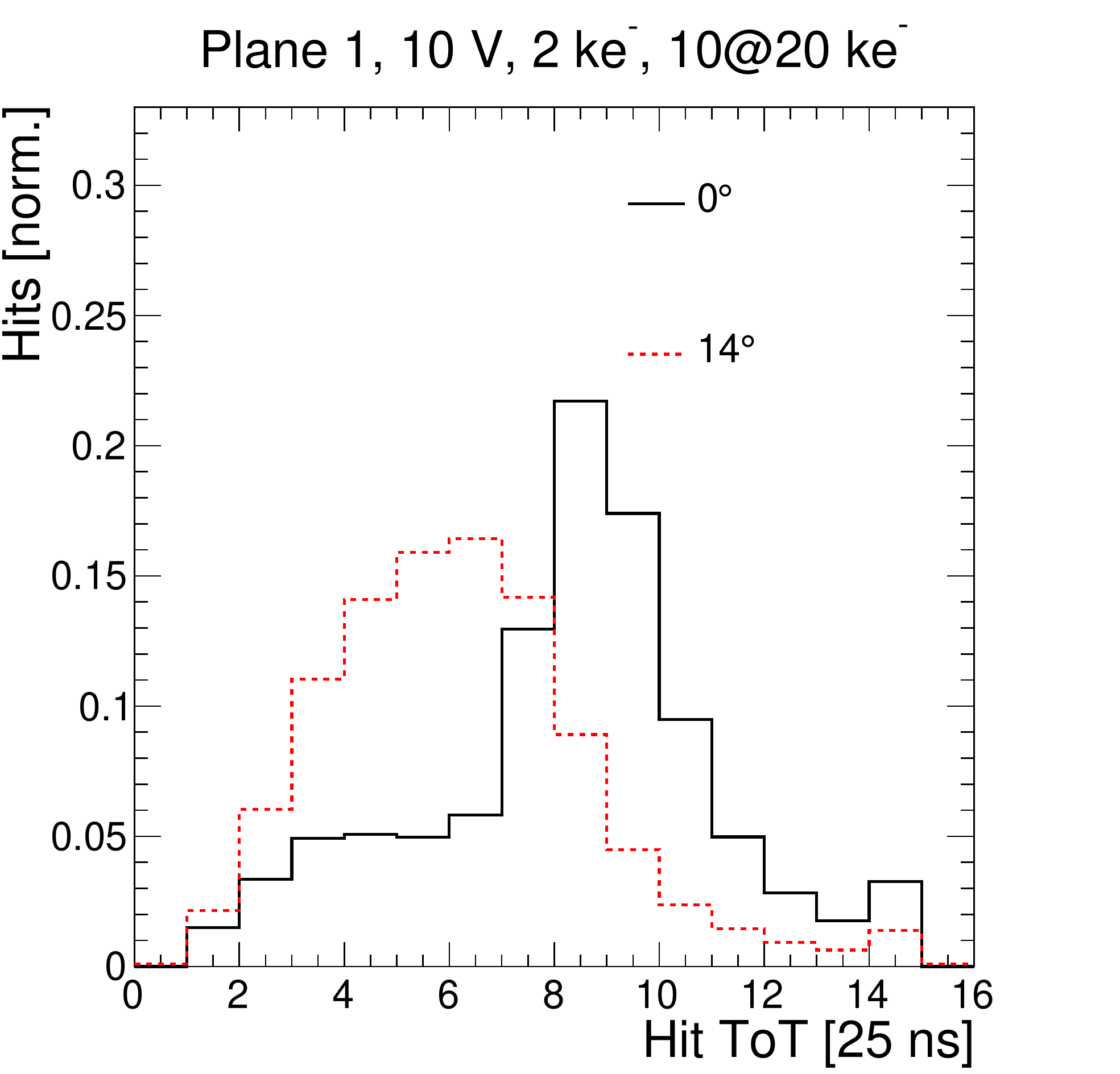}
	\includegraphics[width=7.5cm]{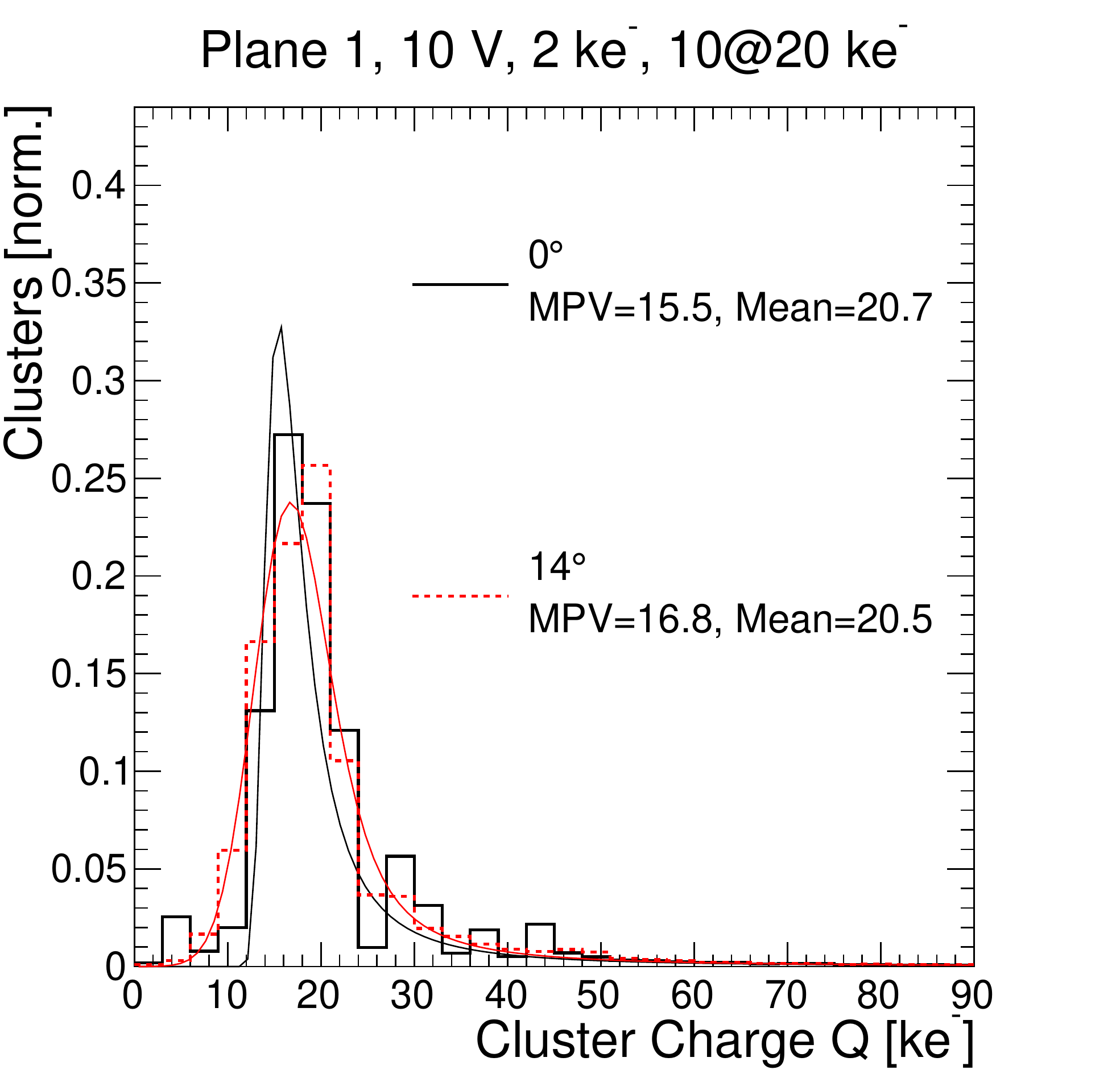}
	\caption{ Hit and cluster properties for plane 1 (CNM) at 0 and 14 degrees at standard operational parameters. }
	\label{fig:hitClusterDifferentAngles}
\end{figure}

\begin{figure}[hp]
	\centering
	\includegraphics[width=7.5cm]{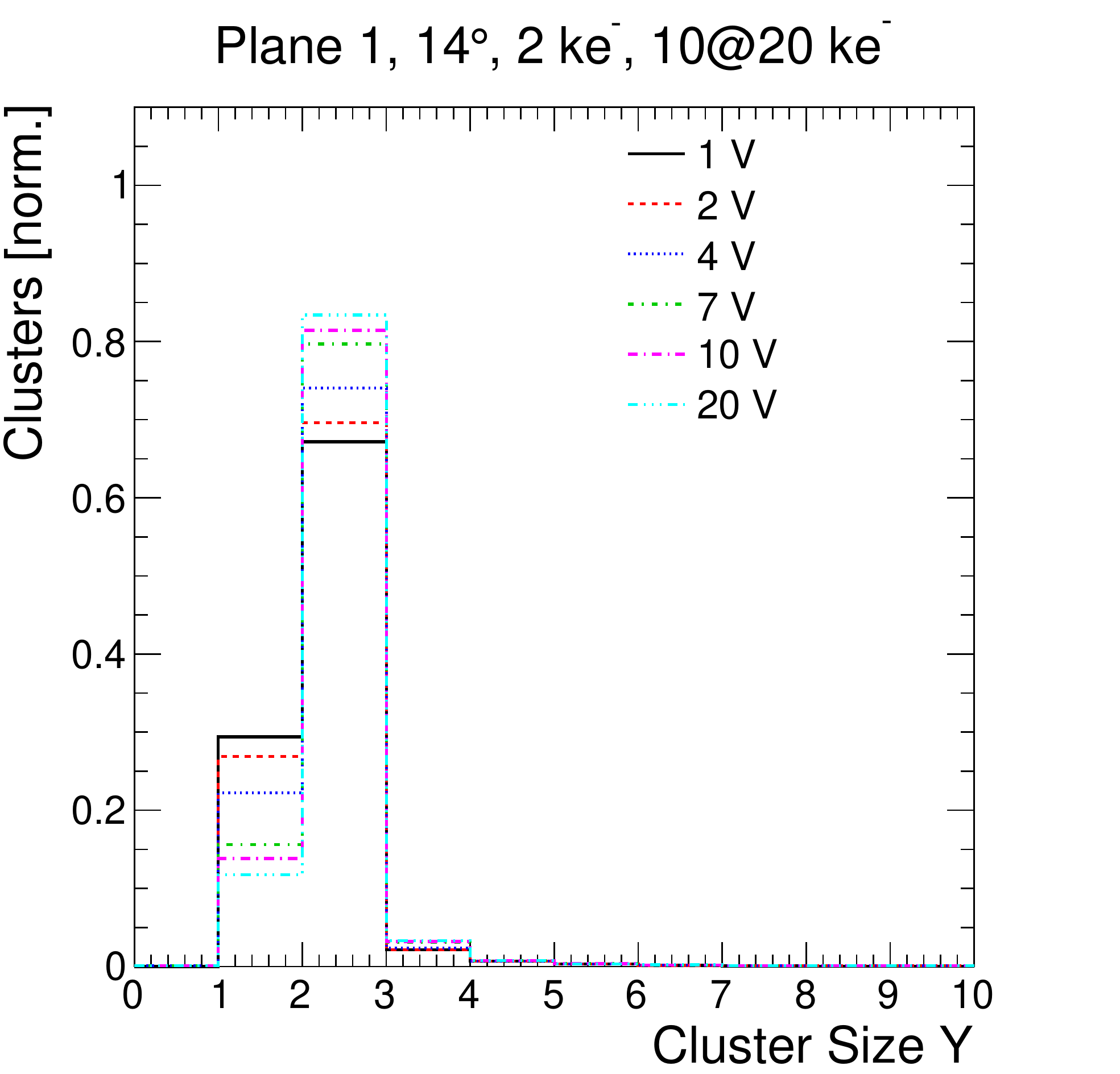}
	\includegraphics[width=7.5cm]{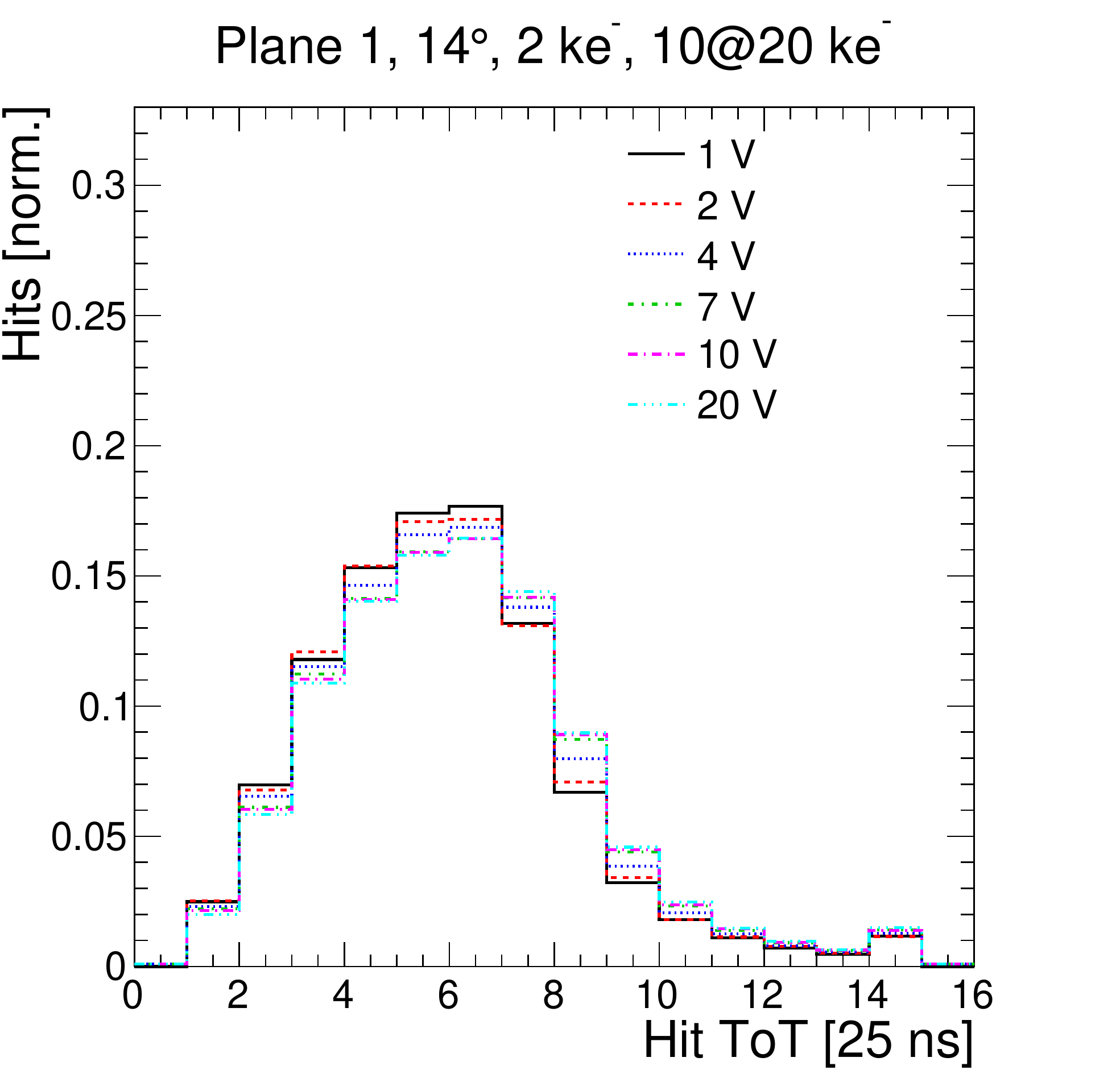}
	\includegraphics[width=7.5cm]{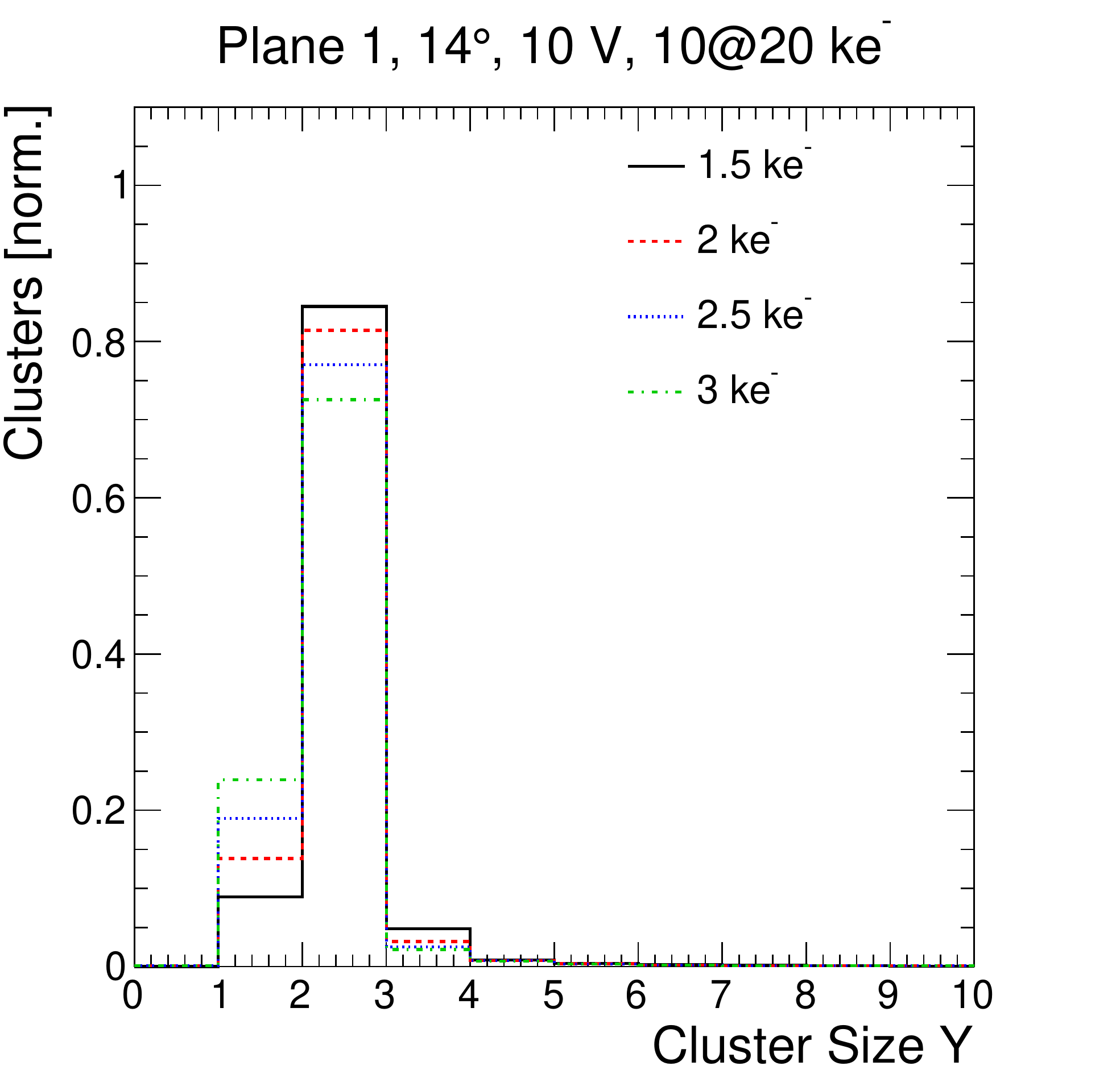}
	\includegraphics[width=7.5cm]{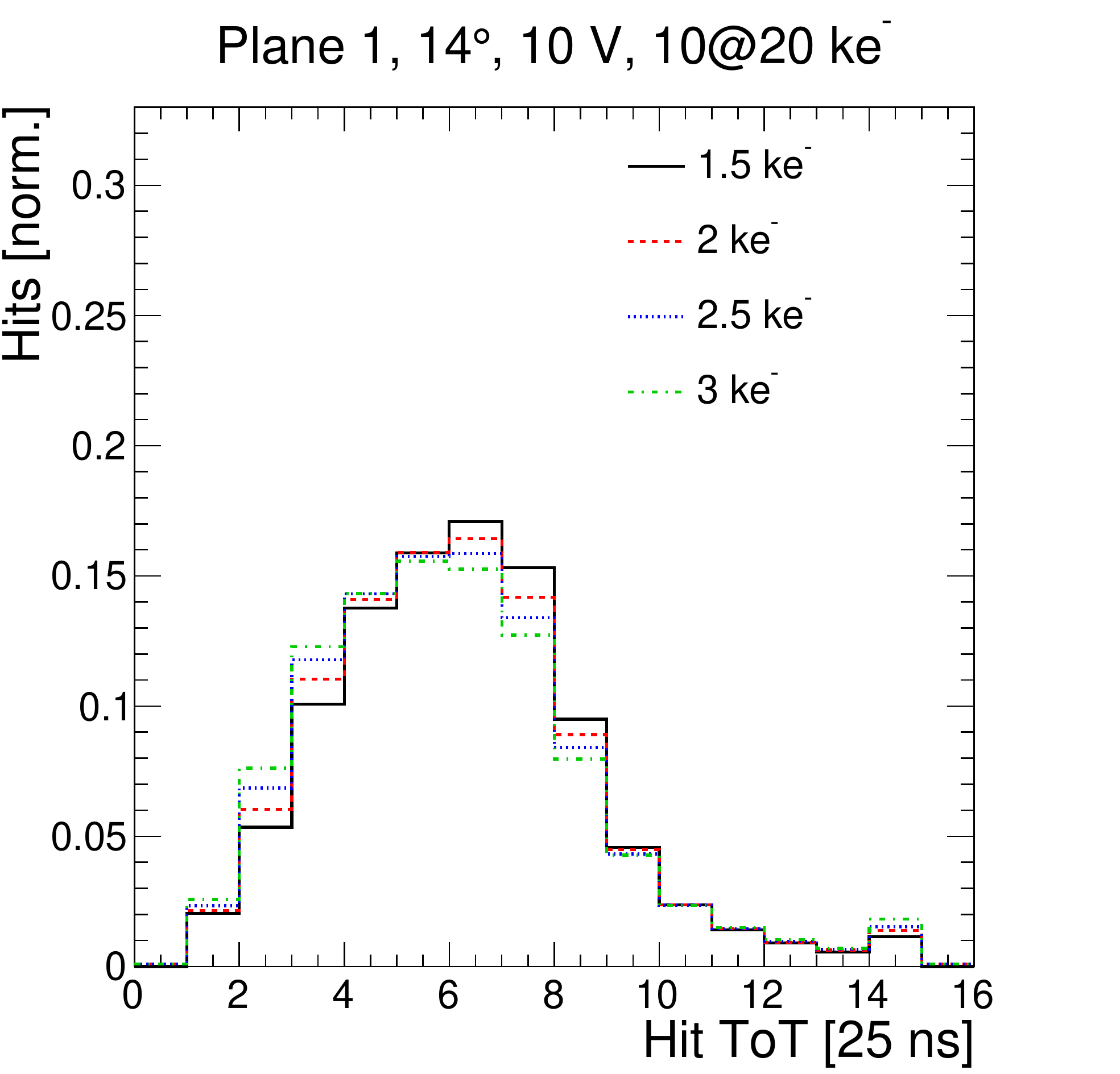}
	\includegraphics[width=7.5cm]{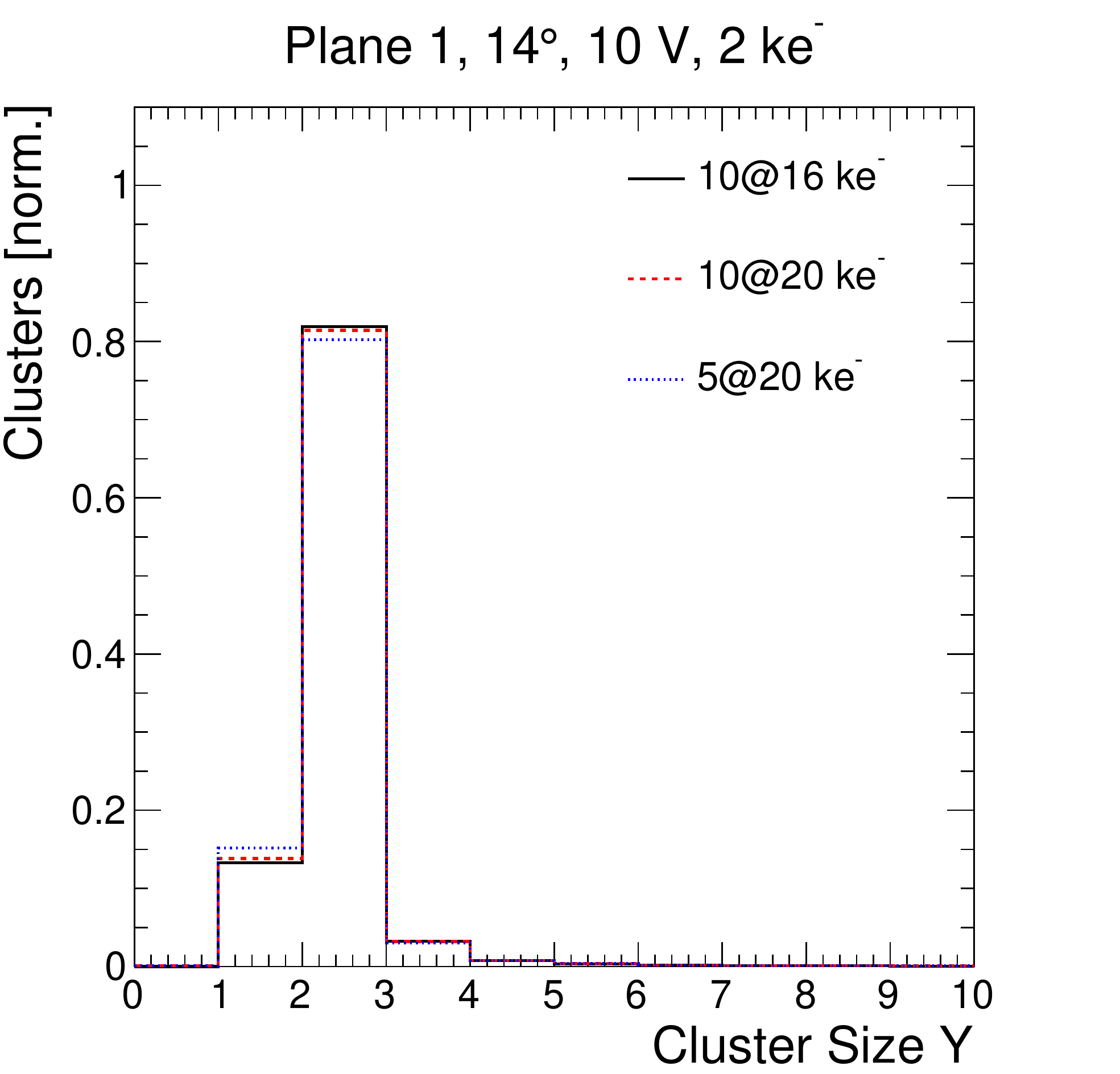}
	\includegraphics[width=7.5cm]{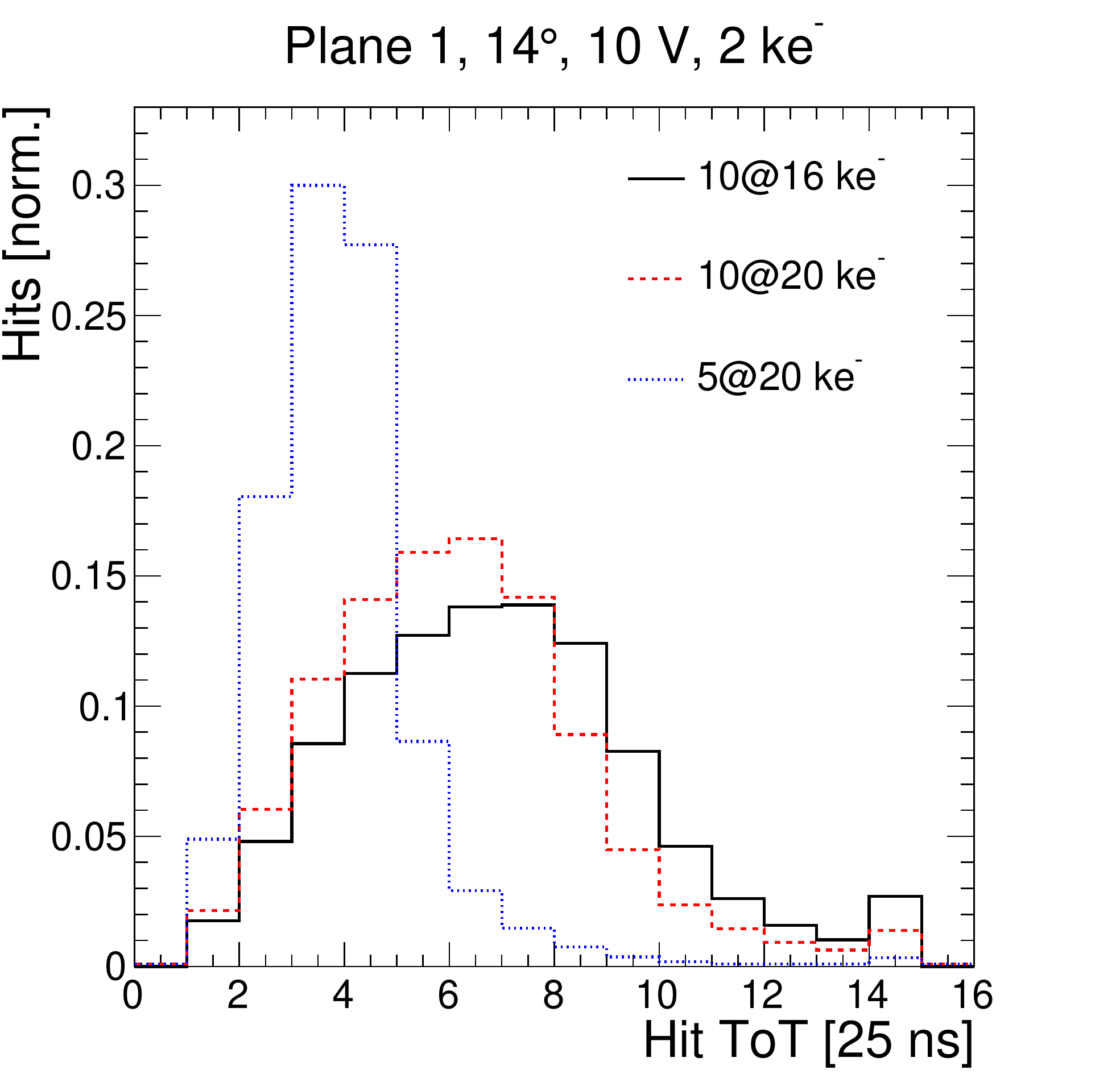}
	\caption{ Cluster size in y direction (left) and hit ToT (right) distribution for plane 1 (CNM) at different operational parameters: bias voltage dependence (top), threshold dependence (centre) and ToT tuning dependence (bottom). }
	\label{fig:hitClusterDifferentParameters}
\end{figure}

\section{Data analysis and detector performance }
\label{sec:performance}
The data taken including the tracker and ToF hit information were stored in a common data format and analysed offline.

\subsection{Tracker reconstruction}
\label{sec:trackReco}

The offline pixel-hit clustering and track reconstruction were performed with the software framework \emph{Judith}~\cite{bib:Judith}. 

\subsubsection{Pixel hit clustering}
\label{sec:clustering}
Neighbouring hit pixels were grouped into hit clusters. The cluster-centre position for each coordinate ($x,y$) was determined either with a ToT-weighted or charge-weighted (\ie after the ToT-to-charge conversion explained in section~\ref{sec:calib}) mean of the single-pixel centres. As default, the ToT-weighted mean was taken since this is a simple algorithm using direct measurement information with a position resolution similar to the charge-weighted mean. More discussion on the resolution and its dependence on the different cluster-position algorithms can be found in section~\ref{sec:resolution}.

Figure~\ref{fig:hitClusterDifferentAngles} shows a collection of important pixel-hit and cluster distributions for plane 1 (a CNM device like in the final AFP detector, which was not included in triggering and hence unbiased) at standard operational parameters, compared for 0$^{\circ}$ and 14$^{\circ}$ tilt: cluster multiplicity, cluster size in both directions (\ie over how many pixels a cluster extends in $x,y$), hit ToT (including all hit pixels before clustering) and cluster charge $Q$ after ToT-to-charge conversion and clustering. For the cluster size $y$ and the hit ToT, the dependence on voltage, threshold and ToT tuning at 14$^{\circ}$ tilt is presented in figure~\ref{fig:hitClusterDifferentParameters}. All these distributions are shown for the minimal-material region defined in section~\ref{sec:material}.

More than 90\% of all events have only one cluster per pixel plane, similar for all operational parameters. Larger cluster multiplicities are mainly from material interactions as discussed in section~\ref{sec:material} and possibly from noise; as shown in section~\ref{sec:trackReco2}, events with more than one reconstructed track are only about 2\%. The fact that at 14$^{\circ}$ almost no events have no clusters (increasing to 3\% at 0$^{\circ}$) indicates already the excellent hit efficiency as presented in section~\ref{sec:material}.

\begin{figure}[tb]
	\centering
	\includegraphics[width=15cm]{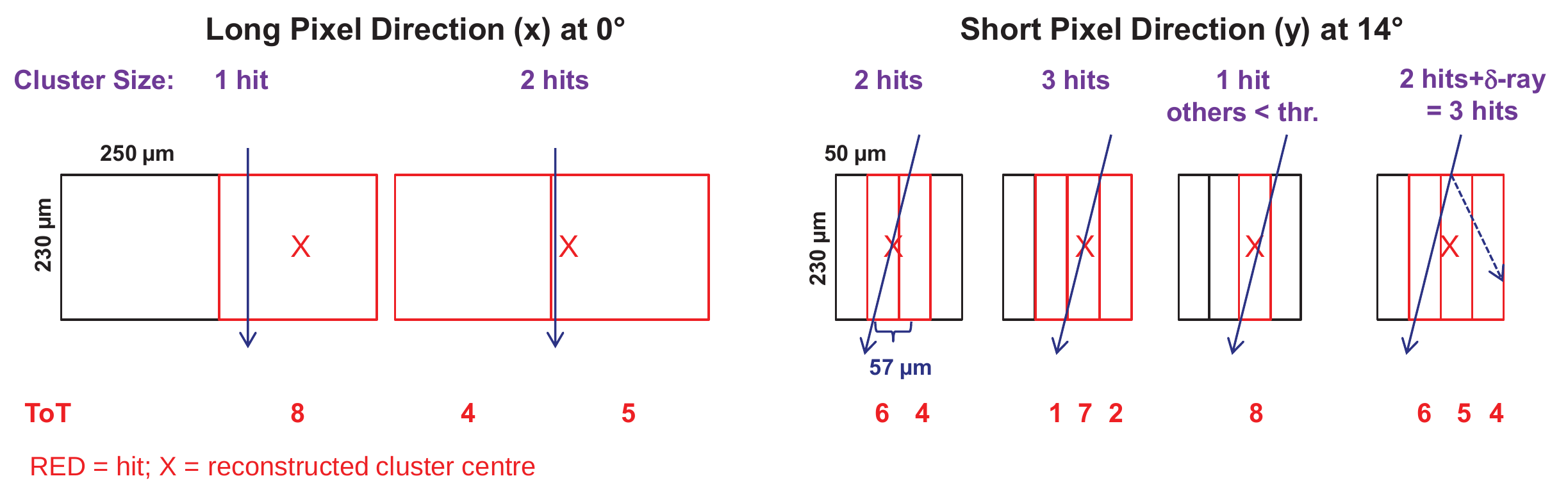}
	\caption{Sketch of pixel hits and clusters for the long pixel direction $x$ at 0$^{\circ}$ (left) and the short pixel direction $y$ at 14$^{\circ}$ (right). The ToT values are examples.}
	\label{fig:clusterSketch}
\end{figure}

The cluster size is an important parameter for the position resolution as discussed in section~\ref{sec:resolution}. The situation is sketched in figure~\ref{fig:clusterSketch}. In the long pixel direction ($x$), charge sharing is minimal and most of the clusters (about 97\%) extend only over one pixel, similar for all operational parameters and angles in $y$. A higher degree of charge sharing takes place in the short pixel direction ($y$). At standard operational parameters (figure~\ref{fig:hitClusterDifferentAngles}) and 0$^{\circ}$, 81\% of the clusters still have a cluster size $y$ of 1, but in 17\% of the cases charge is shared between 2 pixels; a cluster size $y$ of more than 2 happens for 2\% of the clusters and must be due to delta rays and noise since this is geometrically not possible at 0$^{\circ}$. Charge sharing in $y$ is strongly enhanced at 14$^{\circ}$ tilt since the particles travel over 57\,$\mu$m in that direction (at a pixel pitch of 50\,$\mu$m) and hence pass through mostly 2 pixels (measured for 81\% of the clusters) and sometimes 3 (measured for 4\% of the clusters). However, due to the threshold occasionally one pixel does not fire, so that for 14\% of the clusters the cluster size $y$ is found to be 1. From figure~\ref{fig:hitClusterDifferentParameters} it can be seen that at 14$^{\circ}$, charge sharing in plane 1 is increasing with voltage since this is a CNM sensor with non-fully-through-passing columns, which is not yet depleted throughout its whole depth at low voltages. FBK devices with fully-through-passing columns, on the contrary, show hardly any voltage dependence for the cluster size. When increasing the threshold, the events with a cluster size $y$ of 1 are increasing, as expected. No strong dependence on the ToT tuning is observed.

At standard operational parameters (figure~\ref{fig:hitClusterDifferentAngles}), the distribution of the hit ToT peaks at 8 for 0$^{\circ}$, which is consistent with the expectation of 17\,ke$^{-}$ most probable deposited charge and no charge sharing. For 14$^{\circ}$ the ToT distribution is shifted to lower values with a peak at 6 due to enhanced charge sharing. After the ToT-to-charge conversion, the cluster charge (the sum of all pixels hit) is roughly consistent between 0$^{\circ}$ (most probable value MPV=15.5\,ke$^{-}$) and 14$^{\circ}$ with an MPV of 16.8\,ke$^{-}$ (from geometry one would expect 3\% more charge at 14$^{\circ}$ than at 0$^{\circ}$) and expectations. It should be noted that the 4-bit ToT resolution and the non-linearity between ToT and charge (see equation~\ref{eq:ToTtoQ}) are not ideal for precision charge measurements, which is not the aim of these devices. From figure~\ref{fig:hitClusterDifferentParameters} it can be seen that the hit ToT slightly increases with voltage in plane 1 (due to more efficient charge collection and more complete depletion for the CNM device) and decreasing threshold (mainly an effect of the re-definition of the ToT-to-charge relation since \eg ToT=1 is by definition close to the threshold and hence moves when changing it). Obviously, when changing the ToT tuning, the ToT distribution is highly affected and can be shifted to much lower values for 5@20\,ke$^{-}$, as discussed already in section~\ref{sec:TDAQ} for the maximum hit ToT in the event.

The distributions for the other planes are consistent with plane 1 when considering the chip-to-chip charge-calibration spread of 15\%, except for the fact that the FBK sensors are fully depleted already at 1\,V and hence show almost no voltage dependence.

\subsubsection{Track reconstruction}
\label{sec:trackReco2}
Subsequently, tracks were reconstructed from the clusters by fitting a linear function for each direction ($x$ and $y$) after applying a simple track-cluster-finding algorithm. At least three planes are required to have a hit included in the track. As input resolutions for the weights in the $\chi^2$ fit, the resolutions individually measured for different cluster sizes (from section~\ref{sec:resolution}) were used. Alignment was performed in two steps: first a coarse alignment based on the inter-plane correlations between two consecutive pixel layers (see figure~\ref{fig:onlineMon}, top right) was applied; subsequently a fine alignment was performed based on the track residual distributions (\ie the difference between the projected track position on each layer and the cluster position). Shifts in $x$ and $y$ and rotations around the z axis were corrected for. For more details see reference~\cite{bib:Judith}.

Track reconstruction of the AFP-prototype test-beam data has been performed for different scenarios: 

\begin{enumerate}
	\item The \emph{all-plane scenario} includes all five planes into the track fit, which is mostly used for the analysis of the ToF detector as it gives the best precision at the ToF-detector position. 
	\item The \emph{first-four-plane (AFP-like) scenario} takes only the first four equidistant planes into account, thereby being the most realistic with respect to the final AFP configuration. 
	\item In the \emph{DUT scenarios}, specific planes are excluded from the track fitting and thus treated as independent, unbiased devices-under-test (DUT) for efficiency or resolution studies. 
\end{enumerate}

\begin{figure}[h]
	\centering
	\includegraphics[width=7.5cm]{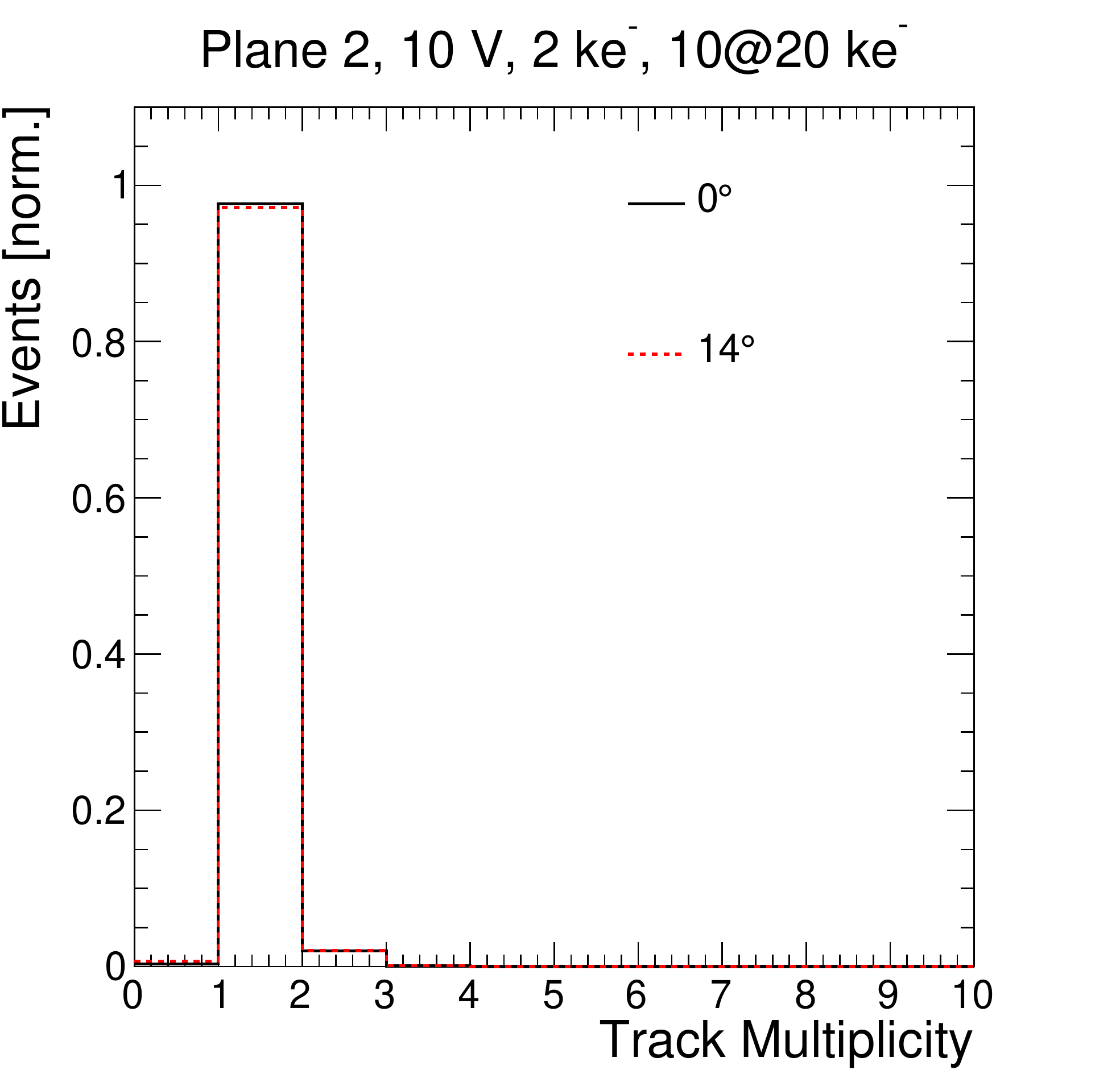}
	\includegraphics[width=7.5cm]{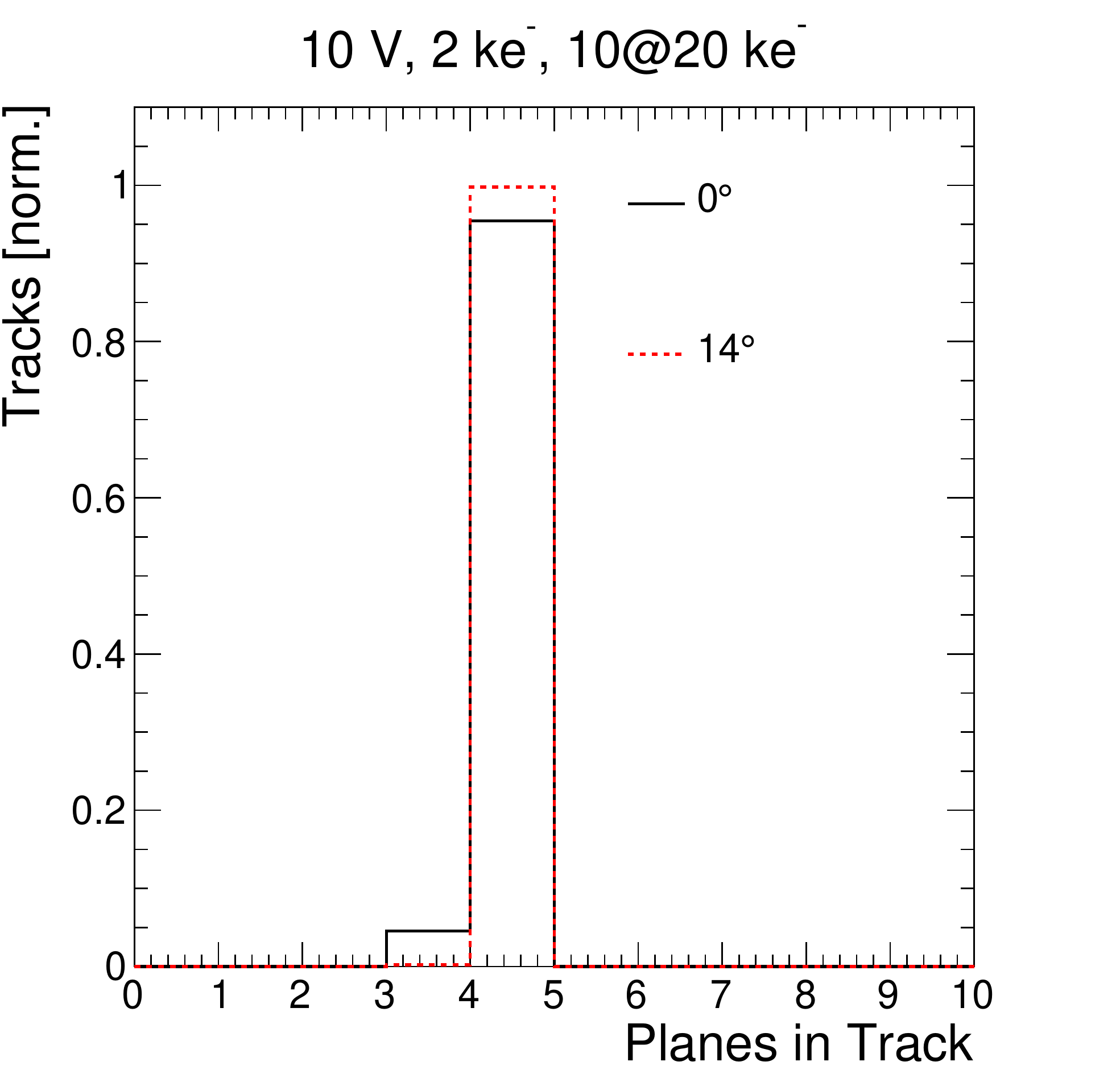}	
	\caption{ Track properties for a track reconstruction from the first four planes (AFP-like scenario) at 0 and 14$^{\circ}$: Track multiplicity (left) and number of pixel planes included in the track (right) for a central region. }
	\label{fig:trackProperties}
\end{figure}

The performance of track reconstruction is found to be similar for all scenarios (the following plots and numbers refer to the AFP-like first-four-plane scenario): for about 97\% of the events exactly one track is reconstructed as seen from figure~\ref{fig:trackProperties} (left). Events with no reconstructed tracks are at the percent level and mostly originate from tracks at the edges of the tracker where due to misalignment not enough planes were hit by the particle. Events with more than one reconstructed track are found in only 2\% of the cases. Figure~\ref{fig:trackProperties} (right) shows the number of planes included in the track. Due to the excellent hit efficiency at 14$^{\circ}$ (see section~\ref{sec:pixelEff}), almost all tracks include all four planes (in the central detector region). At 0$^{\circ}$, about 5\% of the tracks include only three planes, consistent with the two unbiased planes having an efficiency of 97.5\% (the two triggering planes have naturally an efficiency of 100\%). The reconstructed tracks are found to be parallel to the beam axis with an average angle in x and y of about 0.1$^\circ$.

For the analyses presented here, event cleaning cuts have been applied unless stated otherwise. Events with exactly one track and one cluster per plane have been selected to reduce combinatorial background and events with material interactions. 

\subsection{Spatial correlation between tracking and timing detector}

\begin{figure}[bt]
	\centering
	 \includegraphics[width=7.5cm]{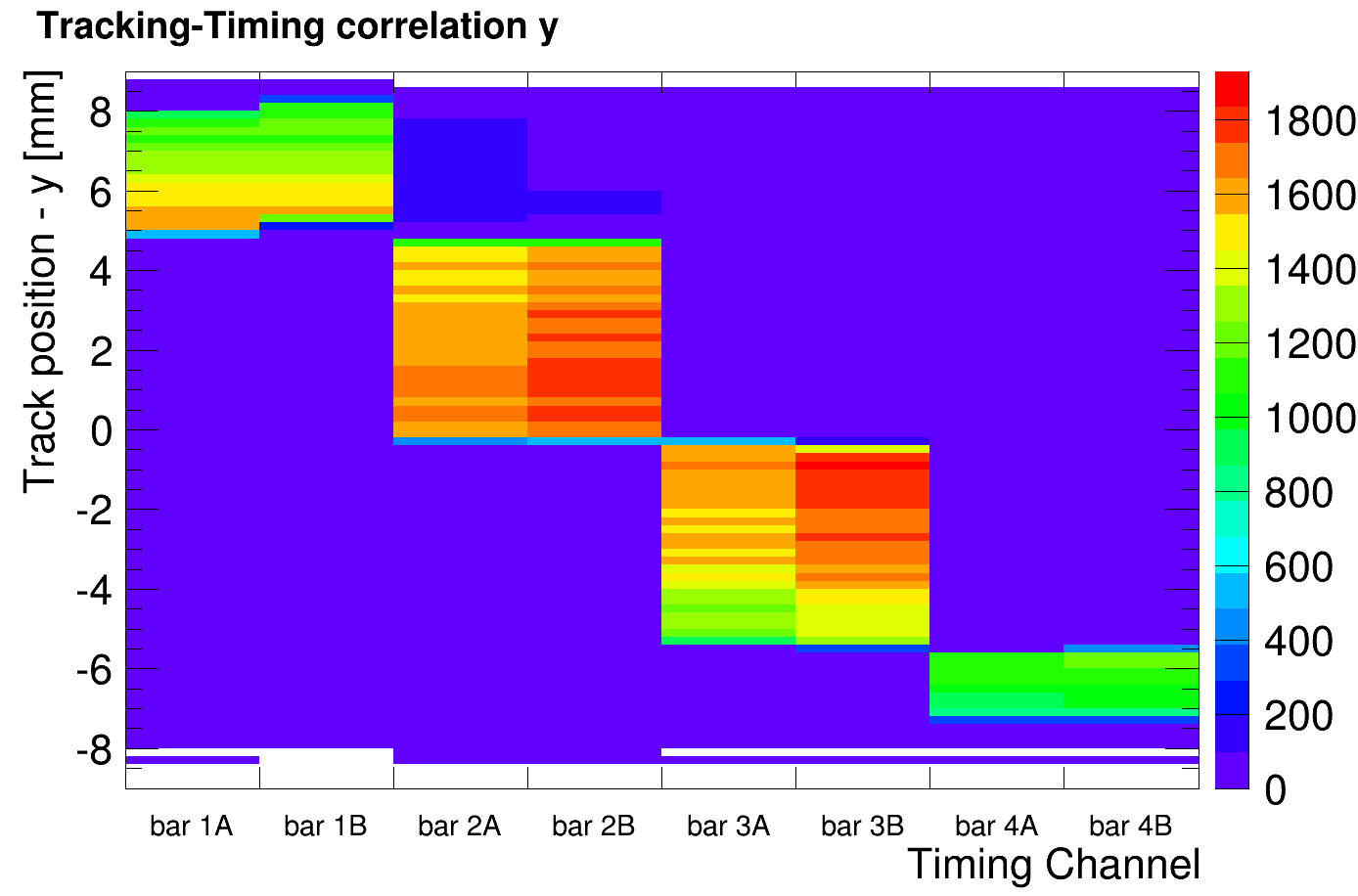}
	 \includegraphics[width=7.5cm]{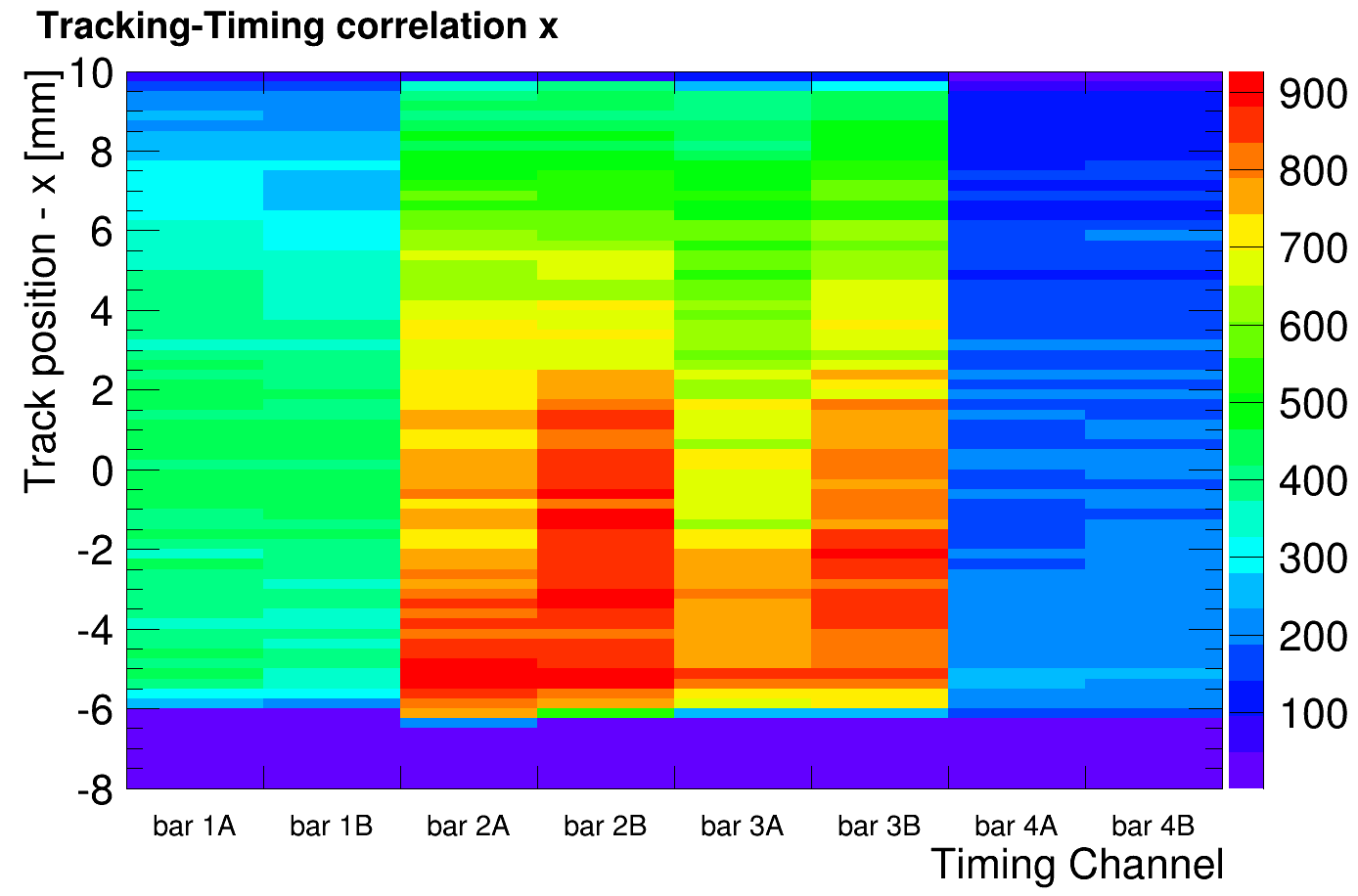}
	\caption{The spatial correlation between reconstructed-track position and ToF channels in the $y$ (left) and $x$ (right) direction at $V_{MCP-PMT}=1800$\,V for the tracker at 0$^{\circ}$. For the bar arrangement and naming scheme see figure~\ref{fig:ToFprototype}.}
	\label{fig:trackingTimingCorr}
\end{figure}

The main objective of the 2014 beam test was the integration of the tracking and timing subsystems with a common trigger and readout. To verify that this integration worked and that the recorded tracking and timing data were inter-related with each other, the spatial tracking-timing correlation was studied. Figure~\ref{fig:trackingTimingCorr} shows the number of events as a function of the track position extrapolated to the timing detector (using the all-plane track-reconstruction scenario with cleaning cuts as described in section~\ref{sec:trackReco2}) and the firing timing channel (the eight LQbars). 


It can be seen that the track position and timing channels that give a signal are clearly correlated with each other in space. In the $y$ direction, the four trains piled on top of each other can be clearly seen. Even a small misalignment between bar A and B of train 1 is visible, which has been confirmed by optical inspection. Train 4 is only partly visible due to the limited pixel-detector and trigger acceptance below -7\,mm. In the $x$ direction it can be seen that the overlap between LQbars and the pixel detectors ranges from -6\,mm (the LQbar cut edge parallel to the beam) to the end of the pixel detector at about 10\,mm. 

These correlation distributions were also used for offline alignment between the timing channels and the tracking detector for the following analyses.

\subsection{Material interactions}
\label{sec:material}

\begin{table}[htb]
			\centering
			\caption{Overview on material interactions, compared for the beam-test setup with and without 2\,mm plastic cover (930\,$\mu$m Si + 500\,$\mu$m Al + 500\,$\mu$m Kapton flex, 1.6\,cm LQbars, 120\,GeV particles) and the final AFP detector (930\,$\mu$m Si + 500\,$\mu$m Al, 3.2\,cm LQbars, 6.5\,TeV particles): the detector material in terms of nuclear interaction length $\lambda_I$ (ToF and tracker) and radiation length $X_0$ (tracker), the mean multiple-scattering angle $\theta_{0,MS}$ and the corresponding offset $d_{MS}$ at the next tracking plane (5\,cm pitch for the beam test, 9\,mm for the final AFP detector). The material constants and formulas are from reference~\cite{bib:PDG}.}
			\label{tab:materialInteractions}
			\begin{tabular}{|l|c|c|c|c|c|c|c|}
			
\hline												
	&	\multicolumn{1}{|c|}{ToF}	&	\multicolumn{4}{|c|}{Tracker (per Plane)}								\\
\cline{2-6}												
Setup	&	$x/\lambda_I$ [\%]	&	$x/\lambda_I$ [\%]	&	$x/X_0$ [\%]	&	$\theta_{0,MS}$ [$\mu$rad]	&	$d_{MS}$ [$\mu$m]		\\
\hline												
Beam Test (no Cover)	&	3.6	&	0.4	&	1.7	&	12.4	&	0.6		\\
Beam Test (Cover)	&	3.6	&	0.7	&	2.3	&	14.6	&	0.7		\\
Final AFP	&	7.3	&	0.3	&	1.5	&	0.2	&	$2\times 10^{-3}$		\\
\hline

\end{tabular} 
	\end{table}

\begin{figure}[hbtp]
	\centering
	 \includegraphics[width=15cm]{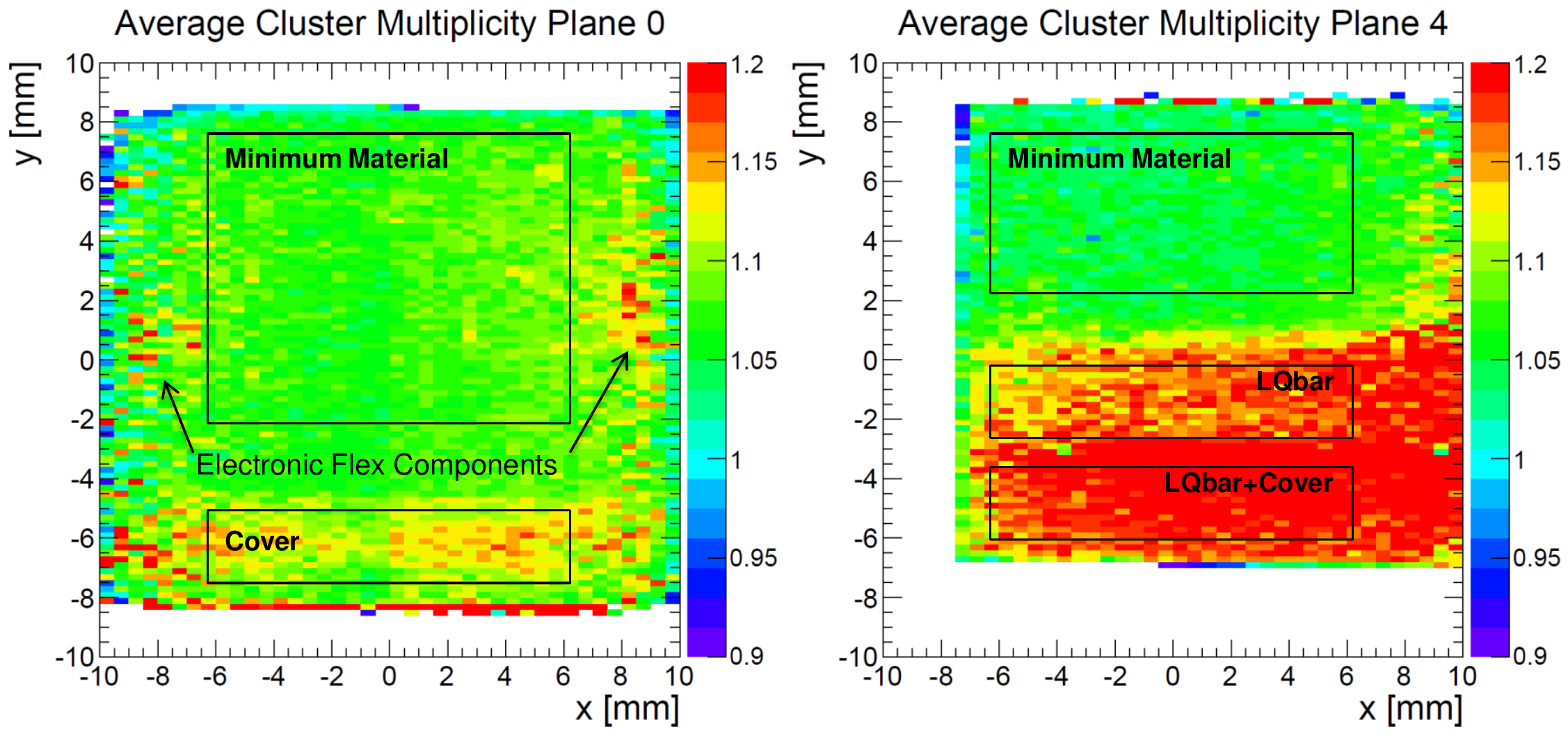}
	  \includegraphics[width=7.5cm]{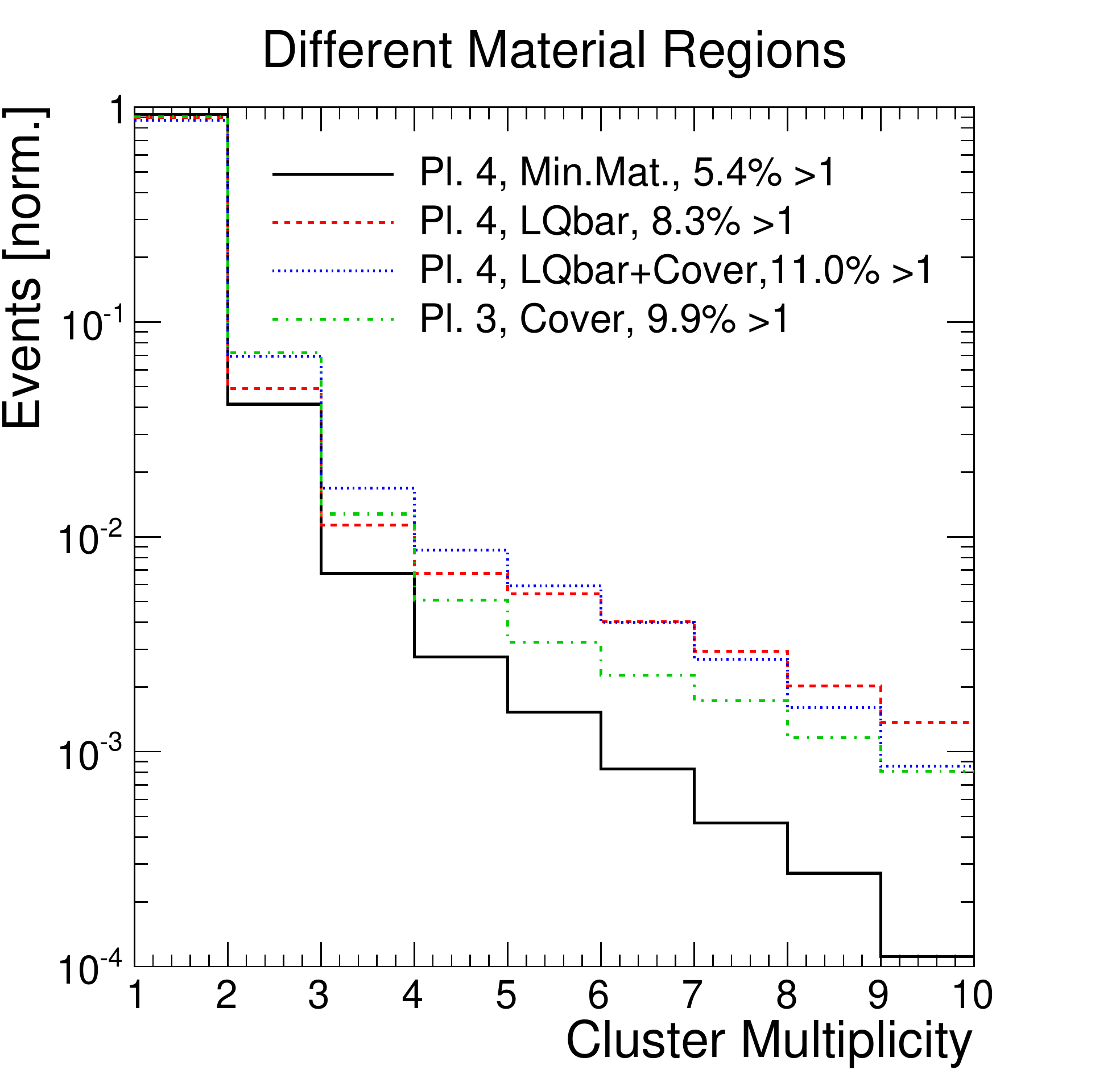}
		\includegraphics[width=7.5cm]{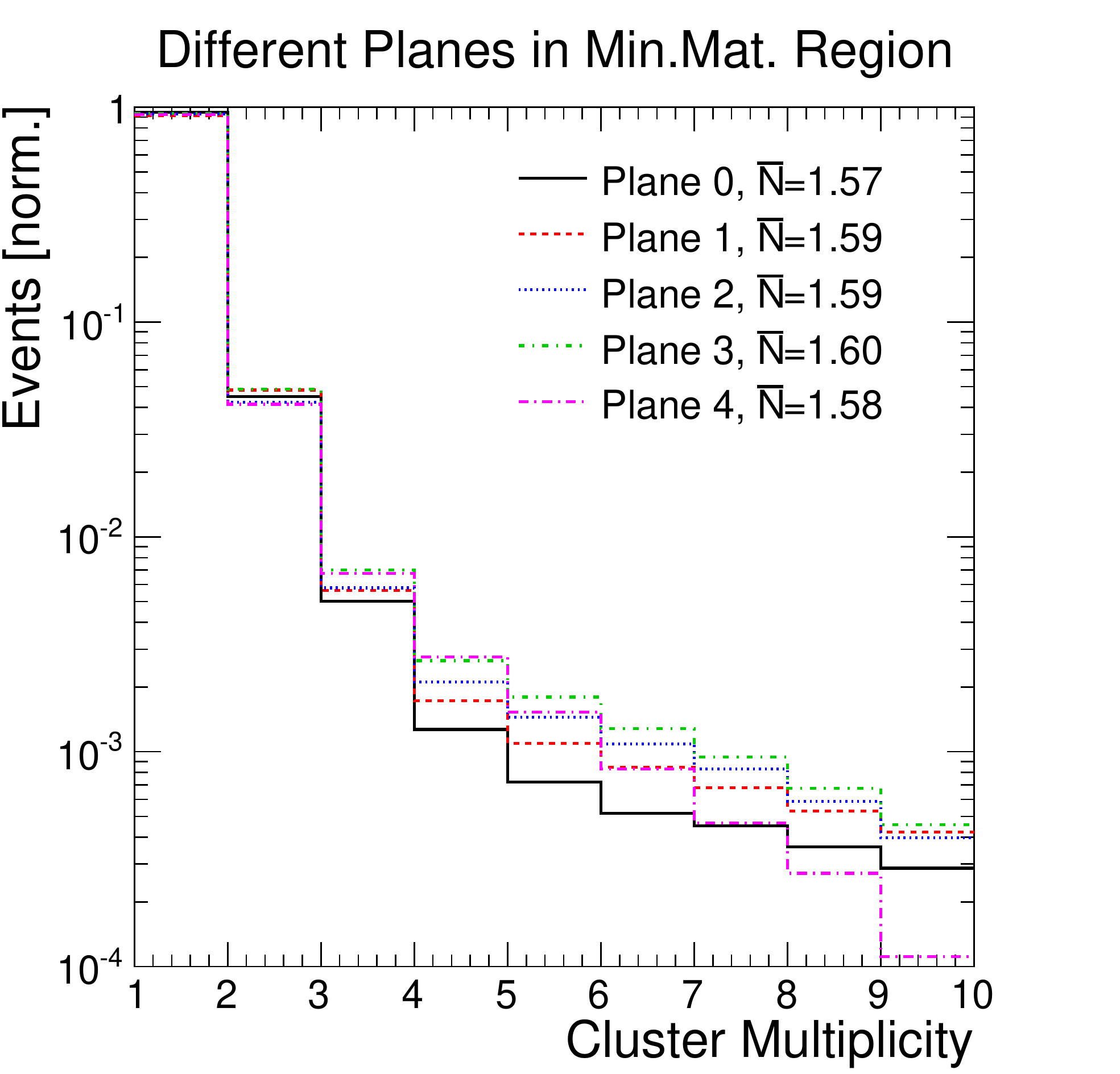}
	\caption{Top: Average cluster-multiplicity map for planes 0 and 4 as a function of track position at 0$^{\circ}$ tilt. Indicated are the different selected material regions of interest (the total areas covered by the plastic or the LQbars are larger). Bottom: The cluster-multiplicity distribution for different regions of planes 4 and 3 (left) and for different planes in the minimal-material region (right).}
	\label{fig:interactions}
\end{figure}


It is important to understand particle interactions in the detector material such as multiple scattering, nuclear interactions and delta rays since they could degrade the performance. Table~\ref{tab:materialInteractions} gives an overview on the detector material in terms of nuclear-interaction and radiation length and the corresponding expected multiple-scattering parameters. This is compared for the beam-test setup (with only two LQbars per train, with a flex on top of the pixel sensor, and with and without plastic cover) and the final AFP detector (with four LQbars per train and no flex or cover).

The tracker material in terms of radiation length $X_0$ is 1.7\% (2.3\%) per tracker plane for the beam-test setup without (with) the plastic cover. This implies a mean multiple-scattering angle $\theta_{0,MS}$ of 12.4~(14.6)\,$\mu$rad for 120\,GeV particles, which leads to a mean offset of 0.6 (0.7)\,$\mu$m at the next tracker plane for the setup with 14$^\circ$ tilt with a plane pitch of 5\,cm. This is an important parameter for the resolution measurements in section~\ref{sec:resolution}. For the final AFP detector without the flex, the material budget in terms of radiation length is reduced to 1.5\%. Together with the high proton momentum of close to 6.5\,TeV, the multiple-scattering effects are negligible with only 0.2\,$\mu$rad scattering angle and a mean offset of $2\times 10^{-3}$\,$\mu$m at the next tracking plane with a distance of 9\,mm.

Nuclear interactions in the Quartz material of the ToF detector could give rise to secondary particles that might affect the time resolution. These additional particles should be measurable in the last tracking plane 4 behind the ToF detector as additional hits and charge deposition. From the nuclear interaction length of $\lambda_I=$44\,cm for Quartz, one would roughly expect an interaction for 3.6\% of the events for a two-LQbar train of 1.6\,cm depth along the beam. For the final AFP detector with four LQbars per plane, this would increase to 7.0\%. On the contrary, for each tracking plane 
one would expect a nuclear interaction in only 0.4\% of the events. This increases to 0.7\% if the sensor is covered by the 2\,mm plastic protection and decreases to 0.3\% for the final AFP detector without the flex on the sensor. 



In the 2014 data, no signs of Quartz material interactions were found from the average cluster multiplicity and deposited charge of plane 4~\cite{bib:AFPreference1}. It was suspected that the 2\,mm thick plastic cover in front of each pixel sensor absorbed part of the secondaries from upstream material, as well as produced new secondary particles, in particular delta rays, that presented a large background for the upstream secondaries. Hence, in 2015 a hole in the plastic cover was made over most of the sensitive pixel area (see section~\ref{sec:prototypeTracking}, in particular figure~\ref{fig:PixelModule}). 

Figure~\ref{fig:interactions} shows the pixel map for the average cluster multiplicity for planes 0 and 4 for a 2015 run in which only the two lower trains 3+4 of the ToF detector were included. One can clearly distinguish the different regions. The cluster multiplicity is significantly higher for the region covered by plastic. However, also in the region without cover, one can clearly identify localised material interactions from the electronic components of the flexible circuit board (cf. figure~\ref{fig:PixelModule}). Since those will not be on top of the sensor for the final AFP pixel modules, a so-called \emph{minimal-material region} is defined avoiding the largest of these components as well as the plastic cover. For plane 4 one can now clearly distinguish the region with the interactions in the LQbars from the region without. 
Figure~\ref{fig:interactions} (bottom left) shows the full cluster-multiplicity distribution for the different regions of plane 4, namely the minimal-material, LQbar and LQbar+cover region. The LQbar region has a significantly higher cluster multiplicity (8.3\% of the events have $>$1 cluster) than the minimal-material region (5.4\% of the events have $>$1 cluster). This is roughly consistent with the expected 3.6\% of the events with nuclear interactions in the LQbars. Adding the plastic cover to the LQbar region increases the number of events with $>$1 cluster to 11.0\%, probably mostly due to delta rays. It is interesting to note that the fraction of events with multiplicities between 2--5 increases (consistent with delta rays that are usually produced with low multiplicity), whereas the multiplicities above 5 decrease, probably due to absorption of secondaries from the LQbars in the plastic cover. The cover region of plane 3 is also added to the figure in order to show that it has a roughly similar multiplicity distribution as the LQbar+cover region of plane 4, which explains why no clear signs of Quartz material interactions were seen in the 2014 data with the plastic cover everywhere.

For the tracker, it is also important to know whether the upstream planes influence the downstream ones. Figure~\ref{fig:interactions} (bottom right) shows the cluster-multiplicity distribution for the minimal-material region compared for all planes. The average cluster multiplicity $\bar{N}$ does not seem to increase systematically from one plane to the other. This is mainly because the number of events with 2 clusters (which is dominating over events with higher multiplicities) is similar for all planes. These events are probably dominated by low-energetic delta rays that are immediately absorbed in one plane. However, the number of events with larger multiplicities, which probably mostly stem from nuclear interactions, does increase with plane number since those secondaries can pass through several planes and can hence accumulate and multiply. Only the last plane, 4, does not follow this trend, possibly due to secondaries with an angle that miss the last plane because of the much larger distance to the previous planes. In any case, these events with high multiplicities still stay at the sub-percent level. 

The impact of the tracker material interactions is expected to be slightly different for the final AFP detector with different tracker holders, the Roman pot outside, no flex with electronic components on top of the sensor and a reduced pitch of 9\,mm between the planes. Hence, these studies will be repeated for the final detector in the next beam test in 2016. For a better understanding of the nature of the interactions, simulations will need to be performed and compared to the data.

In order to minimise the influence of material interactions in these analyses, the tracker studies are typically performed for events with tracks in the minimal-material region, and for most tracker and ToF studies a cluster multiplicity of maximally 1 in each plane is required.

\subsection{Performance of the tracking system}
\label{sec:AFP_TB_tracking}
\subsubsection{Spatial resolution}
\label{sec:resolution}

\begin{figure}[!hbt]
	\centering
	 \includegraphics[width=7.5cm]{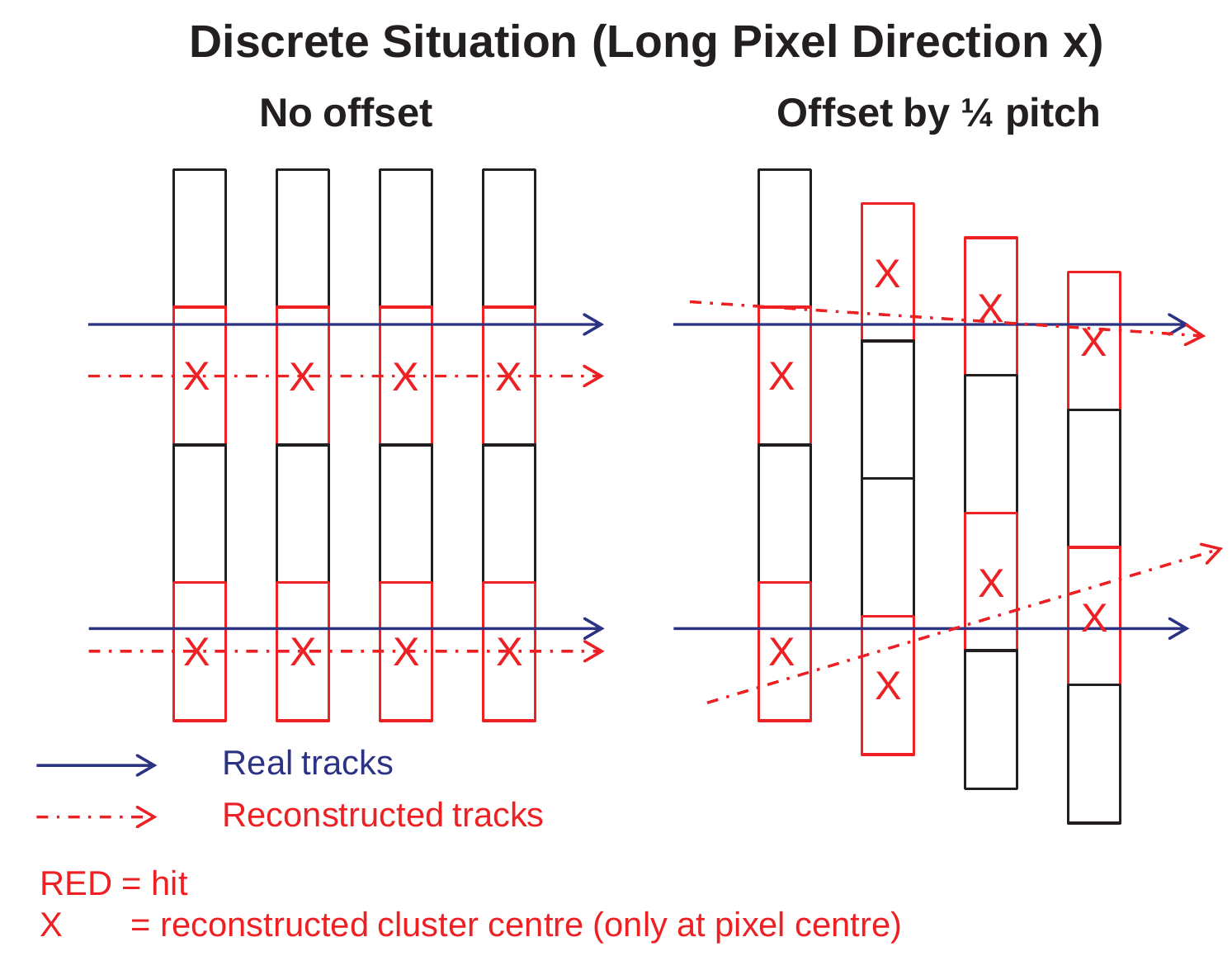}
	 \includegraphics[width=7.5cm]{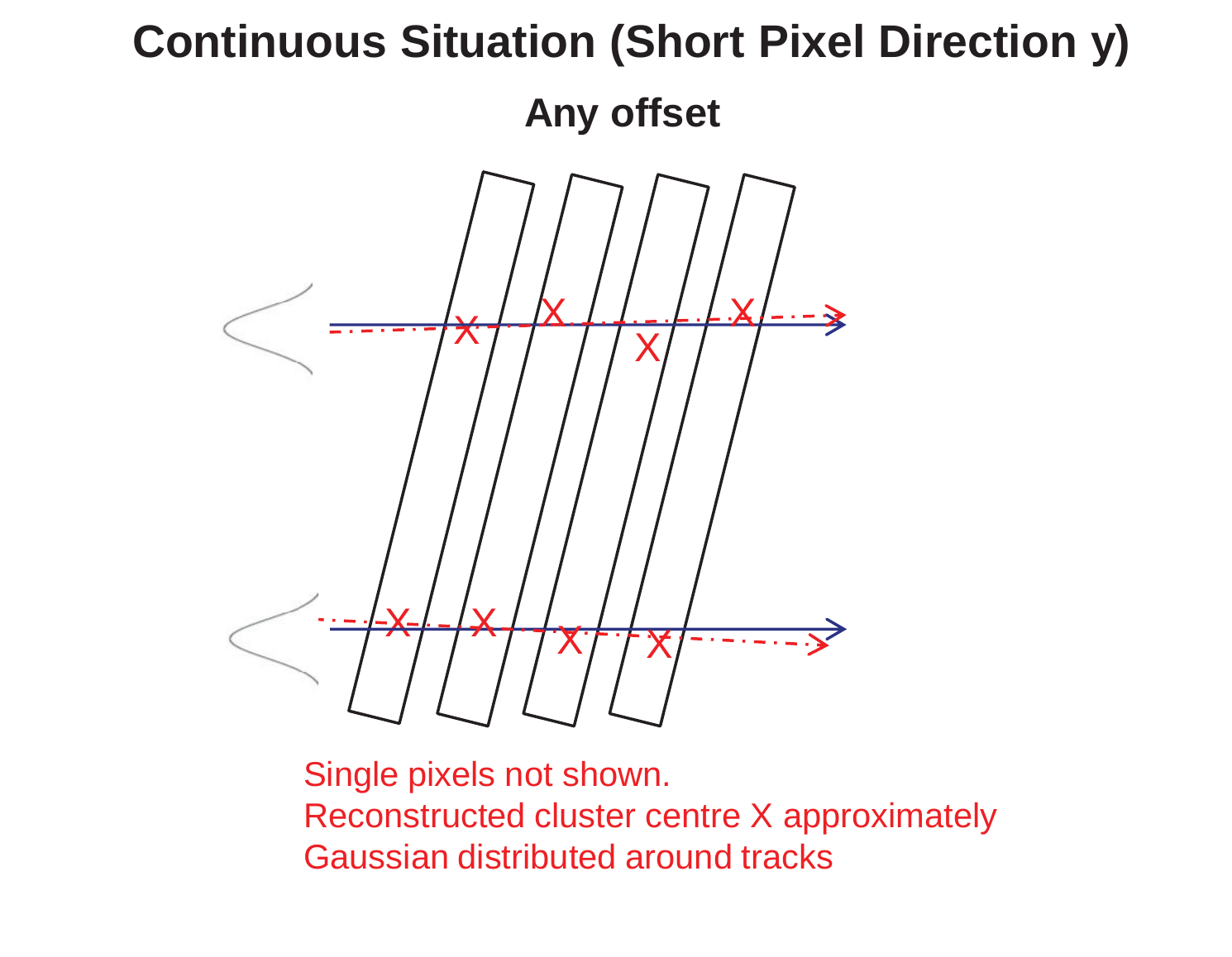}
	 \includegraphics[width=7.5cm]{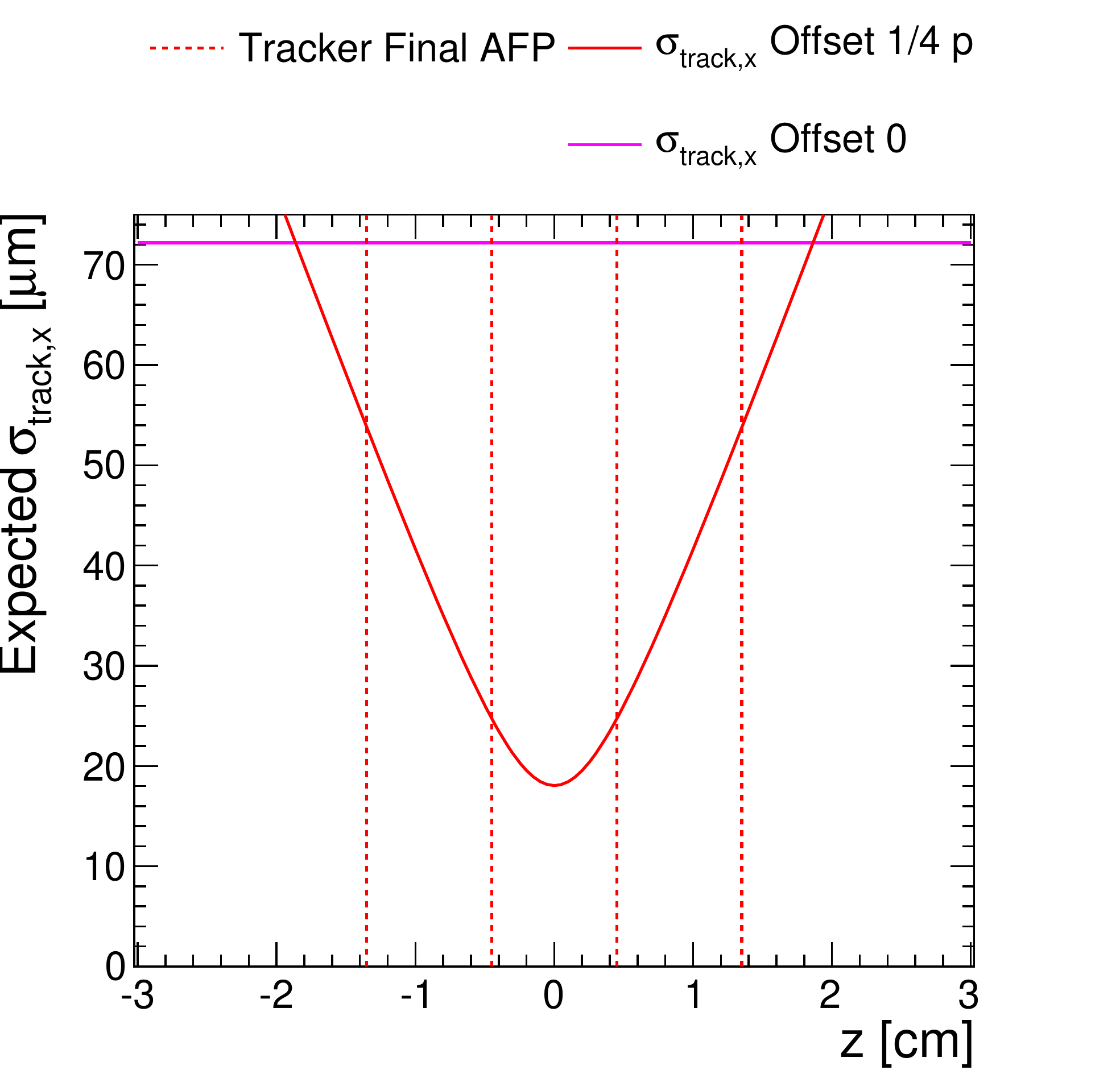}
 	\includegraphics[width=7.5cm]{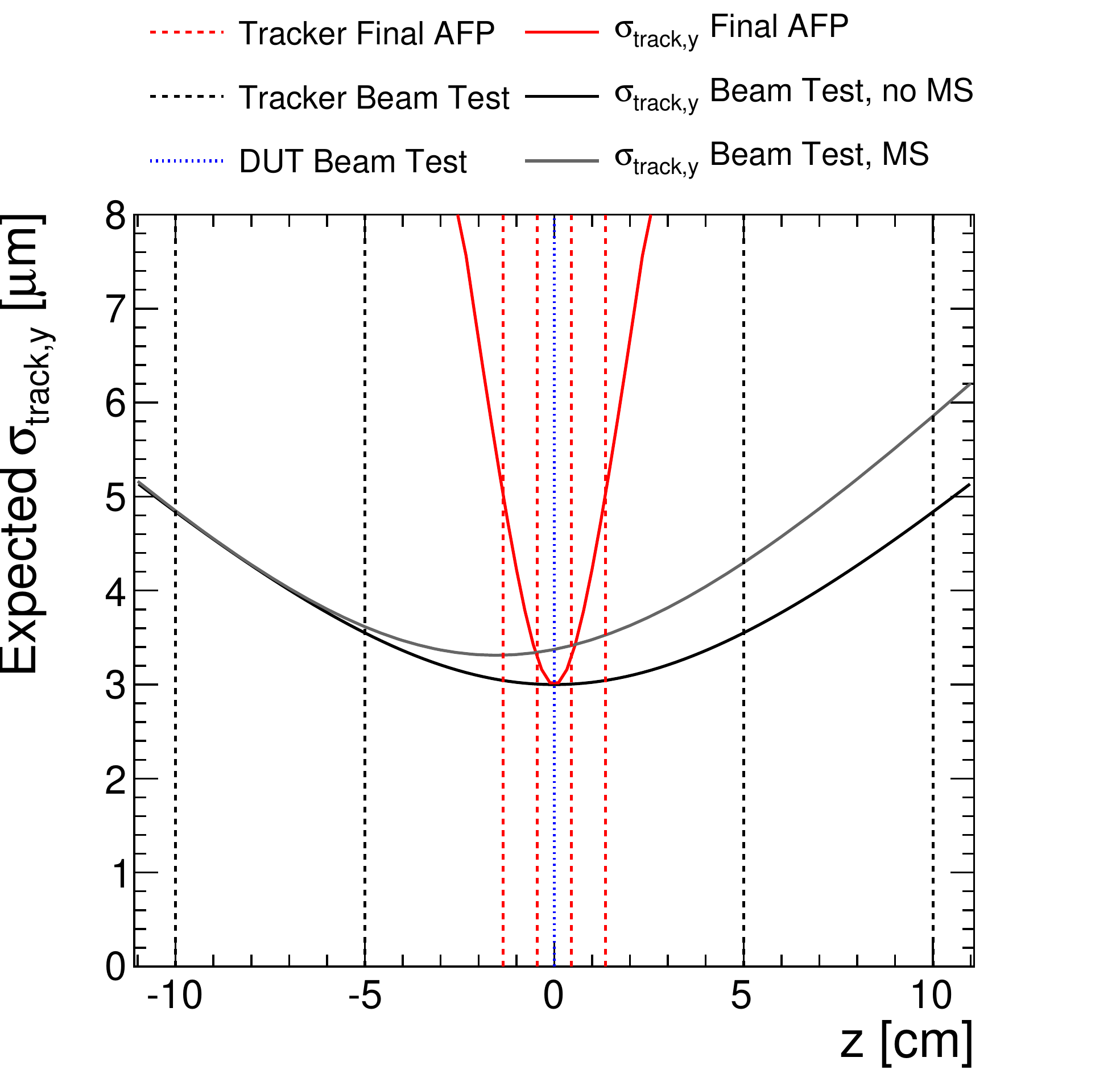}
	\caption{Top: Sketches of the track reconstruction for the discrete situation in the long pixel direction (left), where most of the clusters have only one hit, so that its position is almost always assigned to the pixel centre (this is shown for different staggering scenarios with an offset of 0 or 1/4 pitch between successive planes); and for the continuous situation in the short pixel direction (right), where the reconstructed cluster position approximately follows a Gaussian distribution around the track. Bottom: The corresponding expected track resolution as a function of $z$ in the long pixel direction (left) for different staggering scenarios for the final AFP tracker; and in the short pixel direction (right) at 14$^\circ$ tilt with 6\,$\mu$m input resolution per plane for the final AFP tracker, as well as for the plane-2-as-DUT track-reconstruction scenario in the beam test, with and without MS effects from the beam-test tracker planes.}
	\label{fig:trackPrecision}
\end{figure}

\subsubsection*{General spatial-resolution considerations}

The spatial resolution of each pixel plane as well as of the whole AFP tracker system is a crucial performance parameter. Each AFP station acts as a beam telescope of four planes and can provide a full track (position and angle). However, to measure the angle of the forward proton tracks precisely, the two AFP stations on each side of the IP, separated by 12\,m, are combined, which is not studied here. The most important track-performance parameter for each station individually is the combined position resolution of the four-plane system, which is best in the centre of the four equidistant planes as discussed below.

It has to be distinguished between qualitatively different situations in the short and long pixel direction: 

\begin{enumerate}

\item \emph{Digital/binary/discrete} situation in the long (250\,$\mu$m) pixel direction (beam test $x$/AFP vertical coordinate): 
	
If only one pixel is hit in a certain direction, as is the case in 98\% of the events in the long pixel direction (see figure~\ref{fig:hitClusterDifferentAngles} centre left), the measured hit-cluster position only assumes discrete values, namely the pixel centre as sketched in Figs.~\ref{fig:clusterSketch} and~\ref{fig:trackPrecision}.  In such a case, the digital or binary resolution of pitch/$\sqrt{12}$, \ie 72\,$\mu$m for a pitch of 250\,$\mu$m, is obtained, as confirmed in previous beam tests (see section~\ref{sec:tracker}). 
	
In such a discrete situation, the combined tracker resolution of several pixel planes highly depends on the offset of the planes with respect to each other as shown in figure~\ref{fig:trackPrecision}: \eg in case of perfect alignment (no offset of planes), no improvement over the single-plane resolution is achieved for perpendicular tracks since each pixel plane gives a redundant measurement. In the case of staggering the modules by 1/4 of a pitch, as planned for the final AFP detector, the track resolution for four pixel planes is improved to 1/4 of the binary resolution, \ie 18\,$\mu$m. However, the actual staggering present in this beam test was random since no precision alignment was available. Moreover, the resolution cannot be measured reliably with a setup of a few identical planes in the discrete regime (\eg in case of no offset of planes, the difference in hit position between successive planes is always 0). Hence, a resolution measurement in the long pixel direction is not pursued here but needs to be performed with the final AFP detector including the actual staggering achieved with an external precision telescope in future beam tests.

\item \emph{Analog/continuous} situation in the short (50\,$\mu$m) pixel direction at 14$^\circ$ tilt (beam test $y$/AFP horizontal coordinate): 
	
In the short pixel direction at 14$^\circ$ tilt, due to charge sharing between pixels, the hit position can be interpolated using the ToT or charge information, giving a measurement with approximately continuous values. As explained in section~\ref{sec:prototypeTracking}, such a situation was studied in special runs with all five pixel planes at 14$^\circ$ tilt with an equidistant pitch of 5\,cm without the ToF detector.

In this case, neglecting multiple scattering, the track resolution $\sigma_{track,y}$ as a function of $z$ can be predicted from the uncertainties of the straight-line parameters of the fit (slope $s_{y}$ and offset $b_{y}$) and their covariance $cov_{y}$ as
\begin{equation}
\sigma_{y}^2(z) = \sigma_{b_{y}}^2 + z^2 \cdot \sigma_{s_{y}}^2 + 2z \cdot cov_{y}.
\label{eq:resolution1}
\end{equation}
This is shown in figure~\ref{fig:trackPrecision} with an input resolution of 6\,$\mu$m per plane (as measured below) for the track-reconstruction scenario with the central plane 2 excluded as DUT from the track fitting. This is compared to the final AFP configuration with four planes with 9\,mm pitch. In both cases the track uncertainty is symmetric around its minimum in the centre of the four tracking planes. The minimum track uncertainty is the same in both cases, namely half of the input resolution per plane, \ie 3\,$\mu$m as it scales with $1/\sqrt{N_{planes}}$. 
In the continuous case, the resolution can be measured with a setup of a few identical planes like the one in this beam test. 

The effect of multiple scattering on the track resolution has been assessed in a simple Monte-Carlo simulation, which introduces a Gaussian-distributed multiple-scattering angle $\theta_{0,MS}$ at each plane (see section~\ref{sec:material}) as well as a Gaussian smearing of the hits around the true particle position according to the resolution. Then the simulated hits are fitted with a straight line and compared to the true track position (in the absence of multiple scattering). The expected degradation of track resolution due to multiple scattering in the tracker centre is found to be 0.4\,$\mu$m for the beam test with 120\,GeV particles, as shown in figure~\ref{fig:trackPrecision}. For the final AFP detector with 6.5\,TeV protons, the effect is found to be negligible.

\item \emph{Mixed} situation in the short pixel direction at perpendicular incidence:

For the short pixel direction at 0$^\circ$, there is a mixed situation of cluster size 1 in about 85\% of the cases (digital) and cluster size 2 in about 12\% of the cases (to which analog algorithms can be applied). However, since this configuration is well studied in previous beam tests and has no relevance for the final AFP detector at 14$^\circ$ tilt, this will not be covered here.

\end{enumerate}

In the following, the resolution in the short pixel direction $y$ at 14$^\circ$ tilt (analog/continuous case) will be studied.

\subsubsection*{Spatial-resolution determination methods}

The resolution is determined with two different methods, namely with the triplet method, which gives the average single-plane resolution for three successive planes, and with the track-DUT method, which gives the convoluted resolution of a four-plane track and a single DUT plane.
\begin{enumerate}
	\item Average Single-Plane Resolution with the Triplet Method:
	
	One method to measure the average per-plane resolution of three identical planes without a track fit (only using the alignment constants from a previous track fit) is the \emph{triplet} technique. It defines a residual variable $res_{trip}$ from the hit position $x_i$ of three successive equidistant planes (\eg $i=1,2,3$) as

\[
res_{trip} = \frac{x_1+x_3}{2} - x_2.
\]
An effective average single-plane (SP) resolution can be obtained from the spread (RMS or $\sigma$ of a Gaussian fit) as $\sigma_{SP,trip} = \sqrt{2/3} \cdot \sigma_{trip}$ assuming the resolutions of all planes are equal and neglecting possible misalignment or multiple-scattering effects. In the simulation mentioned above, the effect of multiple-scattering has been assessed to be less than 0.1\,$\mu$m, which is conservatively included as systematic uncertainty.

\item Convoluted Track and Single-Plane Resolution with the Track-DUT method:

Another method to determine the resolution is to exclude one plane (DUT) from the track fit and calculate the residual $res_{track-DUT}$ as the difference between the DUT cluster centre and the track position extrapolated to the DUT. 
The spread ($RMS_{track-DUT}$ or $\sigma_{track-DUT}$) gives the convolution of the DUT single-plane resolution $\sigma_{SP,DUT}$ and the track resolution $\sigma_{track}$ at the position of the DUT plane. Neglecting multiple-scattering and assuming the equality of all planes, both contributions can be disentangled
by using Eq.~\ref{eq:resolution1} as illustrated in figure~\ref{fig:trackPrecision}, \eg using the relation that in the centre of a four-plane system the track resolution is expected to be half of the single-plane resolution. 

As mentioned above, it was found in simulations that for the beam-test conditions with 120\,GeV particles, multiple scattering would degrade the track resolution for a straight-line fit by about 0.4\,$\mu$m with respect to conditions without scattering. Hence, to calculate the track resolution for the beam-test conditions, one would need to correct for this. However, in the end the parameter of interest is the track resolution of the final AFP tracker for 6.5\,TeV particles with negligible multiple scattering. In the simulation, it was found that the convoluted $\sigma_{track-DUT}$ is hardly affected by multiple scattering in the beam test, namely by less than 0.05\,$\mu$m. This can be explained by the fact that the position displacement due to scattering builds up successively from one plane to the next, so that the displacement at the DUT is correlated to the one at other planes, and the effect largely cancels out in the difference between the DUT hit position and the track position extrapolated to the DUT. Hence, the bias for $\sigma_{track}$ under AFP conditions introduced by neglecting scattering in the analysis should be small. Nevertheless, for a conservative estimate of the systematic uncertainty due to multiple scattering, the full 0.4\,$\mu$m difference was taken.

\end{enumerate}

\subsubsection*{Spatial-resolution results}

\begin{figure}[!hbt]
	\centering
	 \includegraphics[width=7.5cm]{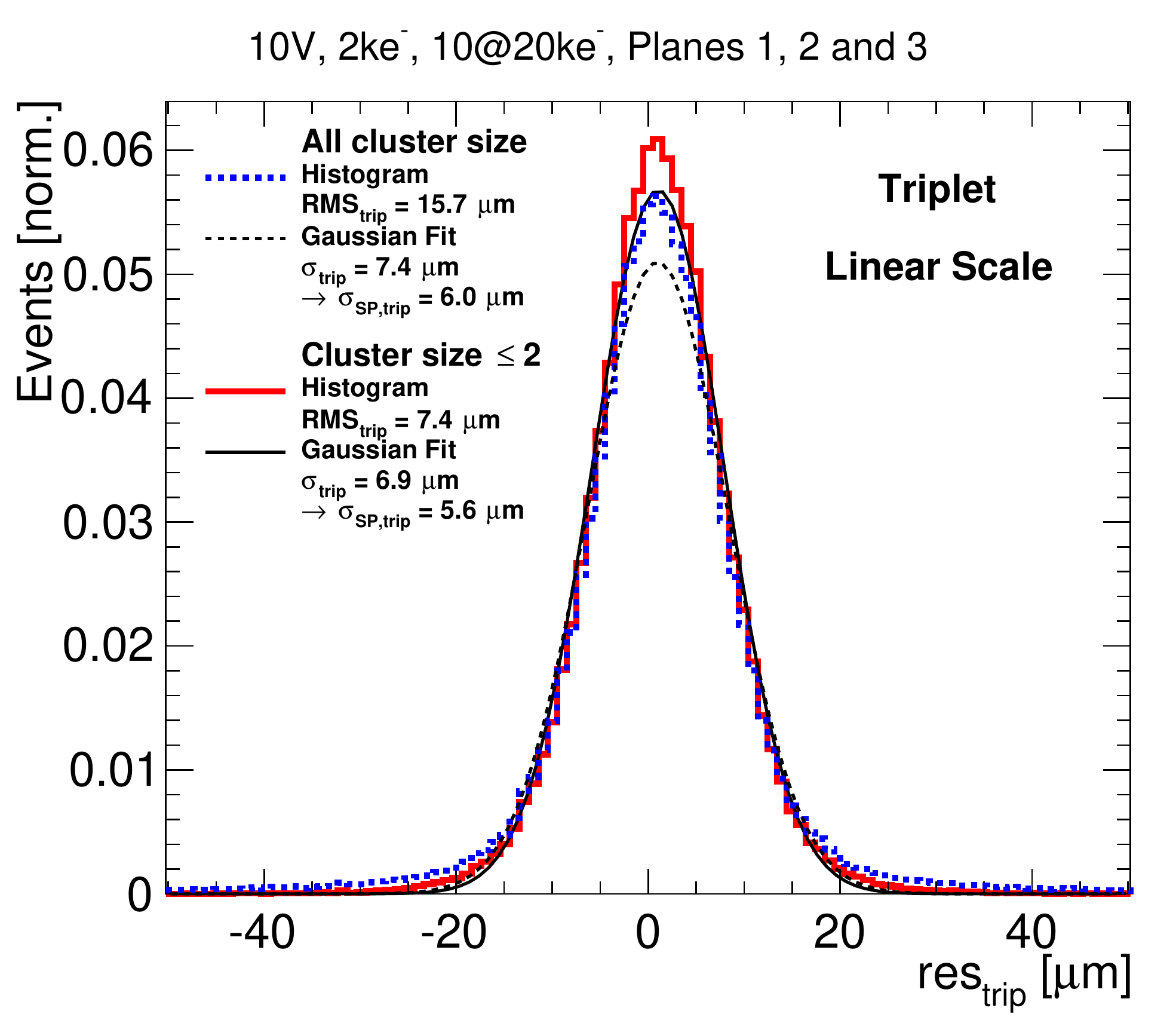}
	 \includegraphics[width=7.5cm]{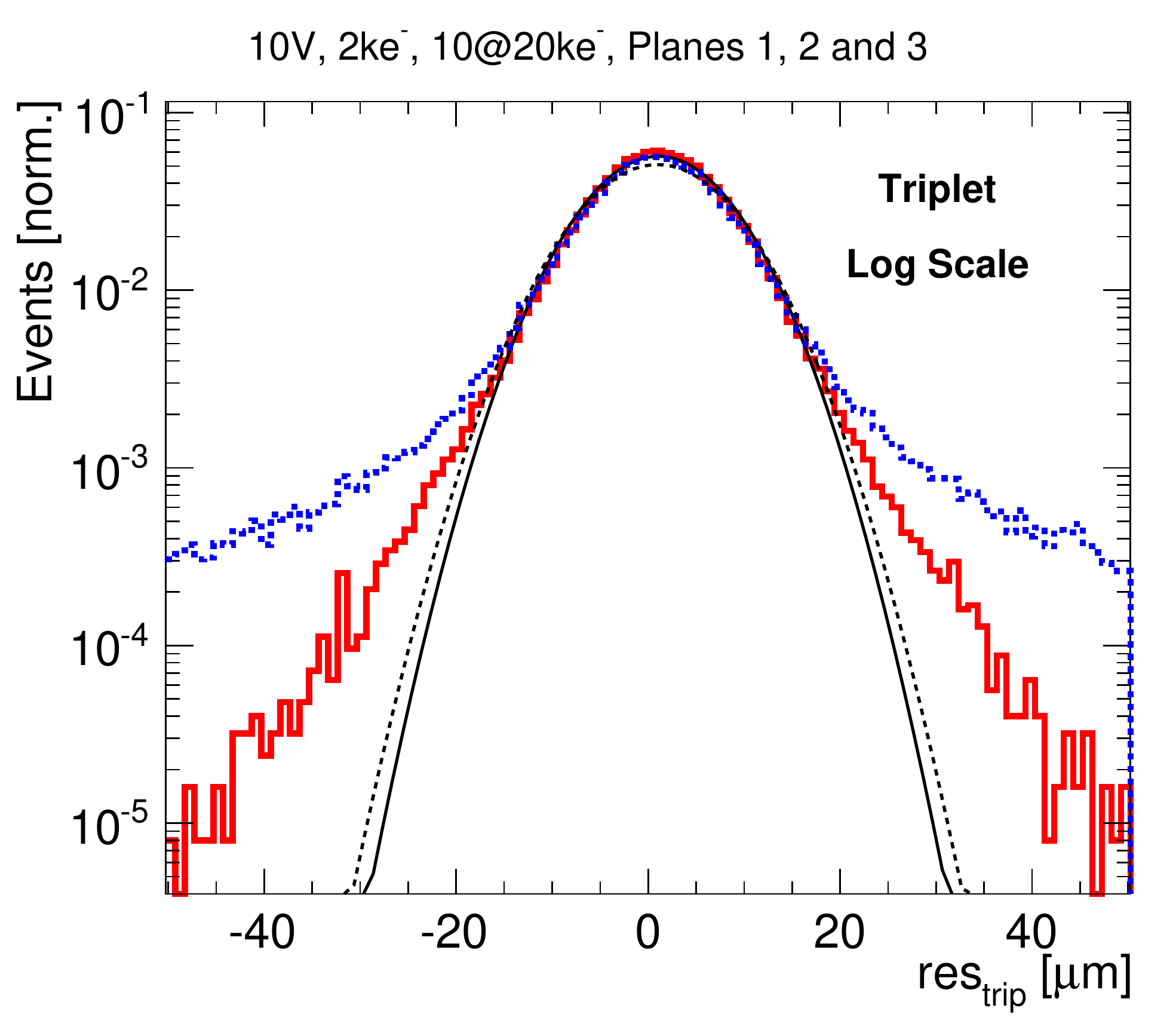}
	\includegraphics[width=7.5cm]{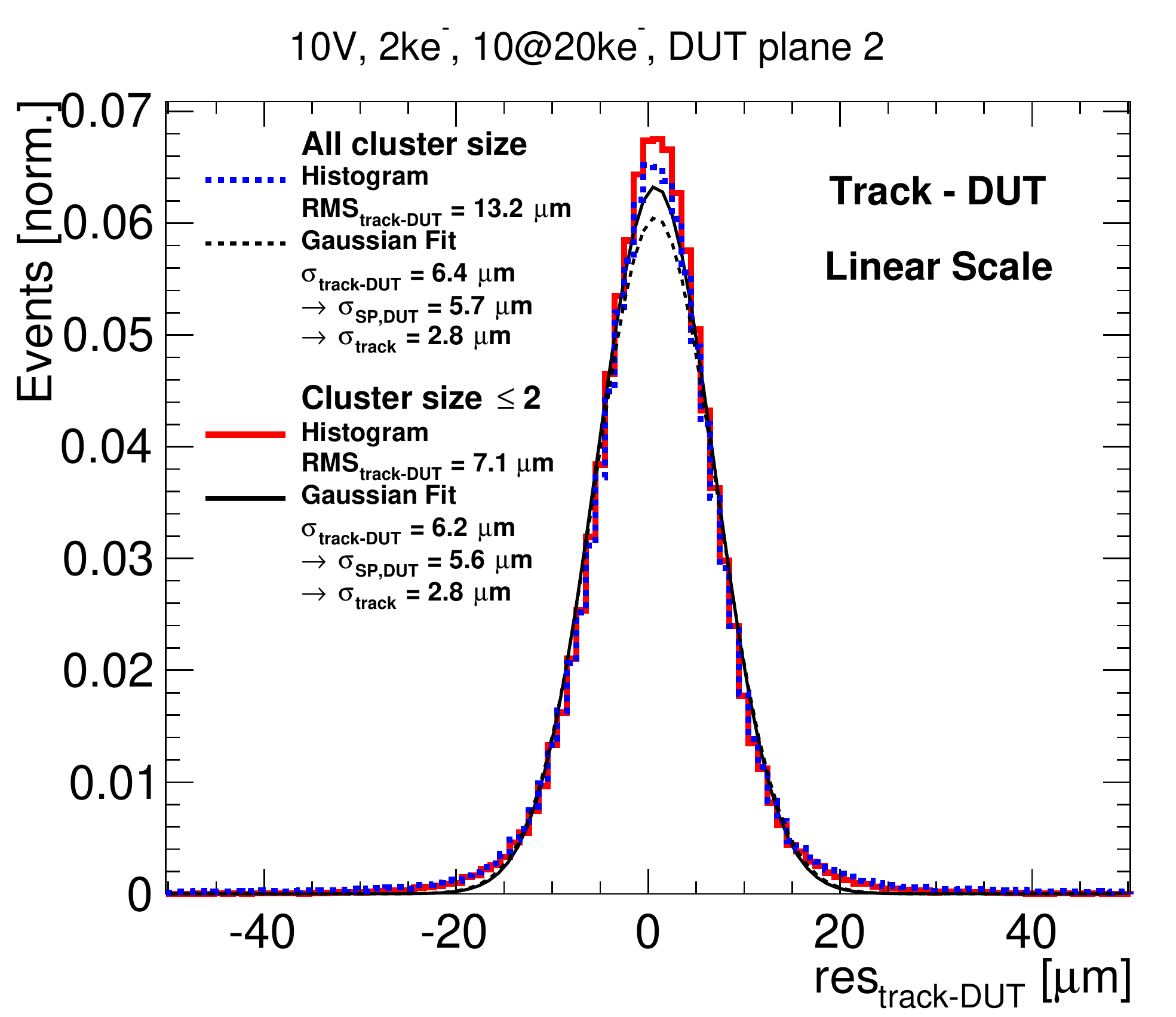}
	\includegraphics[width=7.5cm]{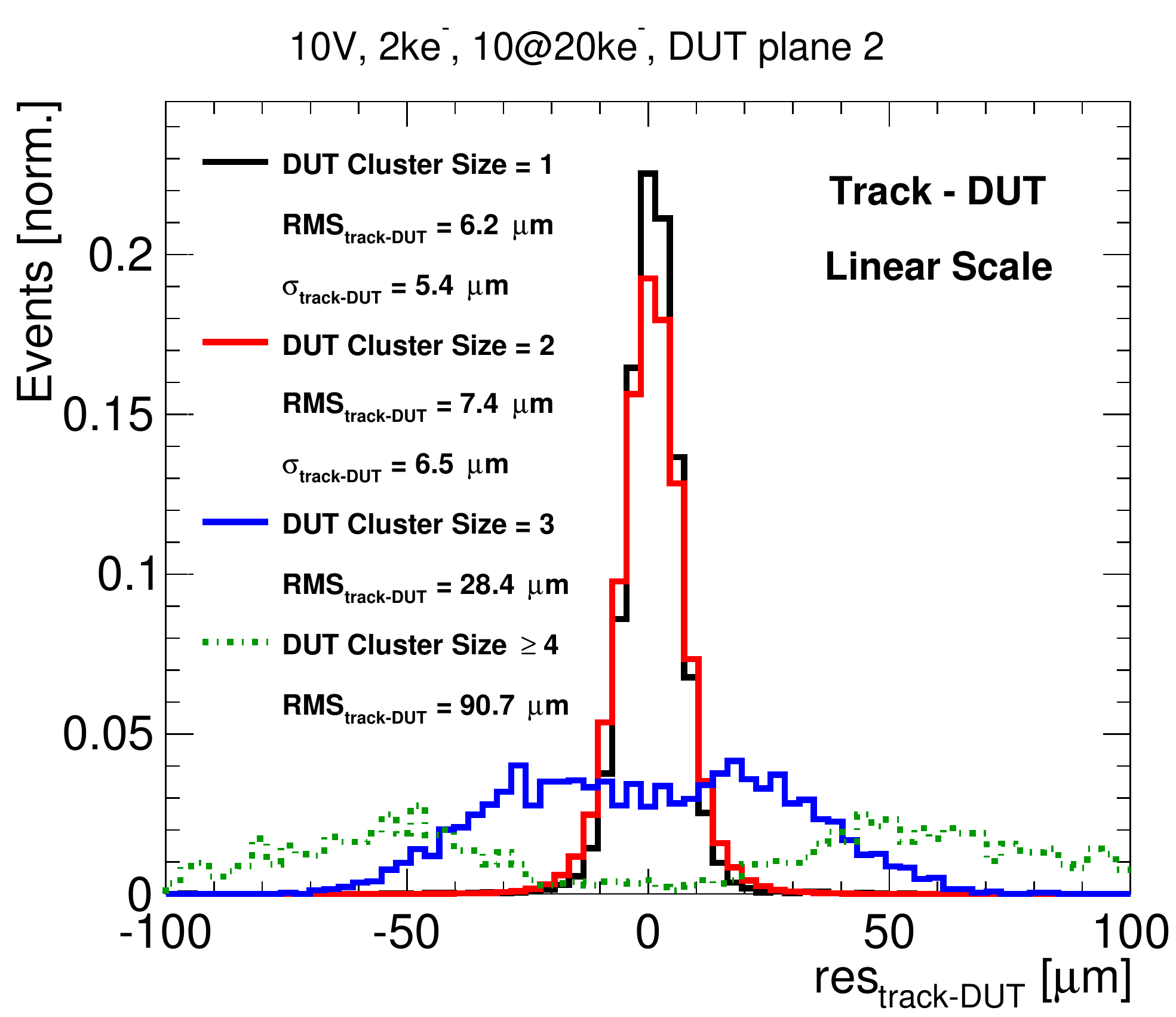}
	\caption{The residual distributions to determine the spatial resolution. Top: The triplet residual $res_{trip}$ using planes 1, 2 and 3, for all cluster sizes and cluster size $\leq2$, in linear (left) and logarithmic (right) scale. Bottom: The track-DUT residual $res_{track-DUT}$ for plane 2 as a DUT, for all cluster sizes and cluster size $\leq2$ of both tracking and DUT planes (left); and for different DUT cluster sizes, whereas the tracking-plane cluster sizes were not restricted (right). All distributions are normalised to unit area. In all cases the standard operational parameters and charge-weighted algorithm were used.}
	\label{fig:residuals}
\end{figure}

Figure~\ref{fig:residuals} shows the residual distributions for the triplet method involving planes 1, 2 and 3 (top) and for the track-DUT method with the central plane 2 as DUT (bottom). Default operational parameters (10\,V, 2\,ke$^-$, 10@20\,ke$^-$) and the charge-weighted cluster centre were taken here. Only events with one track and one cluster per plane were taken into account to eliminate combinatorial background. Each distribution is shown both including all cluster sizes and restricted to cluster sizes of $\leq2$ (96\% of all clusters). It can be seen that including all cluster sizes, non-Gaussian tails are present with an RMS significantly larger than the $\sigma$ of a Gaussian fit. However, these tails are significantly reduced for cluster sizes of $\leq2$ and the RMS approaches the $\sigma$ within maximally 0.9\,$\mu$m. In contrast, the Gaussian $\sigma$ is not strongly affected by the restriction to cluster sizes of $\leq2$. 

More details of this effect can be seen in figure~\ref{fig:residuals} (bottom right) that shows the track-DUT residual distribution separately for different cluster sizes in the DUT (whereas the cluster size of the planes included in the track fit is not restricted). The residuals of DUT cluster size 1 and 2 have a narrow, almost Gaussian peak, whereas for 3 and $\geq4$ a broad double-peak structure is obtained. As discussed in section~\ref{sec:clustering}, a cluster size of 3 is in fact geometrically possible at 14$^\circ$ (although suppressed by the threshold effect). But on the one hand probably the charge-weighted cluster centre is not ideal in this case, and on the other hand the sample of cluster size 3 is expected to also contain events with originally lower cluster size, but with delta rays that travel inside the sensor from one pixel to another and artificially enlarge the cluster size and shift the centre of the cluster (as illustrated in figure~\ref{fig:clusterSketch} right). In fact clusters with a size of at least 4 are geometrically not possible at 14$^\circ$ and are hence expected to be completely dominated by this effect. 

Since 96\% of the clusters in one plane have cluster sizes of $\leq2$, it is simple and efficient to either reject events with larger cluster sizes or to down-weight them in the track fit. Hence, the $\sigma$ of cluster sizes of $\leq2$ will be taken as the figure of merit for the resolution in the following.

For the standard tuning and the charge-weighted cluster-centre algorithm, a single-plane resolution of $\sigma_{SP,trip}=5.6 \pm 0.5\,\mu$m was obtained for  the triplet of planes 1, 2 and 3. The results for other possible combinations of planes (0, 1, 2 and 2, 3, 4) agreed within $0.5\,\mu$m. Hence, any difference in resolution between different planes was not much larger than this level. The statistical uncertainties were 0.03\,$\mu$m only; conservative systematic uncertainties assigned include the $0.5\,\mu$m difference between triplets, $0.1\,\mu$m from a variation of fit ranges (full range vs. restriction to $\pm 3 \sigma$) and the $0.1\,\mu$m multiple-scattering effect estimated from simulation as discussed above.

The convoluted track-DUT resolution for the central plane 2 as DUT was found to be $\sigma_{track-DUT}=6.2 \pm 0.6\,\mu$m. Disentangling both as described above (assuming the resolution of the four-plane system to be half of the single plane) gives $\sigma_{SP,DUT}=5.6 \pm 0.5\,\mu$m (consistent with $\sigma_{SP,trip}$ from the triplet method) and $\sigma_{track}=2.8 \pm 0.5\,\mu$m. Again, the statistical uncertainties were maximally 0.03\,$\mu$m only. For $\sigma_{SP,DUT}$, the systematic uncertainty to account for plane-to-plane variations was taken from the triplet method above (such plane-to-plane variations could not be tested directly for the track-DUT method since taking another, \ie non-central, plane as DUT would change the relation between the single-plane and the track resolution). This was then propagated as well to the systematic uncertainties of $\sigma_{track-DUT}$ and $\sigma_{track}$. The latter includes in addition the conservative 0.4\,$\mu$m systematic uncertainty for multiple-scattering effects as explained above.


\begin{table}[htb]
			\centering
			\caption{The spatial resolutions: the measured convolution of track and DUT resolution ($\sigma_{track-DUT}$); 
			after disentangling the track and single-plane DUT contributions $\sigma_{track}$ and $\sigma_{SP,DUT}$ assuming $\sigma_{track}$ to be half of $\sigma_{SP,DUT}$; and the single-plane resolution $\sigma_{SP,trip}$ obtained from the triplet of planes 1, 2 and 3. The uncertainties are dominated by systematic effects like plane-to-plane variations and multiple scattering as explained in the text. The resolution is compared for different algorithms, voltages and tuning points. If not stated otherwise, the ToT-weighted algorithm and standard operational parameters were used. The values are for cluster size $\leq2$.}
			\label{tab:spatialResolution}
			\begin{tabular}{|l|c|c|c|c|}
			
\cline{2-5}																	
\multicolumn{1}{l}{}	&	\multicolumn{4}{|c|}{Resolution [$\mu$m]}															\\
\hline																	
Variation	&	$\sigma_{track-DUT}$			&	$\sigma_{track}$			&	$\sigma_{SP,DUT}$			&	$\sigma_{SP,trip}$			\\
\hline																	
\hline																	
\multicolumn{5}{|c|}{Different Algorithms}																	\\
\hline																	
Charge-weighted	&	6.2	$\pm$	0.6	&	2.8	$\pm$	0.5	&	5.6	$\pm$	0.5	&	5.6	$\pm$	0.5	\\
ToT-weighted	&	6.5	$\pm$	0.6	&	2.9	$\pm$	0.5	&	5.8	$\pm$	0.5	&	5.9	$\pm$	0.5	\\
\hline																	
\hline																	
\multicolumn{5}{|c|}{Different Voltages}																	\\
\hline																	
1 V	&	6.7	$\pm$	0.6	&	3.0	$\pm$	0.5	&	6.0	$\pm$	0.5	&	6.0	$\pm$	0.5	\\
2 V	&	6.6	$\pm$	0.6	&	2.9	$\pm$	0.5	&	5.9	$\pm$	0.5	&	5.9	$\pm$	0.5	\\
4 V	&	6.5	$\pm$	0.6	&	2.9	$\pm$	0.5	&	5.8	$\pm$	0.5	&	5.8	$\pm$	0.5	\\
7 V	&	6.5	$\pm$	0.6	&	2.9	$\pm$	0.5	&	5.8	$\pm$	0.5	&	5.9	$\pm$	0.5	\\
10 V	&	6.5	$\pm$	0.6	&	2.9	$\pm$	0.5	&	5.8	$\pm$	0.5	&	5.9	$\pm$	0.5	\\
20 V	&	6.5	$\pm$	0.6	&	2.9	$\pm$	0.5	&	5.8	$\pm$	0.5	&	5.9	$\pm$	0.5	\\
\hline																	
\hline																	
\multicolumn{5}{|c|}{Different Thresholds}																	\\
\hline																	
1.5\,ke$^-$	&	7.4	$\pm$	0.7	&	3.3	$\pm$	0.5	&	6.7	$\pm$	0.6	&	6.3	$\pm$	0.6	\\
2.0\,ke$^-$	&	6.5	$\pm$	0.6	&	2.9	$\pm$	0.5	&	5.8	$\pm$	0.5	&	5.9	$\pm$	0.5	\\
2.5\,ke$^-$	&	6.3	$\pm$	0.6	&	2.8	$\pm$	0.5	&	5.7	$\pm$	0.5	&	5.8	$\pm$	0.5	\\
3.0\,ke$^-$	&	6.4	$\pm$	0.6	&	2.9	$\pm$	0.5	&	5.7	$\pm$	0.5	&	6.0	$\pm$	0.5	\\
\hline																	
\hline																	
\multicolumn{5}{|c|}{Different ToT Tunings}																	\\
\hline																	
10@16\,ke$^-$	&	6.8	$\pm$	0.6	&	3.0	$\pm$	0.5	&	6.1	$\pm$	0.5	&	5.8	$\pm$	0.5	\\
10@20\,ke$^-$	&	6.5	$\pm$	0.6	&	2.9	$\pm$	0.5	&	5.8	$\pm$	0.5	&	5.9	$\pm$	0.5	\\
5@20\,ke$^-$	&	7.5	$\pm$	0.7	&	3.3	$\pm$	0.5	&	6.7	$\pm$	0.6	&	6.9	$\pm$	0.6	\\
\hline

\end{tabular} 
	\end{table}

The resolution is compared in table~\ref{tab:spatialResolution} for different cluster-centre algorithms, voltages, thresholds and ToT tunings. 

The ToT-weighted algorithm was found to only slightly degrade the resolution by about 5\%. Hence, due to its simplicity and the lack of the ToT-to-charge calibration for some of the tunings under study, the ToT-weighted algorithm was taken as default without compromising the resolution significantly. However, it should be noted again that the ToT-to-charge calibration was only obtained from a similar device and averaged for all pixels. In the future it is planned to do this for each device and pixel separately, which might improve the performance of the charge-weighted algorithm. 

No strong dependence on voltage and threshold was found. The resolutions at 10@16\,ke$^-$ and 10@20\,ke$^-$ ToT tuning points are similar and degrade by about 15\% for 5@20\,ke$^-$ as expected due to the degrading charge/ToT resolution. 
Hence, tuning at 5@20\,ke$^-$, which improves the dead time of the HitOr trigger signal as discussed in section~\ref{sec:TDAQ}, has a measurable but moderate effect on the resolution, keeping the value well within the AFP requirements.

To conclude, even for the simplest cluster-centre and track-fit algorithms without optimisation, a track resolution of $2.8 \pm 0.5\,\mu$m was found for the short pixel direction, largely surpassing the AFP requirement of 10\,$\mu$m. In the future, this might be even further improved by a more careful ToT-to-charge calibration, more optimised cluster-centre algorithms such as involving the $\eta$ correction or neural networks~\cite{bib:ATLASPixelNN} as it is done for the ATLAS pixel detector, and more advanced track-fit and alignment methods, \eg taking into account multiple scattering. In addition, in future AFP tracker versions one might consider to reduce the tilt slightly from 14$^\circ$ to 12--13$^\circ$ in order to enhance the fraction of cluster sizes of $\leq 2$ above the present 96\% per plane.

\subsubsection{Tracker hit efficiency}
\label{sec:pixelEff}

\begin{figure}[bt]
	\centering
	\includegraphics[width=7.5cm]{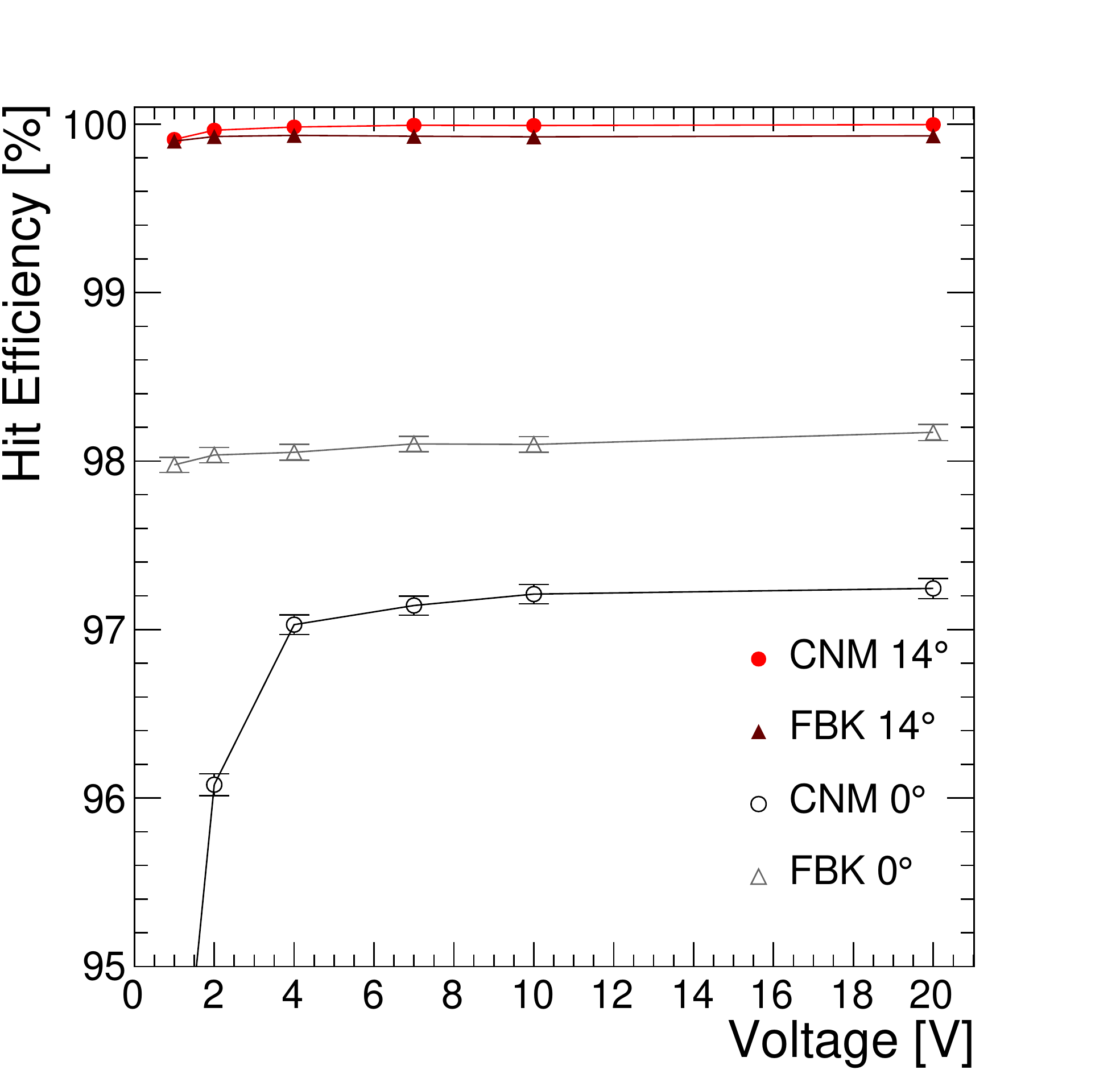}
	\includegraphics[width=7.5cm]{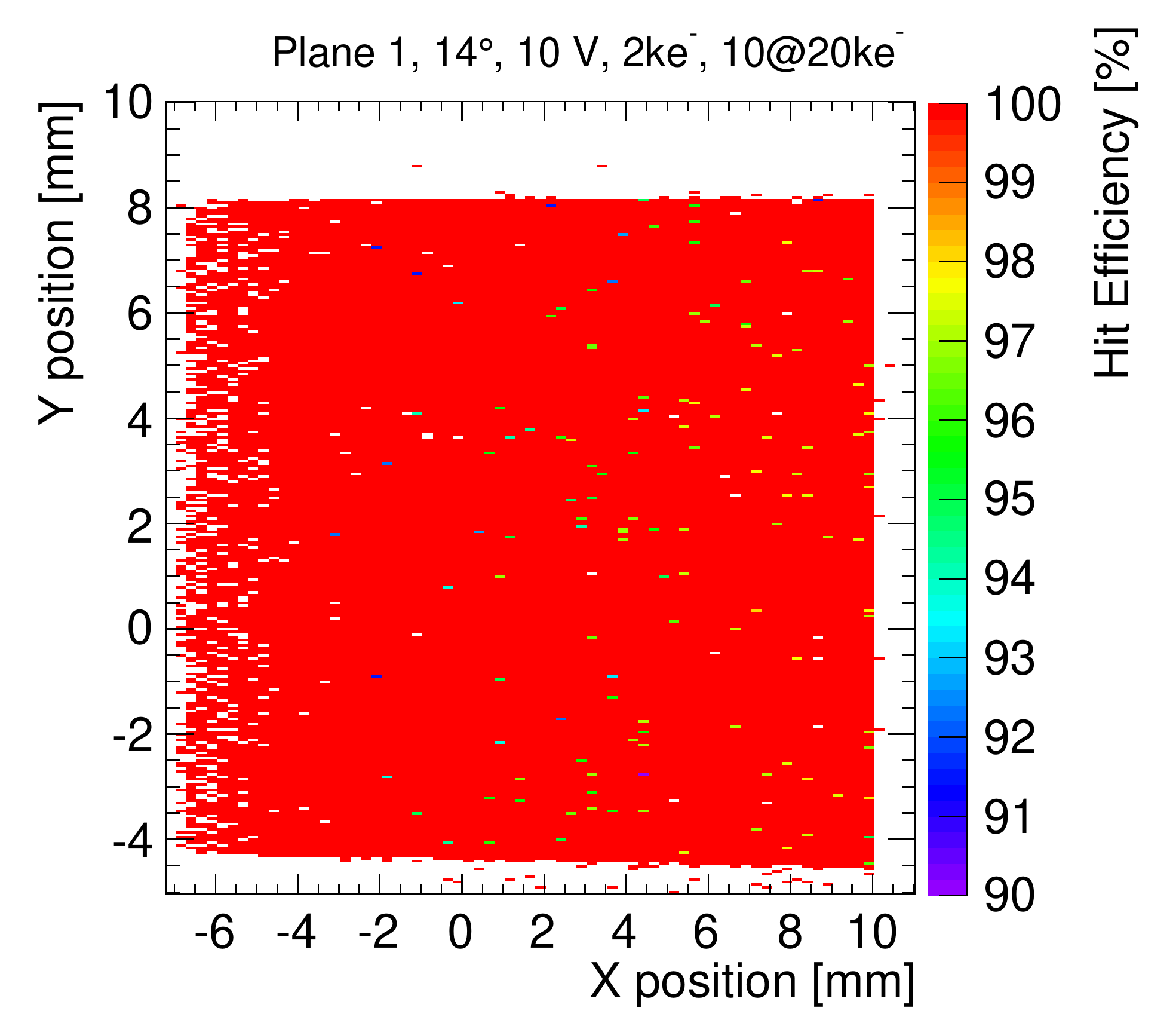}
	\caption{Left: Pixel hit efficiencies as a function of bias voltage for planes 1 (CNM) and 2 (FBK) at 0$^\circ$ and 14$^\circ$. Right: Hit efficiency map for plane 1 at 14$^\circ$ and 10\,V.}
	\label{fig:pixelEfficiency}
\end{figure}

Using the DUT track-reconstruction scenarios, the per-plane hit efficiencies were determined. This was done for the unbiased pixel planes 1 and 2 that were not used for triggering. The DUT was excluded from the track fit and it was checked for each reconstructed track if a hit in the DUT was found close to the track. 

Figure~\ref{fig:pixelEfficiency} (left) shows the hit efficiencies of planes 1 (CNM) and 2 (FBK) at 0$^\circ$ and 14$^\circ$ beam incidence as a function of bias voltage for the default tuning parameters. At 0$^\circ$, the CNM device reaches a plateau at 4\,V, whereas the FBK device reaches its maximum efficiency already at 1\,V. The difference is due to the non-passing-through 3D columns in the CNM device, which needs a slightly higher voltage to reach full lateral depletion. The plateau efficiencies of 97--98\% are reasonable for an IBL-spare quality device (but are lower than for a good-quality IBL device with $>$99\%~\cite{bib:IBLprototypes}). At the AFP tilt of 14$^\circ$, however, the efficiencies increase to $>$99.9\% already from 1\,V for both devices, despite the quality class used here. The efficiency improvement for a tilted device is a well-known effect in 3D sensors which exhibit small localised low-efficiency regions at perpendicular incidence due to the insensitivity of the 3D columns and some low-field regions in between~\cite{bib:IBLprototypes}. Figure~\ref{fig:pixelEfficiency} (right) displays the 3D efficiency map for plane 1 at default operational parameters and a tilt of 14$^\circ$. It can be seen that the high efficiency is uniform over the whole device, apart from single pixels with slightly lower efficiency (white pixels are either masked or regions without entries at the edge due to the beam profile or the lack of overlap with the trigger area).
The efficiency results are found to be insensitive to the threshold and ToT tunings within the range studied here.

\subsection{Performance of the time-of-flight system}

In this section, the performance of the LQbar time-of-flight (ToF) system is presented. The results are from the 2014 data only with the standard setup as described in section~\ref{sec:prototypeToF}. The 2015 data with some setup variations and more detailed studies of the ToF properties are still being analysed. 

The common data format of tracking and timing detectors allowed the event-by-event use of track information to predict independently whether a certain LQbar was traversed by a particle, which was crucial for the studies of efficiency, cross talk, noise and time-resolution. For this, well-reconstructed tracks from the all-plane track-reconstruction scenario after cleaning cuts were used as described in section~\ref{sec:trackReco2}, giving the best track precision at the position of the timing system. 

\subsubsection{LQbar hit efficiency}

\begin{figure}[bt]
	\centering
	\includegraphics[width=7.5cm]{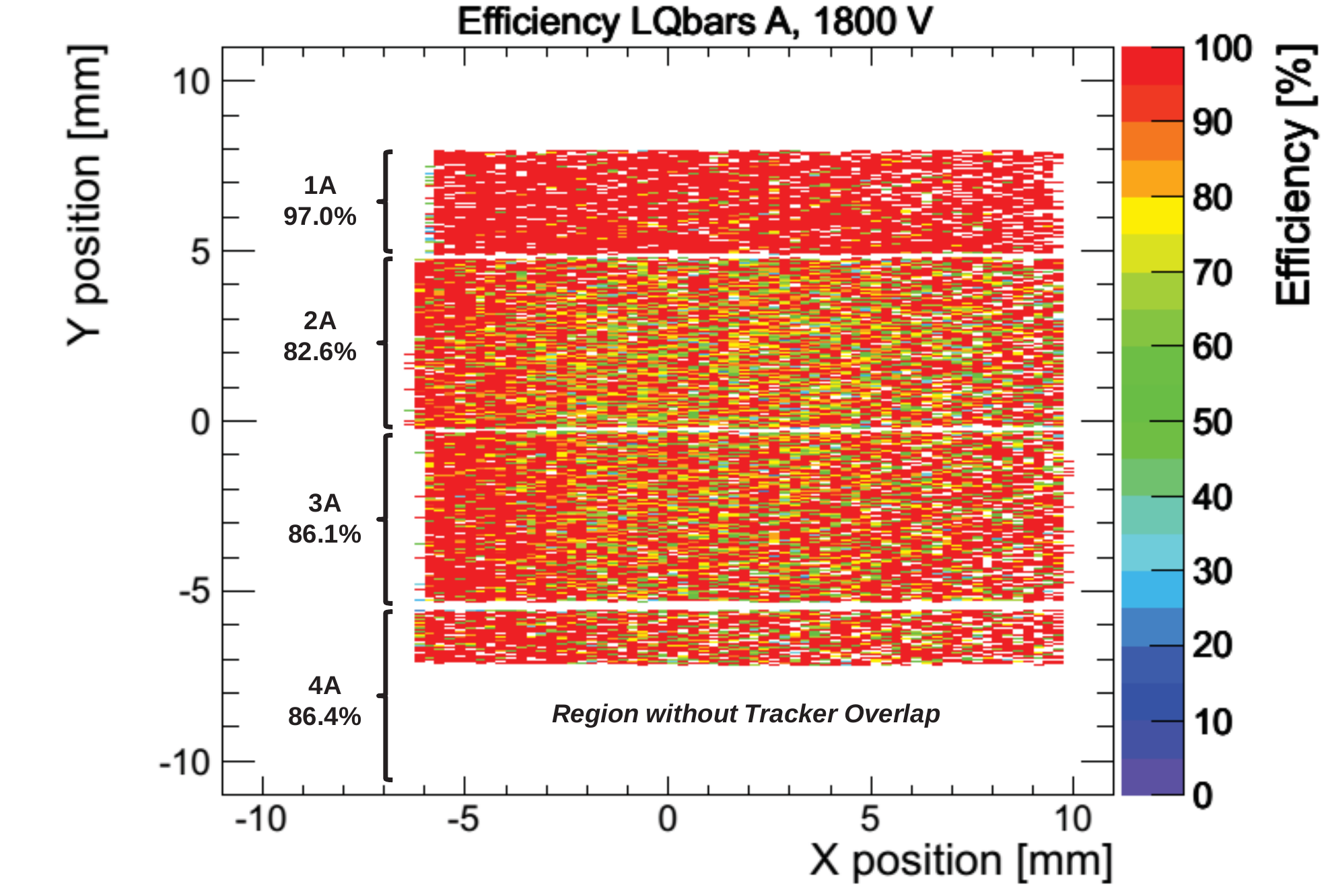}
	\includegraphics[width=7.5cm]{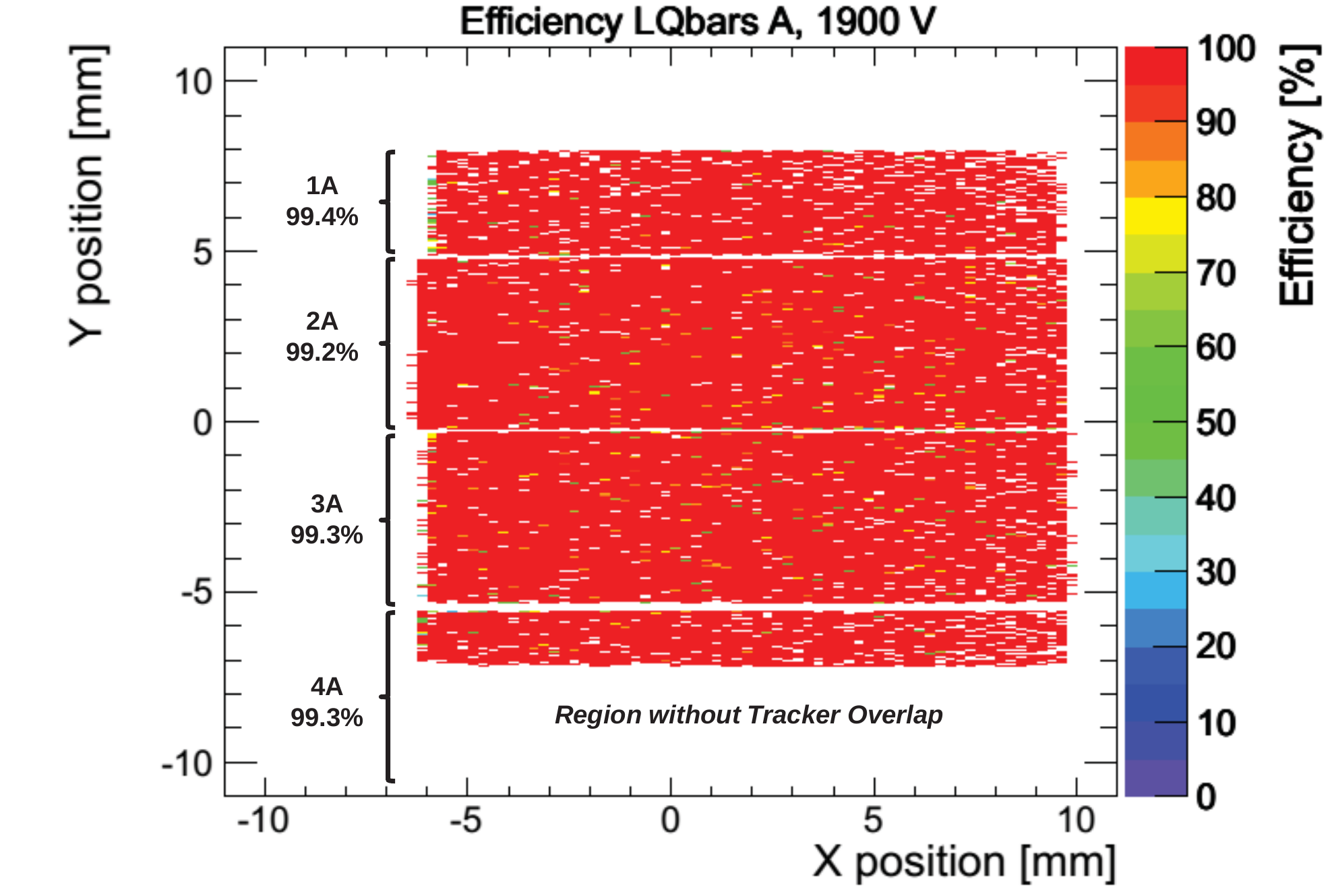}
	\includegraphics[width=7.5cm]{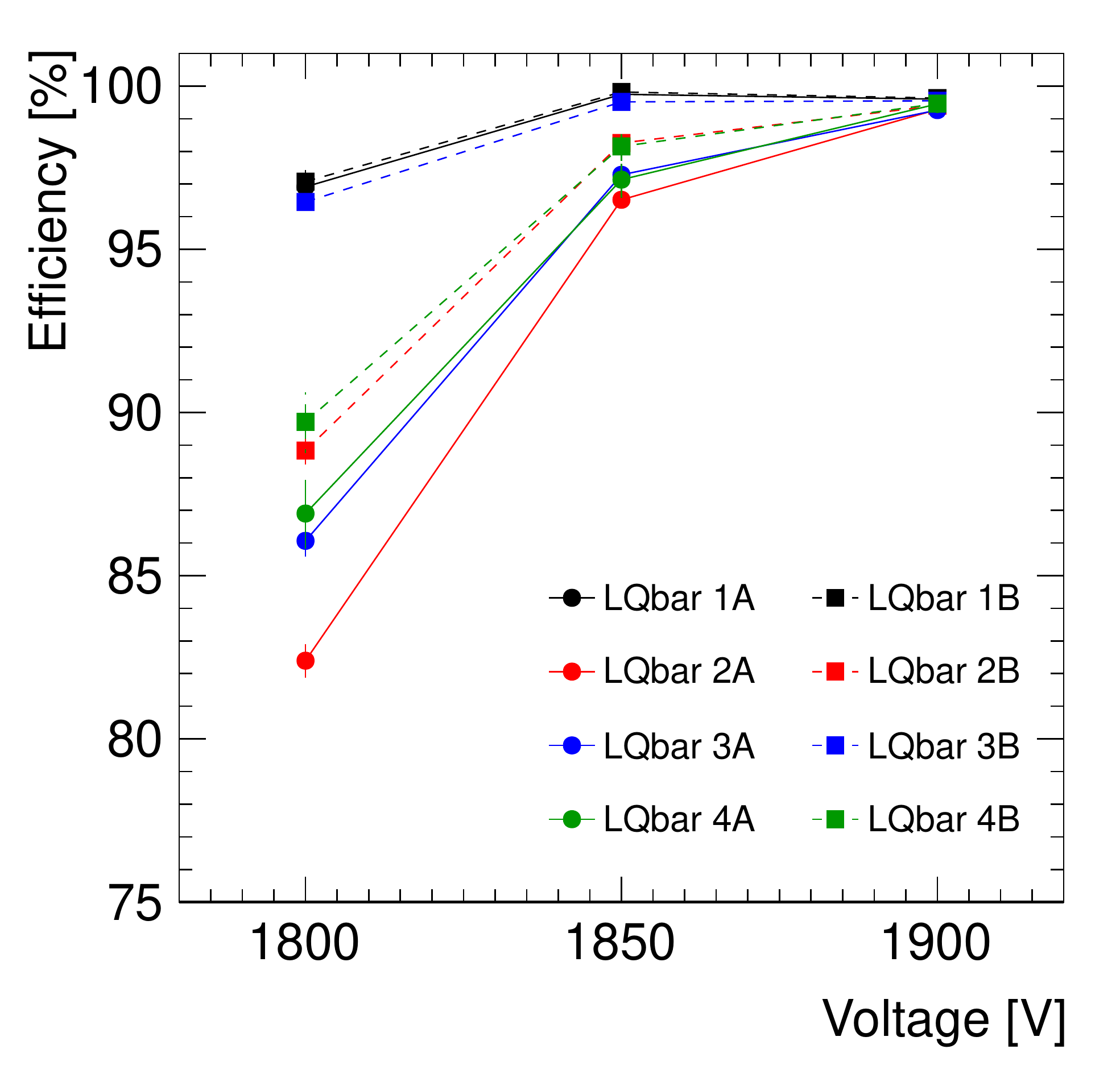}
	\caption{The hit efficiencies of the LQbar ToF channels. Top: The efficiency map as a function of reconstructed track position for the first bar (A) of each train at $V_{MCP-PMT}=1800$\,V (left) and 1900\,V (right). Bottom: The mean efficiencies as a function of $V_{MCP-PMT}$ for all eight bars.}
	\label{fig:TimingEff}
\end{figure}

The hit efficiency of each LQbar was determined using events with tracks passing through the bar of interest and determining the fraction of those events in which this bar gave a signal. Figure~\ref{fig:TimingEff} (top) shows the two-dimensional hit-efficiency map as a function of reconstructed track position for the first bar (A) of each train at $V_{MCP-PMT}=1800$\,V (left) and 1900\,V (right), for the fixed CFD threshold of 100\,mV. Figure~\ref{fig:TimingEff} (bottom) displays the corresponding mean efficiencies as a function of $V_{MCP-PMT}$ for all the bars. It can be seen that the LQbar efficiencies generally increase with $V_{MCP-PMT}$ as expected as the MCP-PMT gain increases. Whereas at 1800\,V, there is a significant spread between the efficiencies of different bars ranging from 83 to 97\%, the efficiencies at 1900\,V are all above 99\%. Another interesting observation at 1800\,V is a slightly higher efficiency close to the cut edge near $x=-6$\,mm compared to areas further away from it. 
This effect was considered in the design, and gives an enhanced light yield, ranging from a few \% at 5\,mm from the edge of the bar to a factor 2 at the edge of the bar where the whole Cherenkov light cone is detected~\cite{bib:qsim}.

\subsubsection{Cross talk between LQbar trains}
Another important ToF-detector parameter is the cross talk between bars of different trains. A high level of inter-train cross talk is disadvantageous for the use of the ToF detector as a position-resolved trigger, as well as for the operation at high pile-up and hence occupancies where there might be several trains hit by particles in one bunch crossing. However, for low occupancies like in the beam test, cross talk was found to have no large influence on the time resolution if the bars really hit by the particle can be selected using track information, as demonstrated in section~\ref{sec:timeRes} below.

The cross talk seen by a bar of interest was determined as the fraction of events in which it gives a signal when the track actually passes through a bar in another train (of the same LQbar column A or B).

\begin{figure}[bt]
	\centering
	\includegraphics[width=7.5cm]{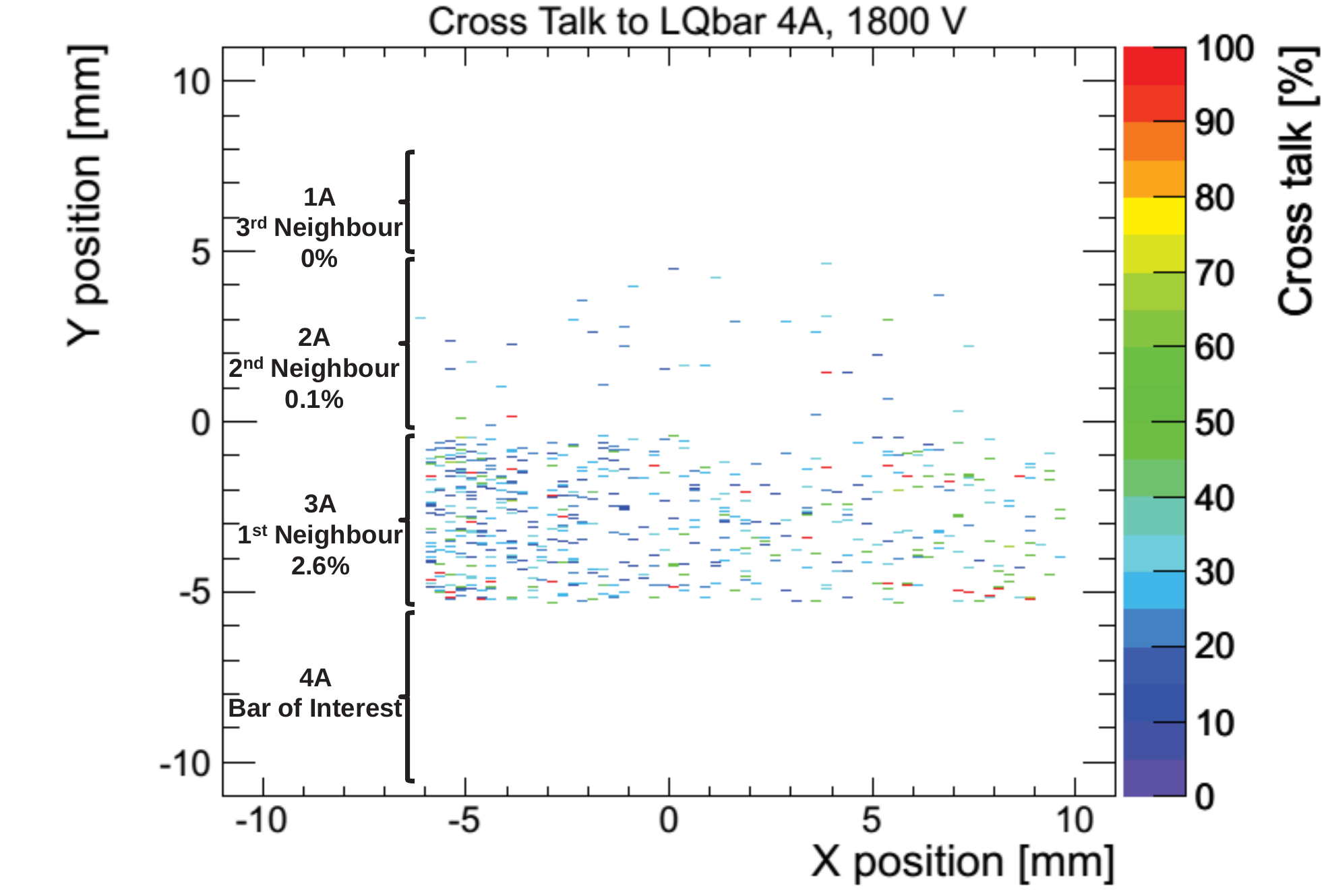}
	\includegraphics[width=7.5cm]{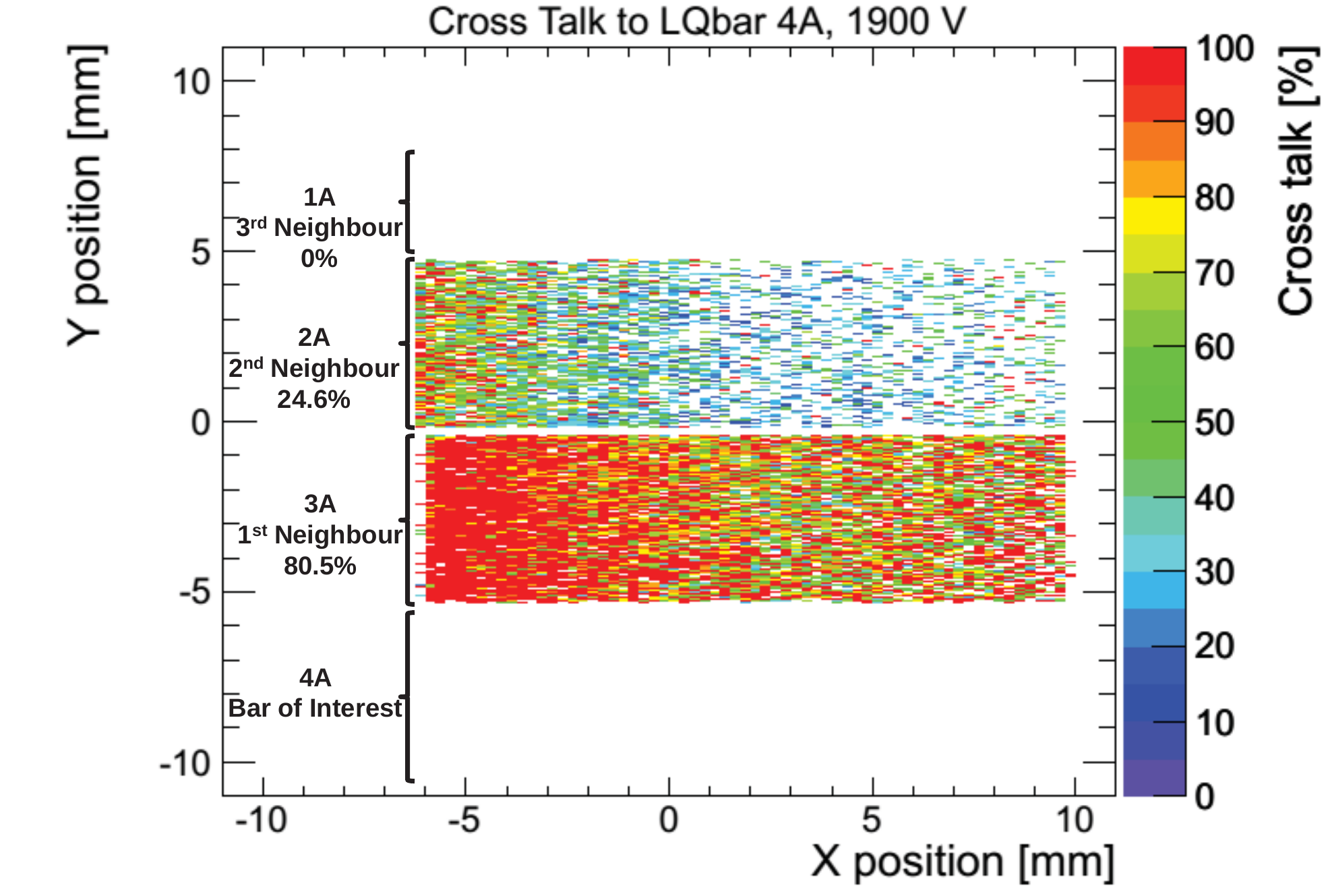}
	\includegraphics[width=7.5cm]{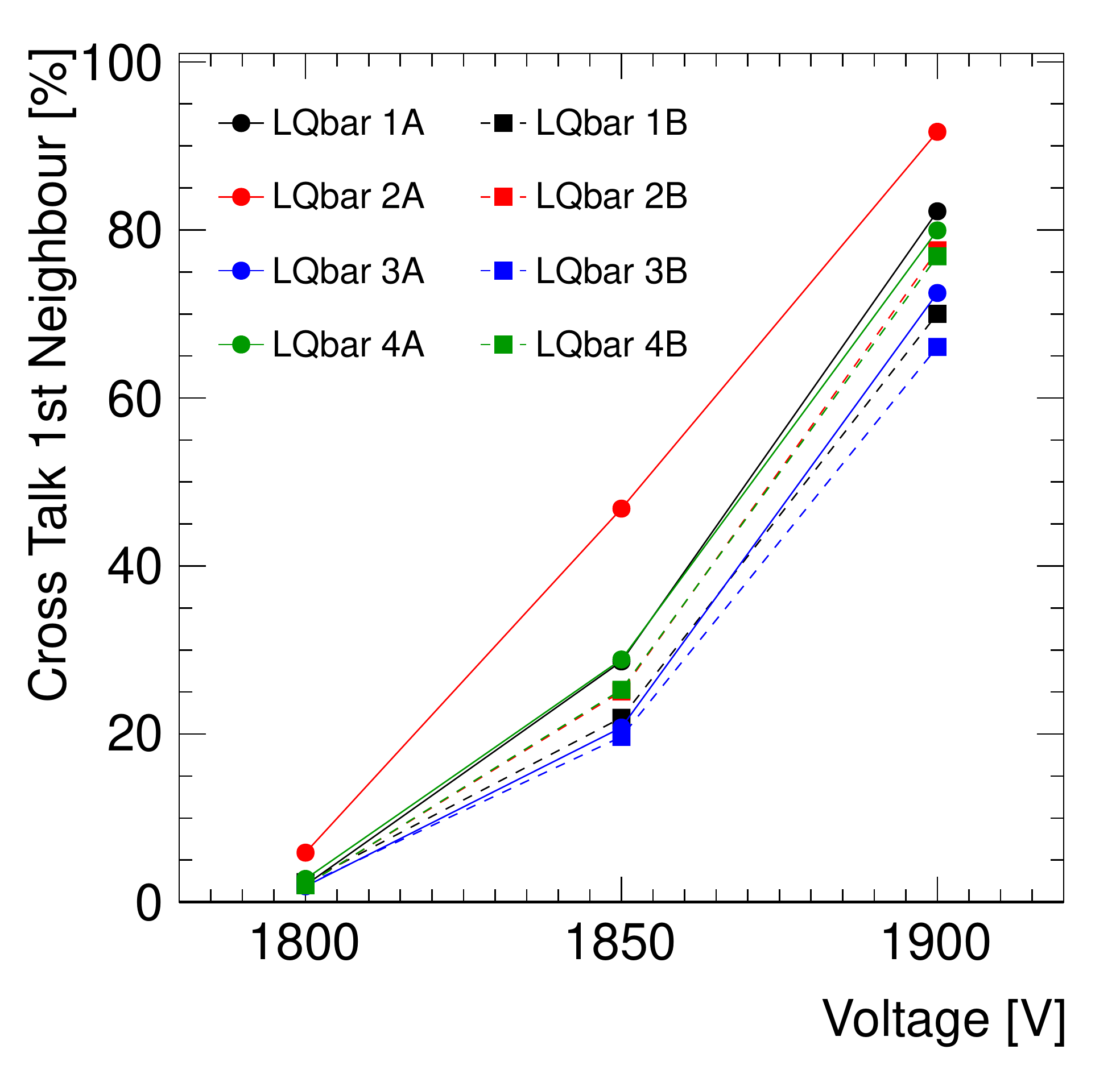}
	\includegraphics[width=7.5cm]{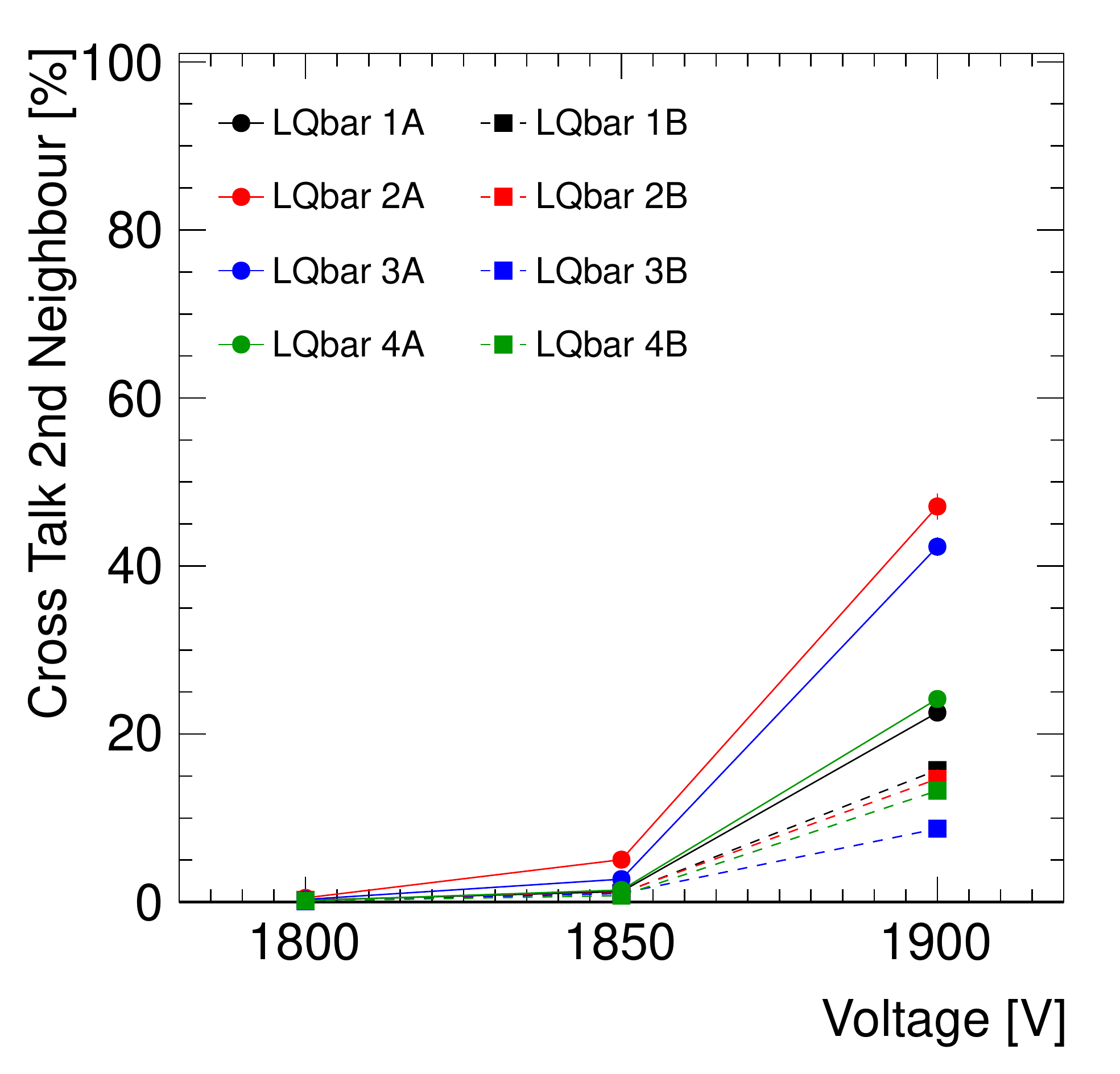}
	\caption{The cross talk between LQbars in neighbouring trains. Top: the map of cross-talk seen by LQbar 4A as a function of track position at $V_{MCP-PMT}=1800$\,V (left) and 1900\,V (right). Bottom: the mean cross talk from bars in directly neighbouring (left) and next-to-directly-neighbouring trains (right) as a function of $V_{MCP-PMT}$ for all eight bars.}
	\label{fig:TimingCrossTalk}
\end{figure}

Figure~\ref{fig:TimingCrossTalk} (top) shows the two-dimensional map of cross talk to LQbar 4A as a function of track position (for tracks passing through other trains) at $V_{MCP-PMT}=1800$\,V (left) and 1900\,V (right), for the fixed CFD threshold of 100\,mV. In figure~\ref{fig:TimingCrossTalk} (bottom) the mean cross talk from the first (left) and second neighbours (right) is displayed for all bars as a function of $V_{MCP-PMT}$. Whereas the cross talk from the first neighbours is at the few-percent level at 1800\,V, it increases steeply up to 66--92\% at 1900\,V. Cross talk from the second neighbour is maximally at the few-percent level up to 1850\,V and increases to 8--47\% at 1900\,V. Cross talk from the third neighbour is not observed. Similar to the efficiency, the cross talk is observed to be higher near the LQbar cut edge at $x=-6$\,mm.

As the LQbars themselves were optically well isolated between adjacent trains with mylar foils at the radiator level and Aluminium at the light-guide level as explained above, most of the cross talk has to originate from the MCP-PMT level. 
Possible explanations include optical leakage at or in the photo-cathode window and the lateral spread of the photo electrons in the MCP-PMT. The level of cross talk also depends on the CFD threshold settings. It is believed that it can be significantly reduced with little loss of efficiency by raising the CFD thresholds to somewhat higher than the current 100\,mV, as the cross-talk pulse height in a neighbouring LQbar is significantly smaller than the signal pulse height of a hit LQbar. Further studies to understand its origin and to optimise it are envisaged.

\subsubsection{LQbar noise}

\begin{table}[h]
			\centering
			\caption{Noise rate per LQbar channel for different voltages.}
			\label{tab:noiserate}
			\begin{tabular}{|c|c|c|c|c|c|c|c|c|}
			
			\hline		
		LQbar Channel & 1A & 1B & 2A & 2B & 3A & 3B & 4A & 4B \\ 
		\hline
		$V_{MCP-PMT}$     & \multicolumn{8}{|c|}{Noise Rate [kHz]}	\\
			
			\hline 
			1800\,V & 0 & 3 & 45 & 7 & 10 & 16 & 29 & 28 \\ 
			\hline 
			1850\,V & 0 & 14 & 57 & 19 & 22 & 22 & 35 & 35 \\ 
			\hline 
			1900\,V & 0 & 20 & 63 & 29 & 47 & 49 & 46 & 47 \\ 
			\hline 
			\end{tabular} 
	\end{table}

The noise rates of the LQbar channels were measured as their mean signal firing rates in events in which the track missed any LQbar ($x<-7$\,mm). It was measured to be at the level of 0 to 63\,kHz for the fixed CFD threshold of 100\,mV, with a large variation between different LQbar channels (which is still under investigation) and increasing with $V_{MCP-PMT}$ as shown in table~\ref{tab:noiserate}. This corresponds to noise occupancies in the order of 10$^{-4}$ to 10$^{-3}$ for a 25\,ns window.

\subsubsection{Time resolution}
\label{sec:timeRes}

\begin{figure}[tbh]
	\centering
	\includegraphics[width=7.5cm]{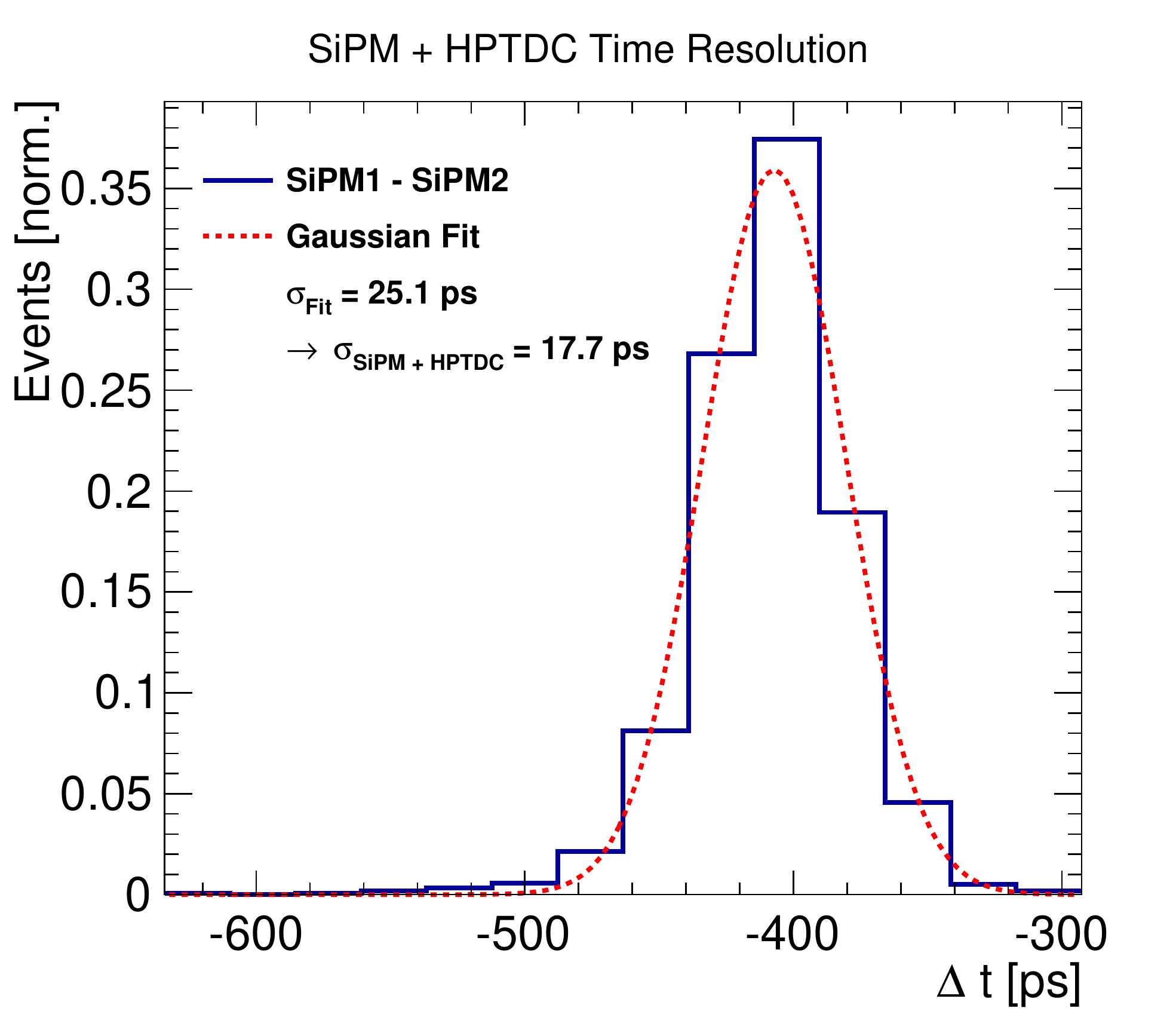}
	\includegraphics[width=7.5cm]{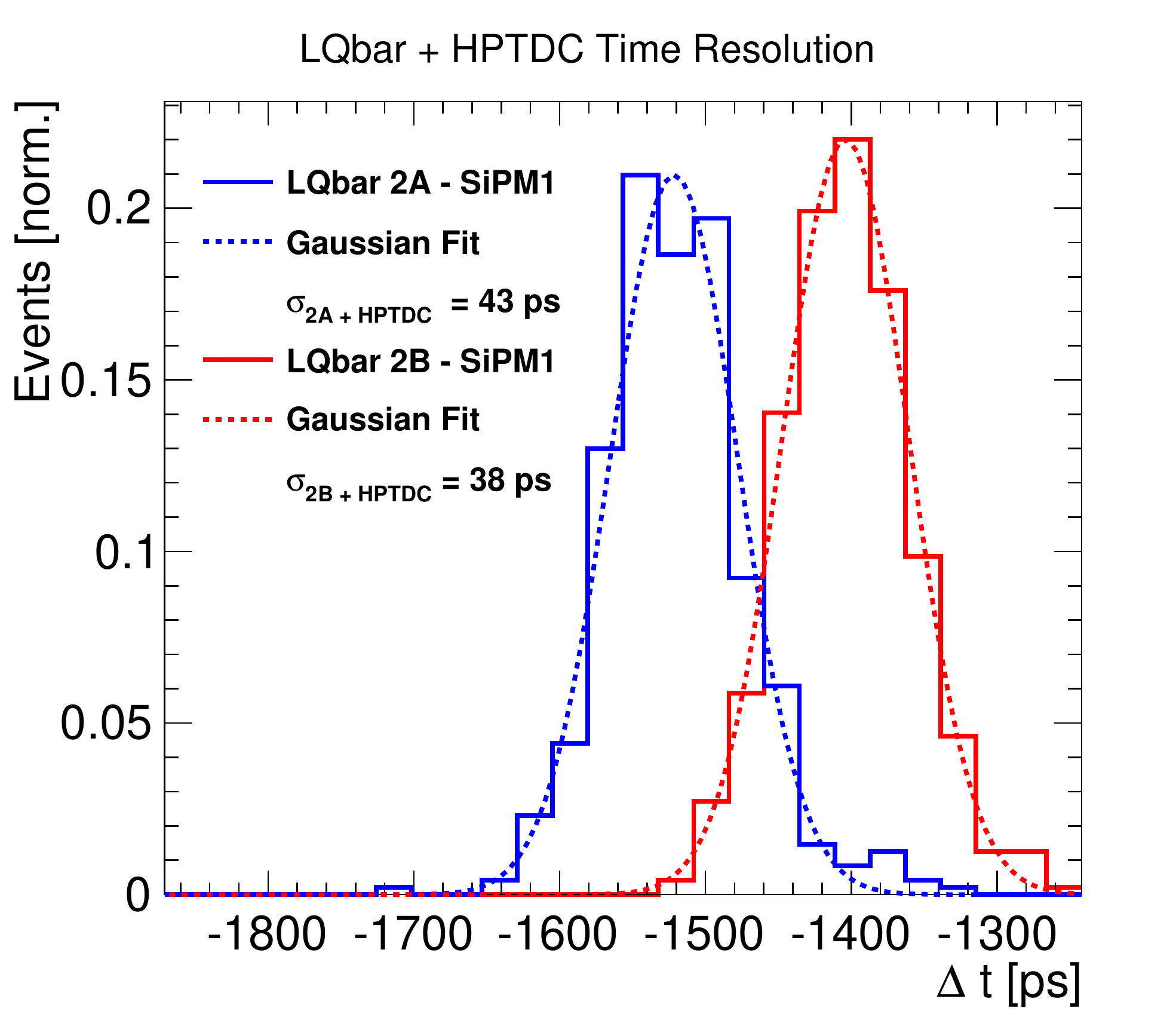}
	\caption{Time differences measured with the HPTDC-RCE readout. Left: the time differences between the two SiPM references with total convoluted resolution $\sigma_{fit}$ and after dividing by $\sqrt{2}$, $\sigma_{SiPM+HPTDC}$. Right: the time differences between the LQbars of the second train (2A and 2B) and SiPM1 at $V_{MCP-PMT}=1900$\,V; the displayed fitted resolution values here have already the SiPM+HPTDC resolution quadratically subtracted. }
	\label{fig:TimingResolutionRCE}
\end{figure}

\begin{table}[tbh]
\caption{The time resolutions of different LQbars and the train average for different $V_{MCP-PMT}$ measured with the HPTDC-RCE readout (\ie including the HPTDC contribution) with respect to the SiPM reference. $\sigma_{SiPM+HPTDC}$ was subtracted. The uncertainties include statistical and systematic uncertainties as described in the text.}
	\centering
	\small
		\begin{tabular}{|l|r|r|r|r|}
		\hline		
		$V_{MCP-PMT}$ [V] &	1750 &	1800 & 1850 &	1900 \\
		\hline
		LQbar     & \multicolumn{4}{|c|}{Time resolution $\sigma_{LQbar+HPTDC}$ [ps]}	\\
		\hline
		\hline
		1A			       &  $78\pm5$	 & $61\pm6$	   & $52\pm6$	  & $46\pm5$ \\
		1B				     & $85\pm6$	 & $60\pm6$	   & $47\pm6$	  & $41\pm6$ \\ 
		Average Train 1& $67\pm7$   & $54\pm12$    & $44\pm6$   & $37\pm6$ \\
		\hline
		2A				      & $94\pm5$    & $80\pm10$	 & $50\pm6$	   & $43\pm7$  \\
		2B				      & $94\pm8$    & $64\pm5$	 & $45\pm6$    & $38\pm6$ \\
		Average Train 2 & $77\pm7$    & $63\pm7$  & $41\pm6$    & $35\pm6$ \\		
		\hline
		\end{tabular}
	
	\label{tab:timeResRCE}
\end{table}

The HPTDC board digitised the arrival time of the rising edge of the CFD output for each LQbar and SiPM reference channel (only the two SiPMs 1 and 2 were included during these tests). This time information was recorded for each event in the RCE output file. 
Time resolutions were measured from the spread of the time differences between two timing channels. For this, it was always required that a well-reconstructed track passed through the overlap region of the sensors related to these channels. The $3\times3$\,mm$^2$ SiPMs were placed between -0.5 to 2.5\,mm in $x$, \ie their centres were about 7\,mm away from the LQbar cut edge, and between 3.5 to 6.5\,mm in $y$, \ie they had an overlap with the LQbar trains 1 and 2 (see figure~\ref{fig:trackingTimingCorr}).

The first step is to understand the time resolutions of the SiPM reference devices. Figure~\ref{fig:TimingResolutionRCE} (left) shows the time difference between SiPM1 and SiPM2. It is approximately Gaussian distributed and has a total convoluted width of $\sigma_{fit}=25.1$\,ps, including the HPTDC contributions of the two SiPM channels, which have been on the same HPTDC chip. This value appeared to be stable for many different runs within less than 1\,ps. Assuming an equal performance of both devices, the resolution of a single SiPM+HPTDC device can be obtained from dividing by $\sqrt{2}$, which gives $\sigma_{SiPM+HPTDC}=17.7$\,ps. A similar analysis has been performed with the oscilloscope instead of the HPTDC, which gave a value of $\sigma_{SiPM}=11.0$\,ps (the contribution of the oscilloscope is considered to be negligible). Thus, the HPTDC resolution can be estimated from quadratic subtraction as $\sigma_{HPTDC}=13.9$\,ps, which is in good agreement with laboratory and previous beam-test measurements. 
The HPTDC contribution might slightly depend on the exact HPTDC channel combinations, which has not been taken into account at this beam test, but will be studied in the future. 

The LQbar time resolutions were measured from the time differences between one bar and one of the SiPM references, respectively. The SiPM and the LQbar channels were connected to different HPTDC chips. This measurement could only be done for the trains 1 and 2 which had an overlap with the SiPMs. The time differences are shown in figure~\ref{fig:TimingResolutionRCE} (right) for the LQbars 2A and 2B with respect to SiPM1 at $V_{MCP-PMT}=1900$\,V. Measurements using SiPM2 are consistent within a few~ps. The resolution values from a Gaussian fit shown there have already the reference resolution $\sigma_{SiPM+HPTDC}=17.7$\,ps as obtained above quadratically subtracted. 

The LQbar resolutions, including the HPTDC contributions, are listed in table~\ref{tab:timeResRCE} for different $V_{MCP-PMT}$ values. The resolution was found to improve with $V_{MCP-PMT}$ and at 1900\,V it reached values between 38 and 46\,ps for the single LQbars. Statistical and fit uncertainties are estimated to be 2\,ps, and in addition systematic uncertainties of typically 6\,ps have been assigned to account for the differences between the two SiPM references as well as for run-to-run and selection-cut variations. 

Also included in table~\ref{tab:timeResRCE} are the results for the resolution of the average time of the two LQbars in one train (A and B). The train average improves the resolution with respect to the best single-bar measurement, \eg to $35\pm6$\,ps for train 2 at 1900\,V, consistent with the required design value of 30\,ps for initial low-luminosity AFP runs. In general, however, the observed improvement is less than expected for uncorrelated measurements (\eg for two bars of the same resolution an improvement by $1/\sqrt{2}$ is expected). Correlation between the bars of the same train has been observed before in previous beam tests and can originate in optical leakage at the photo-cathode window or in the lateral spread of the photo electrons in the MCP-PMT, similarly as for the inter-train cross talk. 
Further detailed studies of the correlations and of the dependence of the time resolution on the number of LQbars are envisaged, with the aim of optimising the LQbar configurations.

\section{Conclusions and outlook}
\label{sec:conclusions}
Beam tests with a first unified AFP prototype detector combining pixel tracking and LQbar ToF sub-detectors and a common readout and data format have been performed at the CERN-SPS in November 2014 and September 2015. The successful tracking-timing integration was demonstrated by the spatial correlation of recorded tracking and timing data. A TDAQ system close to the final AFP-ATLAS readout based on a track trigger was successfully tested.

Moreover, the performances of the tracking and ToF systems were studied. The tracker hit efficiency was found to be $>99.9$\% per plane at a tilt of 14$^{\circ}$ like foreseen for the final AFP detector. The spatial resolution in the short 50\,$\mu$m pixel direction at 14$^{\circ}$ was found to be $5.5 \pm 0.5\,\mu$m per pixel plane and $2.8 \pm 0.5\,\mu$m for the full four-plane tracker in the centre of the four planes. This clearly surpasses the AFP requirement of 10\,$\mu$m for the horizontal AFP direction by a factor of almost 4. Due to the discrete hit behaviour in the long 250\,$\mu$m pixel direction (corresponding to the vertical AFP direction), the resolution could not be measured in that direction with the beam-test setup here and in any case will highly depend on the actual staggering achieved in the final AFP detector.

The hit efficiencies of the LQbar ToF detectors were found to increase with MCP-PMT voltage up to more than 99\% at 1900\,V. However, also the cross talk was observed to increase strongly with voltage for the CFD threshold settings used. The time resolutions of the full LQbar timing detectors including the HPTDC contributions were found to improve with voltage and were measured between $38\pm6$\,ps and $46\pm5$\,ps per LQbar and $35\pm6$\,ps and $37\pm6$\,ps per train at 1900\,V. 

Hence, despite not being fully optimised and including only two LQbars per train, a timing resolution consistent with the low-luminosity target of 30\,ps was obtained.
Improved time resolutions of 10--20\,ps required for high-luminosity operation are believed to be achievable through increasing the number of LQbars per train to the final four (space restrictions in the Roman pot prevent the installation of more), optimising the HPTDC performance, the development of a new HPTDC chip (which is on-going) and a reduction of the mini-Planacon pore size from 10 to 6\,$\mu$m. 

Parts of the ToF data in the 2015 beam test with different LQbar types and configurations are still being analysed. Moreover, in 2016 follow-up beam tests were already performed and more are planned with the final AFP detector and for a further systematic study and optimisation of the LQbar time performance such as cross talk, inter-bar correlations including the dependence of the time resolution on the number of LQbars per train (up to the final four-LQbar-train configuration), and HPTDC contribution.
Also, the properties of the track trigger will be investigated further, and the development of an alternative trigger based on the ToF system will be pursued. 

In addition, a first arm of the AFP detector with two Roman Pots including a tracker each has been successfully installed at one side of the ATLAS IP in the 2015-2016 year-end shutdown, allowing the commissioning and study of the AFP tracker performance and backgrounds with the final detector and the LHC beam.

\acknowledgments 
The authors wish to thank H.\,Wilkens, S.\,Vlachos and the CERN SPS and NA teams for valuable support during the beam tests.

The equipment and detectors were provided by the following institutes and people: the CDFs and HPTDC boards by the University of Alberta; the pre-amplifiers and the mechanical setup by Stony Brook University; the MCP-PMT by the University of Texas at Arlington and Photonis; the SiPM reference detectors by M. Albrow; the LQbars by Palacky University Olomouc; the trigger PCB and pixel modules by IFAE Barcelona and INFN Genova; the RCE systems by SLAC; the pixel holders and cables by the ATLAS ITk pixel test-beam group; the LTB by M. Backhaus; the 6\,GHz WaveRunner Oscilloscope by LeCroy.

This work was partially funded by: the MINECO, Spanish Government, under grants FPA2013-48308 and SEV-2012-0234 (Severo Ochoa excellence programme); the European Union's Horizon 2020 Research and Innovation programme under Grant Agreement no. 654168;
the Polish National Science Centre grant UMO-2012/05/B/ST2/02480;
the Czech projects "MSMT INGO II, $\check{c}$. LG15052", "MSMT $\check{c}$. LO1305 (RCPTM-NPU)" and PrF\_2016\_002 of IGA UP Olomouc; Texas ARP U.S. Dept. of Energy grants (A. Brandt); the U.S. Department of Energy grant DE-SC0007054 and the U.S. National Science Foundation grants PHY1068677 and PHY1119200; 
CERN/FIS-NUC/0005/2015, OE, FCT from Portugal;
previous beam tests by the T-958 Fermilab Test Beam experiment.

\newpage



\providecommand{\href}[2]{#2}\begingroup\raggedright\endgroup

\end{document}